\documentclass[prl,english,letterpaper,superscriptaddress,twocolumn,amsmath,amssymb]{revtex4-1}
\usepackage{times}
\usepackage{amsfonts}
\usepackage{graphicx}
\usepackage{dcolumn}
\usepackage{bm}
\usepackage{color,xcolor}

\usepackage[colorlinks,bookmarks=false,citecolor=blue,linkcolor=red,urlcolor=blue]{hyperref}
\newcommand{\bra}[1]{\langle #1 |}
\newcommand{\ket}[1]{| #1 \rangle}

\newcommand{\bee}{\begin{equation}}
\newcommand{\ee}{\end{equation}}
\newcommand{\bma}{\begin{pmatrix}}
\newcommand{\ita}{\end{pmatrix}}
\newcommand{\balig}{\begin{align}}
\newcommand{\ealig}{\end{align}}

\newcommand{\clb}{\color{blue}}

\newcommand{\bk}{\boldsymbol{k}}

\def\Re {\mbox{Re}}
\def\Im {\mbox{Im}}
\def\be{\begin{equation}}       \def\ee{\end{equation}}
\def\bea{\begin{eqnarray}}      \def\eea{\end{eqnarray}}

\begin{document}

\title{Fermion doubling theorems in 2D non-Hermitian systems for Fermi points and exceptional points}

\author{Zhesen Yang}
\affiliation{Beijing National Laboratory for Condensed Matter Physics, Institute of Physics, Chinese Academy of Sciences, Beijing 100190, China}
\affiliation{Kavli Institute for Theoretical Sciences, University of Chinese Academy of Sciences, Beijing 100190, China}

\author{A. P. Schnyder}
\affiliation{Max-Planck-Institute for Solid State Research, Heisenbergstr. 1, D-70569 Stuttgart, Germany}

\author{Jiangping Hu}
\affiliation{Beijing National Laboratory for Condensed Matter Physics, Institute of Physics, Chinese Academy of Sciences, Beijing 100190, China}
\affiliation{CAS Center of Excellence in Topological Quantum Computation and Kavli Institute of Theoretical Sciences, University of Chinese Academy of Sciences, Beijing 100190, China}
\affiliation{South Bay Interdisciplinary Science Center, Dongguan, Guangdong Province, China}

\author{Ching-Kai Chiu}\email{Corresponding: ching-kai.chiu@riken.jp}
\affiliation{Kavli Institute for Theoretical Sciences, University of Chinese Academy of Sciences, Beijing 100190, China}
\affiliation{RIKEN Interdisciplinary Theoretical and Mathematical Sciences (iTHEMS), Wako, Saitama 351-0198, Japan}

\date{\today}

\begin{abstract}
The fermion doubling theorem plays a pivotal role in Hermitian topological materials. It states, for example, that Weyl points must come in pairs in three-dimensional semimetals. Here, we present an extension of the doubling theorem to non-Hermitian lattice Hamiltonians. We focus on two-dimensional non-Hermitian systems without any symmetry constraints, which can host two different types of topological point nodes, namely, (i)~Fermi points and (ii)~exceptional points. We show that these two types of protected point nodes obey doubling theorems, which require that the point nodes come in pairs. To prove the doubling theorem for exceptional points, we introduce a generalized winding number invariant, which we call the \emph{discriminant number}. Importantly, this invariant is applicable to any two-dimensional non-Hermitian Hamiltonian with exceptional points of arbitrary order, and  moreover can also be used  to characterize non-defective degeneracy points. 
Furthermore, we show that a surface of a three-dimensional system can violate the non-Hermitian doubling theorems, which implies unusual bulk physics.  
\end{abstract}

\maketitle

\emph{Introduction.---}Fermion doubling theorems~\cite{NIELSEN1981219,Nielsen_Ninomiya_1981,NIELSEN1981173} are an important concept in the topological band theory of condensed matter physics~\cite{Fangeaat2374,beriTopologicallyStableGapless2010,RevModPhys.90.015001,RevModPhys.88.035005}. They state that topological point nodes in the energy spectrum of lattice Hamiltonians must come in pairs. Thereby they prevent the occurrence of quantum anomalies in lattices. This is because for a single point node, the low-energy physics is described by a field theory with a quantum anomaly, while for two point nodes the anomalies cancel. Well known examples of doubled point nodes include the two Dirac points of graphene~\cite{RevModPhys.81.109} and the two Weyl points of magnetic Weyl semimetals~\cite{RevModPhys.90.015001,burkovWeylSemimetalTopological2011}. While the doubling theorems must be fulfilled in the bulk of any lattice Hamiltonian, they may be violated on a lattice surface. For example, the topological insulator with time-reversal symmetry exhibits a single Dirac point with parity anomaly on its surface~\cite{fuTopologicalInsulatorsThree2007}. These anomalous surface states lead to unusual physical responses and give a powerful diagnostic of the nontrivial bulk topology~\cite{RyuMooreLudwig2012,PhysRevB.78.195424}.

Recently, topological band theory has been extended to non-Hermitian Hamiltonians~\cite{benderMakingSenseNonHermitian2007b,moiseyevNonHermitianQuantumMechanics2011,rotterReviewProgressPhysics2015a,fengNonHermitianPhotonicsBased2017b,el-ganainyNonHermitianPhysicsPT2018d,ozdemirParityTimeSymmetry2019a,miriExceptionalPointsOptics2019a,doi:10.1002/adma.201903639,martinezalvarezTopologicalStatesNonHermitian2018b,Ghatak_2019,PhysRevLett.80.5243,PhysRevLett.89.270401,PhysRevLett.76.4472,PhysRevLett.77.570,PhysRevLett.102.065703,PhysRevLett.105.013903,leeAnomalousEdgeState2016a,leykamEdgeModesDegeneracies2017e,PhysRevLett.120.146601,2017arXiv170805841K,PhysRevB.99.201107,PhysRevLett.120.146402,PhysRevLett.121.026403,yaoEdgeStatesTopological2018b,yaoNonHermitianChernBands2018b,PhysRevLett.123.170401,PhysRevLett.123.246801,kunstBiorthogonalBulkBoundaryCorrespondence2018a,yokomizoNonBlochBandTheory2019a,2019arXiv191205499Y,zhangCorrespondenceWindingNumbers2019,PhysRevLett.124.086801,PhysRevLett.118.045701,	PhysRevB.97.075128, PhysRevA.98.042114, PhysRevB.99.081102, PhysRevLett.124.186402,PhysRevLett.123.066405,	PhysRevLett.122.076801,PhysRevLett.122.195501,PhysRevLett.123.073601,PhysRevLett.123.016805,PhysRevX.8.031079,PhysRevB.99.235112,PhysRevX.9.041015,PhysRevB.99.125103,PhysRevLett.123.206404,PhysRevLett.123.090603,PhysRevLett.122.237601,PhysRevLett.123.190403,PhysRevX.9.041015,PhysRevX.8.031079,kawabataParitytimesymmetricTopologicalSuperconductor2018,2020arXiv200302219Y}, which can be realized in, e.g.,  photonic cavity arrays~\cite{harari_topological_2018,bandres_topological_2018,Bahari636,PhysRevB.99.121101}, and provide effective descriptions of open quantum systems~\cite{rotterReviewProgressPhysics2015a,fengNonHermitianPhotonicsBased2017b,el-ganainyNonHermitianPhysicsPT2018d,ozdemirParityTimeSymmetry2019a,miriExceptionalPointsOptics2019a,doi:10.1002/adma.201903639,PhysRevLett.123.170401,PhysRevLett.124.196401,PhysRevLett.124.040401}, where energy is not conserved due to, e.g., dissipation or particle gain and loss. In contrast to the Hermitian case, two-dimensional  (2D) non-Hermitian Hamiltonians can exhibit three different types of point nodes, namely (i) Fermi points (FPs), (ii) exceptional points (EPs), and (iii) non-defective degeneracy points (NDPs). Both at an EP and at an NDP two (or more) energy bands become degenerate at a degeneracy point (DP). However, at an EP the corresponding eigenstates coalesce~\cite{kankiExactDescriptionCoalescing2017} (become identical), while at an NDP the eigenstates remain distinct. For this reason, non-Hermitian Hamiltonians at EPs are non-diagonalizable and can only be reduced to   Jordan block forms~\cite{kankiExactDescriptionCoalescing2017}. In the absence of symmetry, both FPs and EPs can be topologically stable in 2D, meaning that these point nodes cannot be removed by perturbations~\cite{leykamEdgeModesDegeneracies2017e,PhysRevLett.118.045701,PhysRevLett.120.146402,PhysRevLett.123.066405}. 

\begin {figure*}[t!]
\centerline{\includegraphics[height=2.5cm]{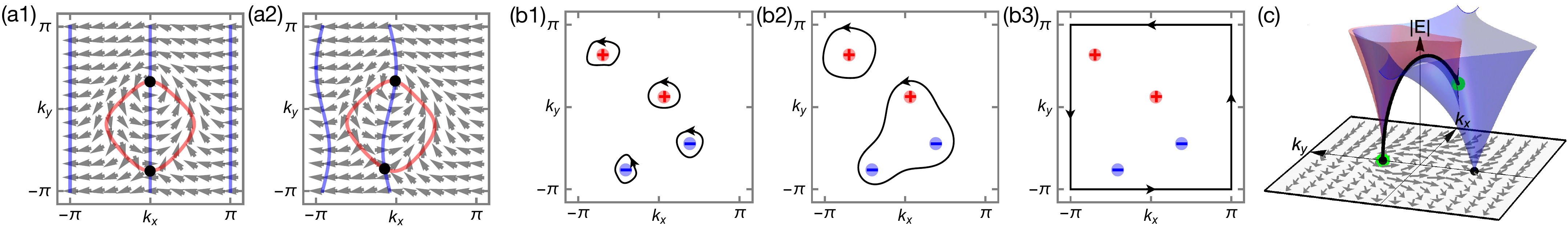}}
\caption{(a) In the 2D BZ, Fermi points (black points) are located at the crossings of the red line ($\Re f_\mu (\bm{k})=0$) and the blue line ($\Im f_\mu (\bm{k})=0$). The grey arrows display the orientation of the vector field $(\Re f_\mu (\bm{k}), \Im f_\mu (\bm{k}) )$. {\clb Panel (a2)} shows that a small perturbation does not destroy the Fermi points. (b) The integration paths  $\Gamma (\bk_F^j)$ [panel~(b1)] of Eq.~\eqref{invariant FP} [or Eq.~\eqref{invariant disc}] can be continuously deformed to the boundary of the BZ [panel~(b3)], without passing through any singular point [panel~(b2)]. The blue (red) dots represent Fermi points, or degeneracy points, with positive (negative) topological charge. (c) The absolute value of the complex energy bands $| E ( {\bk} ) |$ (red and blue surfaces) and vector field of the characteristic polynomial $(\Re f_\mu (\bm{k}), \Im f_\mu (\bm{k}) )$ (grey arrows) for the two-band model (\ref{2EP}). The black and green points are Fermi points and exceptional points, respectively. The black arc represents the branch cut that connects the two exceptional points.
	\label{P1} }
\end{figure*}

The physics of EPs has recently attracted attention~\cite{heissPhysicsExceptionalPoints2012,heissExceptionalPointsNonHermitian2004,PhysRevLett.86.787,PhysRevLett.101.080402,PhysRevLett.108.173901,wiersigEnhancingSensitivityFrequency2014,tanakaExoticQuantumHolonomy2014,dopplerDynamicallyEncirclingExceptional2016,hodaeiDarkstateLasersMode2016,kangChiralExceptionalPoints2016,kimDirectObservationExceptional2016,gaoObservationNonHermitianDegeneracies2015,PhysRevX.6.021007,shiAccessingExceptionalPoints2016,xuDetectingTopologicalExceptional2017,hodaeiEnhancedSensitivityHigherorder2017a,pickGeneralTheorySpontaneous2017,jingHighorderExceptionalPoints2017,PhysRevApplied.8.044020,cerjanEffectsNonHermitianPerturbations2018,lyubarovExceptionalPointsPhoton2018,yiPairExceptionalPoints2018,zhouObservationBulkFermi2018,yoshidaNonHermitianPerspectiveBand2018a,wangArbitraryOrderExceptional2019,sweeneyPerfectlyAbsorbingExceptional2019,PhysRevB.99.201107}, in particular in the context of photonic platforms, where they have many interesting applications, for example, as optical omnipolarizers~\cite{PhysRevLett.118.093002}, or as sensors with enhanced sensitivity~\cite{hodaeiEnhancedSensitivityHigherorder2017a,wiersigEnhancingSensitivityFrequency2014}. While the occurrence and stability of FPs and EPs have been studied in various settings, the existence of doubling theorems for these topological point nodes remains unknown. In this Letter we derive doubling theorems for FPs and EPs in 2D periodic lattice Hamiltonians without any symmetry constraints. For that purpose, we consider a non-Hermitian Hamiltonian $\mathcal{H} ( \bk )$ possessing well-separated complex energy bands~\cite{leykamEdgeModesDegeneracies2017e,PhysRevLett.120.146402}, except for some possible DPs, which are either EPs or NDPs. The proof of the doubling theorems relies on the fact that topological points carry nonzero topological charges, whose sum must vanish in the entire Brillouin zone (BZ)~\cite{NIELSEN1981219,Nielsen_Ninomiya_1981,NIELSEN1981173} due to its periodicity, i.e.,
\bee
\sum_{\bk_i \in \rm{BZ}} C(\bk_i)=0, \label{nogo}
\ee
where $C(\bk_i)$ is the topological charge of a topological point located at $\bk_i$ in the BZ. Particularly, for 2D non-Hermitian system, topological points can be FPs or EPs. The charges $C(\bk_i)$ are defined in terms of an integral of some topological charge density, along a closed contour that counterclockwise encircles the FP or EP. Equation~\eqref{nogo} then follows by continuously deforming the integration contours to the boundary of the BZ. We find that for FPs the appropriate charge density is the logarithmic derivative of $\det [ \mu - \mathcal{H} ( \bk ) ]$, while for EPs it is the logarithmic derivative of the \emph{discriminant} of $ \mathcal{H} ( \bk )$ with respect to the energy $E$. We demonstrate by several examples that EPs do not need to be branch point singularities, contrary to  previous reports~\cite{miriExceptionalPointsOptics2019a}. Finally, we show that the doubling theorems can be violated at surfaces in 3D systems, which implies
unusual properties of the bulk bands. In particular, inversion-symmetric or reflection-symmetric systems can host single FPs or EPs on their surfaces, which, however, must be accompanied by Fermi lines or exceptional lines in the bulk.

\emph{Doubling theorem for FPs.---}We start with the doubling theorem for FPs of generic 2D non-Hermitian  Hamiltonians $\mathcal{H}(\bk)$ with complex energy bands $E_i(\bk)$. The FPs of $\mathcal{H}(\bk)$ are defined as those points $\bk_F^j$ in the BZ, where the complex chemical potential $\mu$ intersects with one of the energy bands $E_i(\bk)$, i.e., $\mu - E_i (\bk_F^j ) =0 $ for some $i$. By choosing a proper basis, each entry of $\mathcal{H}(\bk)$ is single-valued in the entire BZ. The location of the FPs can then be obtained from the characteristic polynomial of $\mathcal{H}(\bk)$
\bee
f_\mu ( \bk)\equiv \det [\mu- \mathcal{H}(\bk)]= \prod_i [\mu-E_i(\bk)]. \label{chemical potential}
\ee
That is, the FPs are located at those $\bk_F^j$ where $f_\mu (\bk_F^j)=0$. Since $f_\mu ( \bk)$ is a complex function, this gives the two conditions ($\Re f_\mu ( \bk_F^j ) = 0$ and $\Im f_\mu ( \bk_F^j ) = 0$), whose solutions yield two line loops in the 2D BZ, see Fig.~\ref{P1}(a1). The crossings of these two loops give the positions of the FPs. Pictorially, we can see that the two loops must cross each other an even number of times, thereby suggesting a doubling theorem for FPs~\cite{newtwoloops}. Moreover, we observe from Fig.~\ref{P1}(a2) that small perturbations only shift the paths of the  loops, but do not remove the FPs.

These observations can be made more precise using the mathematical formalism of topological invariants. For this purpose we define the global winding number invariant~\cite{PhysRevLett.124.086801}
\bee
W (\boldsymbol{k}_{F}^j )=\frac{i}{2 \pi} \oint_{\Gamma (\boldsymbol{k}_{F}^j )} d\bm{k} \cdot \nabla_{\bm{k}} \ln f_\mu ( \bk), \label{invariant FP}
\ee
where the integration path $\Gamma (\bk_F^j)$   is a loop encircling $\bk_F^j$ counterclockwise, as shown in Fig.~\ref{P1}(b1). Since $f_\mu ( \bk)$ is single-valued in the entire BZ, the winding number $W (\boldsymbol{k}_{F}^j ) $ is quantized to an integer, which endows the FP at $\boldsymbol{k}_{F}^j$ with a topological charge. If the winding number is nonzero, the integration path $\Gamma (\bk_F^j)$ in Eq.~\eqref{invariant FP} cannot be smoothly shrunk to a single point, due to the presence of a singularity at the FP. This guarantees the topological stability of the FP and protects it against gap opening, even in the presence of perturbations. To derive the doubling theorem for FPs, we sum over the winding numbers of all FPs in the BZ 
\bee
\sum_{\bk_{F}^j \in \rm{BZ}} W (\boldsymbol{k}_{F}^j )=\frac{i}{2 \pi} \oint_{\partial\rm{BZ}} d\bm{k} \cdot \nabla_{\bm{k}} \ln f_\mu ( \bk) =0. \label{FP thm}
\ee
The above sum must vanish, because the integration paths of Eq.~\eqref{invariant FP}  can be continuously deformed to the BZ boundary  $\partial \rm{BZ}$, as the FPs are the only singularities in the integrand [see Fig.~\ref{P1}(b)]. Hence, each FP with a positive topological charge must be accompanied by an FP with negative topological charge. This proves the doubling theorem \eqref{nogo} for FPs in 2D non-Hermitian systems.  

\emph{An example of doubled FPs.---}We use an example to demonstrate the topological properties of the FPs by studying a two-band Hamiltonian, given by 
\begin{equation}
\mathcal{H}(\boldsymbol{k})=h_0(\bm{k})\sigma_0+\boldsymbol{h}(\boldsymbol{k}) \cdot \boldsymbol{\sigma}, \label{2EP}
\end{equation}
where $h_0(\bm{k})= \tfrac{1}{2} \sin k_y , h_x(\bm{k})=\sin k_x-i, h_y(\bm{k})=\sin k_y$, and $h_z(\bm{k})=\cos k_x+\cos k_y -2$. With chemical potential $\mu=\sqrt{3}/4$, this example has two FPs located at  $\bk_F^-=(0,-0.479\pi)$ and $\bk_F^+=(0,\pi/3)$ with winding numbers  $W(\bk_F^\pm)=\mp1$, such that the doubling theorem is satisfied [Fig.~\ref{P1}(c)]. The energy spectrum of $\mathcal{H}(\boldsymbol{k})$  is given by $ E_{\pm}(\bm{k}) = \tfrac{s_y}{2} \pm\sqrt{5-4(c_x+c_y)+c_{x+y}+c_{x-y}-2is_x}$, where $c_{x/y}=\cos k_{x/y}$, $s_{x/y}=\sin k_{x/y}$, and $c_{x\pm y}=\cos( k_x\pm k_y)$. This spectrum is multi-valued and exhibits a branch cut that is terminated by two EPs, located at $(0,\pm \pi/3)$ with energies $E - \mu =0$ and  $E - \mu = - \sqrt{3}/2$, respectively. The fact that one of the EPs coincides with one of the FPs is purely accidental. In general EPs and FPs of generic non-Hermitian Hamiltonians are at different positions.

\emph{Discriminant and DPs.---}Next, we turn to DPs  of generic non-Hermitian Hamiltonians and explain how they can be found in an efficient manner using the  discriminant of the characteristic polynomial $f_E ( {\bk} )$. Here,  $f_E ( {\bk} )$ is defined as in Eq.~\eqref{chemical potential}, but the chemical potential $\mu$ is replaced by the energy $E$. A DP occurs at $\bk_D$ when $E_i(\bk_D)=E_j(\bk_D)$ for some $i\neq j$. Hence, the polynomial $f_E ( {\bk})$  must have a double (or multiple) root at $\bk_D$. Moreover, the discriminant of $f_E ( {\bk})$, which is defined as
\begin{align}
\operatorname{Disc}_{E}[\mathcal{H}](\bm{k})=& \prod_{i<j}\left[E_i(\bm{k})-E_j(\bm{k})\right]^{2},
\label{Dis}
\end{align}
must vanish at $\bk_D$. i.e., there is a DP at  $\bk_D$, if and only if $\operatorname{Disc}_{E}[\mathcal{H}](\bm{k}_D)=0$. Apparently, the DPs are computed more efficiently from the zeros of the discriminant, rather than by explicitly calculating all  roots of $f_E ( {\bk})$. This is because the discriminant can be computed directly from the determinant of the Sylvester matrix of $f_E(\bk)$ and $\partial_E f_E( \bk)$, see supplemental materials (SM)~\cite{supp_info}. Hence, determining the zeros of $\operatorname{Disc}_{E}[\mathcal{H}](\bm{k})$ is an efficient way to find all DPs in the entire BZ at any energy \footnote{Wo note that this procedure can also be applied to find DPs of Hermitian Hamiltonians.}. 

The discriminant has the additional advantage of being single valued, since the coefficients of $f_E (\bk)$ are single-valued~\footnote{This is because in a proper basis, all the matrix elements of $\mathcal{H}(\bm{k})$ are periodic functions of $\bm{k}$, which is equivalent to the single-valued condition. Therefore, the corresponding characteristic polynomial $f_E(\bm{k})=\det[E-\mathcal{H}(\bm{k})]$, whose coefficients are algebraic functions of these matrix elements, must also be single-valued.}. This property is   key to define a {\it quantized} invariant in terms of $\operatorname{Disc}_{E}[\mathcal{H}](\bm{k})$ and to prove the doubling theorem of DPs. Before doing so, let us first give an illustrative argument for why DPs must satisfy a doubling theorem. We note that zeros of the discriminant must satisfy the two constraints ${\rm{Re}}(\operatorname{Disc}_{E}[\mathcal{H}](\bm{k}))=0$ and ${\rm{Im}}(\operatorname{Disc}_{E}[\mathcal{H}](\bm{k}))=0$. The solutions to these two equations are two line loops in the 2D BZ, whose crossings give the positions of the DPs. Since two loops in a periodic BZ generally cross each other an even number of times, DPs must come in pairs and satisfy a doubling  theorem
~\footnote{Similar to the FPs, it is possible to have single crossing between $\operatorname{Re}\left(\operatorname{Disc}_{E}[\mathcal{H}](\boldsymbol{k})\right)=0$ and $\operatorname{Im}\left(\operatorname{Disc}_{E}[\mathcal{H}](\boldsymbol{k})\right)=0$.  However, the topological charge (which will be defined in the following contents) at the crossing point must also be zero, which is unstable to external weak perturbations.}. 

Finally, we remark that in the absence of extra symmetries, DPs have in general only two-fold degeneracy. That is, any DP with higher degeneracy can be split into multiple two-fold degenerate DPs  by an infinitesimally small perturbation~\cite{supp_info}. For this reason we focus only on two-fold degenerate DPs.

\emph{Doubling theorem for DPs.---}The invariant that characterizes the topology of DPs, which can be EPs or NDPs, is given in terms of a contour integral over the discriminant, 
\begin{align}
\nu(\bk_D^l)=& \frac{i}{2\pi}\oint_{\Gamma(\bk_{D}^l)} d\bm{k} \cdot \nabla_{\bm{k}} \ln \operatorname{Disc}_{E}[\mathcal{H}](\bk),
\label{invariant disc}
\end{align}	
where $\Gamma(\bk_D^l)$ is a loop counterclockwise encircling the DP at $\bk_{D}^l$. Since $\operatorname{Disc}_{E}[\mathcal{H}](\bk)$ is single valued, this invariant is a quantized winding number, which we call the \emph{discriminant number}. The mathematical structure is almost identical to the winding number~\eqref{invariant FP} characterizing FPs. The only difference is that $\det[\mu -\mathcal{H}(\bm{k})]$ in the integrand of Eq.~\eqref{invariant FP} is replaced by $\operatorname{Disc}_{E}[\mathcal{H}](\bk)$. A nonzero quantized value of $\nu$ guarantees the stability of DPs  against gap opening. Using Eq.~\eqref{invariant disc}  we can identify, in a computationally efficient manner, any stable DP between any pair of bands~\cite{supp_info}. To obtain the doubling theorem for DPs, we sum over the discriminant numbers of all DPs in the BZ
\bee
\sum_{\bk_D^l\in \rm{BZ}} \nu(\bk_D^l)=\frac{i}{2\pi}\oint_{\partial \rm{BZ}} d\bm{k} \cdot \nabla_{\bm{k}} \ln \operatorname{Disc}_{E}[\mathcal{H}](\bk)=0 \label{nogo degeneracy} .
\ee
Since the DPs are the only singularities in the integrand of Eq.~\eqref{invariant disc}, the integration paths in this sum can be continuously deformed  to the BZ boundary $\partial \rm{BZ}$, see Fig.~\ref{P1}. Hence, the above sum must vanish and, therefore, the discriminant number of the DPs must cancel pairwise. This proves the doubling theorem~\eqref{nogo} for DPs. 

We note that the topology of DPs formed by  two bands has previously been characterized using the vorticity invariant~\cite{PhysRevLett.120.146402,PhysRevLett.123.066405}, 
\begin{equation} 
\nu_{ij}(\bk_D^l)= - \frac{1}{2 \pi} \oint_{\Gamma(\bk_D^l)} \nabla_{\bm{k}} \arg \left[E_{i}(\bm{k})-E_{j}(\bm{k})\right] \cdot d \bm{k},  \label{Fu}
\end{equation}
where $i$ and $j$ label the two bands and $\arg (z)=-i\ln (z/|z|)$. We find, after some algebra~\cite{supp_info}, that the discriminant number is equal to the vorticity invariant summed over all pairs of distinct bands, i.e., $\nu(\bk_D^l)=\sum_{i\neq j}\nu_{ij}(\bk_D^l)$. 
 
\begin {figure}[t]
\centerline{\includegraphics[height=2.3cm]{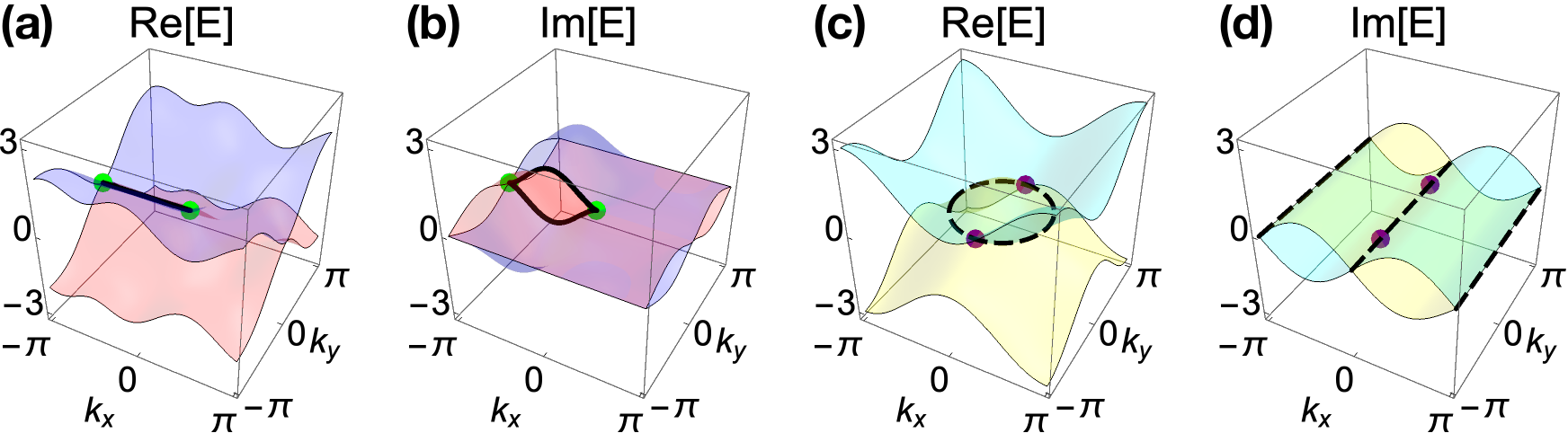}}
\caption{ The complex energy spectra for the DP examples. (a,b) show that the Hamiltonian (\ref{EP1}) possesses two NDPs, which are the ends of the branch cut (solid lines). (c,d) show that in the Hamiltonian (\ref{EP2}) an EP and an NDP do not connect to any branch cut, although these points are located at the crossing of the energy bands (dashed lines). 
	\label{EP_example}}
\end{figure}

\emph{Examples of doubled DPs.---}We use two examples to illustrate the doubling theorem for DPs and to show that only EPs are generically stable, while NDPs can be deformed to EPs by arbitrarily small perturbations. 

The first example contains two NDPs and is given by the two-band Hamiltonian
\begin{eqnarray} \label{EP1}
\mathcal{H} ( {\bk} )
=
\begin{pmatrix}
0 & F({\bk})  \cr
G({\bk}) & 0 \cr
\end{pmatrix},
\end{eqnarray}
where $F({\bk}) = \sin^2 k_x - \tfrac{1}{2} \sin^2 k_y  + 2 i \sin k_x \sin k_y + \cos k_y -1$ and $G({\bk}) = \sin k_x  - i \sin k_y + \cos k_y - 1$. The energy spectrum is $E_{\pm} ({\bk} )= \pm \sqrt{F ({\bk})  G ({\bk} )}$ and the characteristic polynomial reads $f_E ( {\bk} ) = E^2 - F ( {\bk} ) G ( {\bk} )$. From this we obtain the discriminant $\operatorname{Disc}_{E}[\mathcal{H}](\bk)= 4 F({\bk}) G({\bk} )$, which has zeros at $(0,0)$ and $(\pi,0)$ corresponding to two NDPs~\cite{example2}. Using Eq.~\eqref{invariant disc}, we find that these two NDPs have $\nu = \pm 1$, thereby satisfying the doubling theorem. These two NDPs are end points of a branch cut, demonstrating that branch cuts do not need to be terminated by EPs  as shown in Fig.~\ref{EP_example}(a,b). However, this is an unstable situation, since the infinitesimally small perturbation $\delta \sigma_z$ turns the NDPs into EPs. 

The second example contains one NDP and one EP and is described by the following Hamiltonian
\begin{equation} \label{EP2}
\mathcal{H}(\bm{k})=\left(\begin{array}{cc}{A(\bk)} & B(\bk) \\ {0} & {-A(\bk)}\end{array}\right), 
\end{equation}
where $A(\bk)=1-\cos k_x-\cos k_y+i \sin k_x$ and $B(\bk)=1-\sin k_y$. The spectrum is $E_\pm ( {\bk} )= \pm A ( {\bk} )$ and the characteristic polynomial reads $f_E ( {\bk} ) = E^2- A^2 ( {\bk} )$, from which we obtain the discriminant $\operatorname{Disc}_{E}[\mathcal{H}](\bk)=4A^2(\bk)$. Solving for the zeros of the discriminant, we find  two DPs located at  $(0,\pm \pi/2)$. The DP at $(0,- \pi/2)$ is an EP with discriminant number $\nu = -2$, while the DP at $(0,+ \pi/2)$ is an NDP with $\nu = +2$, such that the doubling theorem is satisfied. We observe that the EP does not terminate a branch cut as shown in Fig.~\ref{EP_example}(c,d), since the spectrum $E_\pm ({\bk})$ is single-valued in the entire BZ. But this is a fine-tuned situation, which is destabilized by the infinitesimally small deformation $\eta \sigma_x$. This perturbation splits the NDP  and the EP each into two EPs  with $\nu = \pm 1$, which become end points of branch cuts. 

In general, a two-fold degenerate EP with $\nu = \pm 1$ is always a branch cut termination of the energy spectrum (See the proof in SM~\cite{supp_info} and cf.~[\onlinecite{miriExceptionalPointsOptics2019a}]).  From the insights gained by the above examples, we prove that in two dimensions NDPs are unstable in SM~\cite{supp_info}, i.e., they can be deformed into EPs by generic perturbations. Furthermore, EPs with $| \nu | >1$ are split into several EPs with $| \nu | =1$ by small perturbations and importantly only EPs with $\nu = \pm 1$ are stable.

\emph{Doubling theorem for EPs.---}Taking together the above results, we conclude that in two dimensions the only stable DPs are EPs with $\nu = \pm 1$. Allowing for generic perturbations, these EPs must come in pairs with opposite discriminant number $\nu$.

\emph{Anomalous FPs and EPs at surfaces.---}We close this Letter by discussing anomalous FPs and EPs (or DPs) at surfaces, which violate the doubling theorems. Surfaces of 3D systems can be viewed, in a sense, as one half of  2D bulk systems. As a consequence, surfaces can host, in principle, an odd number of stable FPs or EPs, thereby breaking the doubling theorems. In the presence of these anomalous surfaces, the bulk exhibits unusual topological properties, which depends on the crystalline symmetries.
In this regard, we separately discuss the anomalous physics with and without the symmetries. 

\textit{(i)} First, we consider a symmetry which relates the top and bottom surfaces of a given 3D system. For concreteness, let us consider two surfaces that are related by reflection or inversion, which act on the two surface Hamiltonians as 
\bee \label{sym_op}
P_\pm \mathcal{H}_{\rm{top}}(k_x,k_y)P_{\pm} ^{-1}= \mathcal{H}_{\rm{bot}}(\pm k_x, \pm k_y),
\ee
where $P_\pm$ is a unitary operator implementing reflection (+) or inversion (-). Focusing on surface EPs, we now let symmetry~\eqref{sym_op} act on the discriminant number $\nu$, Eq.~\eqref{invariant disc}. From this we find that $\nu$ summed over all EPs at the top surface is equal to $\nu$ summed over all EPs at the bottom surface, i.e., 
\bee
\sum_{\bk_D^l \in \rm{BZ_{top}}}\nu(\bk_D^l)=\sum_{\bk_D^l \in \rm{BZ_{bot}}}\nu(\bk_D^l) .
\ee
Hence, as opposed to 3D topological insulators, the topological charges of the EPs on the top and bottom surfaces do not cancel. Therefore, there must exist additional EPs with non-zero $\nu$ in the 3D bulk which compensate the nonvanishing sum of the discriminant numbers on the two surfaces. In fact, since non-zero $\nu$ is defined in terms of a line integral, that cannot  be deformed to vanish, there must exist entire lines of EPs in the  3D bulk~\footnote{Here, we implicitly assume that there is no skin effect, in which case the bulk spectrum would drastically depend on the boundary conditions. Note that in the presence of certain symmetry (e.g.,  reflection in spinless systems~\cite{2020arXiv200302219Y}), the skin effect is always absent.}. Therefore, if there are surface EPs violating the doubling theorem, the 3D bulk must contain at least one exceptional line topologically protected by the non-zero discriminant number. Hence, the entire system contains two exotic non-Hermitian physics phenomena  --- anomalous surfaces and exceptional lines in the bulk~\cite{PhysRevLett.118.045701,	PhysRevB.97.075128, PhysRevA.98.042114, PhysRevB.99.081102, PhysRevLett.124.186402,PhysRevLett.123.066405}.

By a similar derivation, we can prove the same property also for surface FPs. That is, if there are surface FPs violating the doubling theorem in an inversion or reflection-symmetric system, the bulk must be gapless (no point gap at the FP energy level) and must have at least one Fermi line. In the SM~\cite{supp_info} we present an example of a 3D non-Hermitian lattice model, which exhibits these anomalous surface EPs and FPs together with bulk Fermi lines and exceptional lines.

(ii) Second, we consider a 3D Hamiltonian without FPs and EPs in the bulk, which breaks both  reflection and inversion symmetry. 
We find that the surface of such a system can still violate the non-Hermitian doubling theorem~\cite{2020arXiv200801090D}. 
In the SM~\cite{supp_info}, we provide an example of a 3D Hamiltonian with a bulk point gap and surfaces that violate the FP doubling theorem.
Since the top and bottom surfaces  have opposite  topological charges, the FP doubling theorem is only violated for an individual surface
but is satisfied for the entire 3D system.  
(This is similar to Hermitian topological gapped systems, for example,   surfaces of 3D topological insulators, which exhibit single Dirac points, both at the top and the bottom surfaces, with opposite topological charges~\cite{fuTopologicalInsulatorsThree2007}.)
Gapped bulk system with such anomalous surfaces exhibit a non-trivial non-Hermitian topology, which is described by the 3D winding number~\cite{2020arXiv200801090D,PhysRevX.8.031079,PhysRevX.9.041015,Ghatak_2019}. Thus, the breaking of the non-Hermitian doubling theorem at the surface allows to identify the non-trivial topology in the bulk.

\emph{Conclusion.---}In summary, we have derived doubling theorems for FPs and EPs in generic 2D non-Hermitian lattice Hamiltonians. To derive the doubling theorem for exceptional points we have introduced a new topological invariant, which we call the \emph{discriminant number}. This discriminant number endows EPs with a quantized topological charge. The doubling theorem ensures that a single EP of first order must be accompanied by another first-order EP with opposite topological charge as all of the energy bands are taken into account. We have shown that 2D non-Hermitian Hamiltonians with fine-tuned parameters can also exhibit higher-order EPs or non-defective degeneracy points. These are, however, unstable and can be removed by arbitrarily small perturbations. Furthermore, we have clarified the relation between EPs and branch cuts, namely only two-fold degenerate EPs of first order   
must necessarily be   branch cut terminations.  Finally, we studied anomalous surfaces of 3D bulk systems that violate the non-Hermitian doubling theorems. 
In the presence of inversion or reflection symmetry, this violation leads to the emergence of topologically protected Fermi or exceptional lines in the bulk. In the absence of these symmetries, the bulk is gapped and must exhibit  non-trivial topology.

\begin{acknowledgements}
\emph{Note added.---}Upon submission of this manuscript we became aware of a related preprint~\cite{wojcik2019topological}, which makes
use of the discriminant to  classify non-Hermitian Hamiltonians.  Before resubmitting this revision to PRL, we also noticed a relevant study~\cite{2020arXiv200801090D}, which first proposed a non-Hermitian model possessing a point gap and surfaces violating the FP doubling theorem. 
\end{acknowledgements}

\bibliography{NH,TOPO_v14,aGBZ}

\begin{thebibliography}{110}%
\makeatletter
\providecommand \@ifxundefined [1]{%
 \@ifx{#1\undefined}
}%
\providecommand \@ifnum [1]{%
 \ifnum #1\expandafter \@firstoftwo
 \else \expandafter \@secondoftwo
 \fi
}%
\providecommand \@ifx [1]{%
 \ifx #1\expandafter \@firstoftwo
 \else \expandafter \@secondoftwo
 \fi
}%
\providecommand \natexlab [1]{#1}%
\providecommand \enquote  [1]{``#1''}%
\providecommand \bibnamefont  [1]{#1}%
\providecommand \bibfnamefont [1]{#1}%
\providecommand \citenamefont [1]{#1}%
\providecommand \href@noop [0]{\@secondoftwo}%
\providecommand \href [0]{\begingroup \@sanitize@url \@href}%
\providecommand \@href[1]{\@@startlink{#1}\@@href}%
\providecommand \@@href[1]{\endgroup#1\@@endlink}%
\providecommand \@sanitize@url [0]{\catcode `\\12\catcode `\$12\catcode
  `\&12\catcode `\#12\catcode `\^12\catcode `\_12\catcode `\%12\relax}%
\providecommand \@@startlink[1]{}%
\providecommand \@@endlink[0]{}%
\providecommand \url  [0]{\begingroup\@sanitize@url \@url }%
\providecommand \@url [1]{\endgroup\@href {#1}{\urlprefix }}%
\providecommand \urlprefix  [0]{URL }%
\providecommand \Eprint [0]{\href }%
\providecommand \doibase [0]{http://dx.doi.org/}%
\providecommand \selectlanguage [0]{\@gobble}%
\providecommand \bibinfo  [0]{\@secondoftwo}%
\providecommand \bibfield  [0]{\@secondoftwo}%
\providecommand \translation [1]{[#1]}%
\providecommand \BibitemOpen [0]{}%
\providecommand \bibitemStop [0]{}%
\providecommand \bibitemNoStop [0]{.\EOS\space}%
\providecommand \EOS [0]{\spacefactor3000\relax}%
\providecommand \BibitemShut  [1]{\csname bibitem#1\endcsname}%
\let\auto@bib@innerbib\@empty
\bibitem [{\citenamefont {Nielsen}\ and\ \citenamefont
  {Ninomiya}(1981{\natexlab{a}})}]{NIELSEN1981219}%
  \BibitemOpen
  \bibfield  {author} {\bibinfo {author} {\bibfnamefont {H.}~\bibnamefont
  {Nielsen}}\ and\ \bibinfo {author} {\bibfnamefont {M.}~\bibnamefont
  {Ninomiya}},\ }\href {\doibase https://doi.org/10.1016/0370-2693(81)91026-1}
  {\bibfield  {journal} {\bibinfo  {journal} {Physics Letters B}\ }\textbf
  {\bibinfo {volume} {105}},\ \bibinfo {pages} {219 } (\bibinfo {year}
  {1981}{\natexlab{a}})}\BibitemShut {NoStop}%
\bibitem [{\citenamefont {Nielsen}\ and\ \citenamefont
  {Ninomiya}(1981{\natexlab{b}})}]{Nielsen_Ninomiya_1981}%
  \BibitemOpen
  \bibfield  {author} {\bibinfo {author} {\bibfnamefont {H.}~\bibnamefont
  {Nielsen}}\ and\ \bibinfo {author} {\bibfnamefont {M.}~\bibnamefont
  {Ninomiya}},\ }\href {\doibase https://doi.org/10.1016/0550-3213(81)90361-8}
  {\bibfield  {journal} {\bibinfo  {journal} {Nuclear Physics B}\ }\textbf
  {\bibinfo {volume} {185}},\ \bibinfo {pages} {20 } (\bibinfo {year}
  {1981}{\natexlab{b}})}\BibitemShut {NoStop}%
\bibitem [{\citenamefont {Nielsen}\ and\ \citenamefont
  {Ninomiya}(1981{\natexlab{c}})}]{NIELSEN1981173}%
  \BibitemOpen
  \bibfield  {author} {\bibinfo {author} {\bibfnamefont {H.}~\bibnamefont
  {Nielsen}}\ and\ \bibinfo {author} {\bibfnamefont {M.}~\bibnamefont
  {Ninomiya}},\ }\href {\doibase https://doi.org/10.1016/0550-3213(81)90524-1}
  {\bibfield  {journal} {\bibinfo  {journal} {Nuclear Physics B}\ }\textbf
  {\bibinfo {volume} {193}},\ \bibinfo {pages} {173 } (\bibinfo {year}
  {1981}{\natexlab{c}})}\BibitemShut {NoStop}%
\bibitem [{\citenamefont {Fang}\ and\ \citenamefont {Fu}(2019)}]{Fangeaat2374}%
  \BibitemOpen
  \bibfield  {author} {\bibinfo {author} {\bibfnamefont {C.}~\bibnamefont
  {Fang}}\ and\ \bibinfo {author} {\bibfnamefont {L.}~\bibnamefont {Fu}},\
  }\href {\doibase 10.1126/sciadv.aat2374} {\bibfield  {journal} {\bibinfo
  {journal} {Science Advances}\ }\textbf {\bibinfo {volume} {5}} (\bibinfo
  {year} {2019}),\ 10.1126/sciadv.aat2374}\BibitemShut {NoStop}%
\bibitem [{\citenamefont
  {B{\'e}ri}(2010)}]{beriTopologicallyStableGapless2010}%
  \BibitemOpen
  \bibfield  {author} {\bibinfo {author} {\bibfnamefont {B.}~\bibnamefont
  {B{\'e}ri}},\ }\href {\doibase 10.1103/PhysRevB.81.134515} {\bibfield
  {journal} {\bibinfo  {journal} {Phys. Rev. B}\ }\textbf {\bibinfo {volume}
  {81}},\ \bibinfo {pages} {134515} (\bibinfo {year} {2010})}\BibitemShut
  {NoStop}%
\bibitem [{\citenamefont {Armitage}\ \emph {et~al.}(2018)\citenamefont
  {Armitage}, \citenamefont {Mele},\ and\ \citenamefont
  {Vishwanath}}]{RevModPhys.90.015001}%
  \BibitemOpen
  \bibfield  {author} {\bibinfo {author} {\bibfnamefont {N.~P.}\ \bibnamefont
  {Armitage}}, \bibinfo {author} {\bibfnamefont {E.~J.}\ \bibnamefont {Mele}},
  \ and\ \bibinfo {author} {\bibfnamefont {A.}~\bibnamefont {Vishwanath}},\
  }\href {\doibase 10.1103/RevModPhys.90.015001} {\bibfield  {journal}
  {\bibinfo  {journal} {Rev. Mod. Phys.}\ }\textbf {\bibinfo {volume} {90}},\
  \bibinfo {pages} {015001} (\bibinfo {year} {2018})}\BibitemShut {NoStop}%
\bibitem [{\citenamefont {Chiu}\ \emph {et~al.}(2016)\citenamefont {Chiu},
  \citenamefont {Teo}, \citenamefont {Schnyder},\ and\ \citenamefont
  {Ryu}}]{RevModPhys.88.035005}%
  \BibitemOpen
  \bibfield  {author} {\bibinfo {author} {\bibfnamefont {C.-K.}\ \bibnamefont
  {Chiu}}, \bibinfo {author} {\bibfnamefont {J.~C.~Y.}\ \bibnamefont {Teo}},
  \bibinfo {author} {\bibfnamefont {A.~P.}\ \bibnamefont {Schnyder}}, \ and\
  \bibinfo {author} {\bibfnamefont {S.}~\bibnamefont {Ryu}},\ }\href {\doibase
  10.1103/RevModPhys.88.035005} {\bibfield  {journal} {\bibinfo  {journal}
  {Rev. Mod. Phys.}\ }\textbf {\bibinfo {volume} {88}},\ \bibinfo {pages}
  {035005} (\bibinfo {year} {2016})}\BibitemShut {NoStop}%
\bibitem [{\citenamefont {Castro~Neto}\ \emph {et~al.}(2009)\citenamefont
  {Castro~Neto}, \citenamefont {Guinea}, \citenamefont {Peres}, \citenamefont
  {Novoselov},\ and\ \citenamefont {Geim}}]{RevModPhys.81.109}%
  \BibitemOpen
  \bibfield  {author} {\bibinfo {author} {\bibfnamefont {A.~H.}\ \bibnamefont
  {Castro~Neto}}, \bibinfo {author} {\bibfnamefont {F.}~\bibnamefont {Guinea}},
  \bibinfo {author} {\bibfnamefont {N.~M.~R.}\ \bibnamefont {Peres}}, \bibinfo
  {author} {\bibfnamefont {K.~S.}\ \bibnamefont {Novoselov}}, \ and\ \bibinfo
  {author} {\bibfnamefont {A.~K.}\ \bibnamefont {Geim}},\ }\href {\doibase
  10.1103/RevModPhys.81.109} {\bibfield  {journal} {\bibinfo  {journal} {Rev.
  Mod. Phys.}\ }\textbf {\bibinfo {volume} {81}},\ \bibinfo {pages} {109}
  (\bibinfo {year} {2009})}\BibitemShut {NoStop}%
\bibitem [{\citenamefont {Burkov}\ and\ \citenamefont
  {Balents}(2011)}]{burkovWeylSemimetalTopological2011}%
  \BibitemOpen
  \bibfield  {author} {\bibinfo {author} {\bibfnamefont {A.~A.}\ \bibnamefont
  {Burkov}}\ and\ \bibinfo {author} {\bibfnamefont {L.}~\bibnamefont
  {Balents}},\ }\href {\doibase 10.1103/PhysRevLett.107.127205} {\bibfield
  {journal} {\bibinfo  {journal} {Phys. Rev. Lett.}\ }\textbf {\bibinfo
  {volume} {107}},\ \bibinfo {pages} {127205} (\bibinfo {year}
  {2011})}\BibitemShut {NoStop}%
\bibitem [{\citenamefont {Fu}\ \emph {et~al.}(2007)\citenamefont {Fu},
  \citenamefont {Kane},\ and\ \citenamefont
  {Mele}}]{fuTopologicalInsulatorsThree2007}%
  \BibitemOpen
  \bibfield  {author} {\bibinfo {author} {\bibfnamefont {L.}~\bibnamefont
  {Fu}}, \bibinfo {author} {\bibfnamefont {C.~L.}\ \bibnamefont {Kane}}, \ and\
  \bibinfo {author} {\bibfnamefont {E.~J.}\ \bibnamefont {Mele}},\ }\href
  {\doibase 10.1103/PhysRevLett.98.106803} {\bibfield  {journal} {\bibinfo
  {journal} {Phys. Rev. Lett.}\ }\textbf {\bibinfo {volume} {98}},\ \bibinfo
  {pages} {106803} (\bibinfo {year} {2007})}\BibitemShut {NoStop}%
\bibitem [{\citenamefont {{Ryu}}\ \emph {et~al.}(2012)\citenamefont {{Ryu}},
  \citenamefont {{Moore}},\ and\ \citenamefont
  {{Ludwig}}}]{RyuMooreLudwig2012}%
  \BibitemOpen
  \bibfield  {author} {\bibinfo {author} {\bibfnamefont {S.}~\bibnamefont
  {{Ryu}}}, \bibinfo {author} {\bibfnamefont {J.~E.}\ \bibnamefont {{Moore}}},
  \ and\ \bibinfo {author} {\bibfnamefont {A.~W.~W.}\ \bibnamefont
  {{Ludwig}}},\ }\href {\doibase 10.1103/PhysRevB.85.045104} {\bibfield
  {journal} {\bibinfo  {journal} {\prb}\ }\textbf {\bibinfo {volume} {85}},\
  \bibinfo {eid} {045104} (\bibinfo {year} {2012})}\BibitemShut {NoStop}%
\bibitem [{\citenamefont {Qi}\ \emph {et~al.}(2008)\citenamefont {Qi},
  \citenamefont {Hughes},\ and\ \citenamefont {Zhang}}]{PhysRevB.78.195424}%
  \BibitemOpen
  \bibfield  {author} {\bibinfo {author} {\bibfnamefont {X.-L.}\ \bibnamefont
  {Qi}}, \bibinfo {author} {\bibfnamefont {T.~L.}\ \bibnamefont {Hughes}}, \
  and\ \bibinfo {author} {\bibfnamefont {S.-C.}\ \bibnamefont {Zhang}},\ }\href
  {\doibase 10.1103/PhysRevB.78.195424} {\bibfield  {journal} {\bibinfo
  {journal} {Phys. Rev. B}\ }\textbf {\bibinfo {volume} {78}},\ \bibinfo
  {pages} {195424} (\bibinfo {year} {2008})}\BibitemShut {NoStop}%
\bibitem [{\citenamefont {Bender}(2007)}]{benderMakingSenseNonHermitian2007b}%
  \BibitemOpen
  \bibfield  {author} {\bibinfo {author} {\bibfnamefont {C.~M.}\ \bibnamefont
  {Bender}},\ }\href {\doibase 10.1088/0034-4885/70/6/R03} {\bibfield
  {journal} {\bibinfo  {journal} {Rep. Prog. Phys.}\ }\textbf {\bibinfo
  {volume} {70}},\ \bibinfo {pages} {947} (\bibinfo {year} {2007})}\BibitemShut
  {NoStop}%
\bibitem [{\citenamefont
  {Moiseyev}(2011)}]{moiseyevNonHermitianQuantumMechanics2011}%
  \BibitemOpen
  \bibfield  {author} {\bibinfo {author} {\bibfnamefont {N.}~\bibnamefont
  {Moiseyev}},\ }\href@noop {} {\emph {\bibinfo {title} {Non-{{Hermitian
  Quantum Mechanics}}}}}\ (\bibinfo {year} {2011})\BibitemShut {NoStop}%
\bibitem [{\citenamefont {Rotter}\ and\ \citenamefont
  {Bird}(2015)}]{rotterReviewProgressPhysics2015a}%
  \BibitemOpen
  \bibfield  {author} {\bibinfo {author} {\bibfnamefont {I.}~\bibnamefont
  {Rotter}}\ and\ \bibinfo {author} {\bibfnamefont {J.~P.}\ \bibnamefont
  {Bird}},\ }\href {\doibase 10.1088/0034-4885/78/11/114001} {\bibfield
  {journal} {\bibinfo  {journal} {Rep. Prog. Phys.}\ }\textbf {\bibinfo
  {volume} {78}},\ \bibinfo {pages} {114001} (\bibinfo {year}
  {2015})}\BibitemShut {NoStop}%
\bibitem [{\citenamefont {Feng}\ \emph {et~al.}(2017)\citenamefont {Feng},
  \citenamefont {{El-Ganainy}},\ and\ \citenamefont
  {Ge}}]{fengNonHermitianPhotonicsBased2017b}%
  \BibitemOpen
  \bibfield  {author} {\bibinfo {author} {\bibfnamefont {L.}~\bibnamefont
  {Feng}}, \bibinfo {author} {\bibfnamefont {R.}~\bibnamefont {{El-Ganainy}}},
  \ and\ \bibinfo {author} {\bibfnamefont {L.}~\bibnamefont {Ge}},\ }\href
  {\doibase 10.1038/s41566-017-0031-1} {\bibfield  {journal} {\bibinfo
  {journal} {Nat. Photonics}\ }\textbf {\bibinfo {volume} {11}},\ \bibinfo
  {pages} {752} (\bibinfo {year} {2017})}\BibitemShut {NoStop}%
\bibitem [{\citenamefont {{El-Ganainy}}\ \emph {et~al.}(2018)\citenamefont
  {{El-Ganainy}}, \citenamefont {Makris}, \citenamefont {Khajavikhan},
  \citenamefont {Musslimani}, \citenamefont {Rotter},\ and\ \citenamefont
  {Christodoulides}}]{el-ganainyNonHermitianPhysicsPT2018d}%
  \BibitemOpen
  \bibfield  {author} {\bibinfo {author} {\bibfnamefont {R.}~\bibnamefont
  {{El-Ganainy}}}, \bibinfo {author} {\bibfnamefont {K.~G.}\ \bibnamefont
  {Makris}}, \bibinfo {author} {\bibfnamefont {M.}~\bibnamefont {Khajavikhan}},
  \bibinfo {author} {\bibfnamefont {Z.~H.}\ \bibnamefont {Musslimani}},
  \bibinfo {author} {\bibfnamefont {S.}~\bibnamefont {Rotter}}, \ and\ \bibinfo
  {author} {\bibfnamefont {D.~N.}\ \bibnamefont {Christodoulides}},\ }\href
  {\doibase 10.1038/nphys4323} {\bibfield  {journal} {\bibinfo  {journal} {Nat.
  Phys.}\ }\textbf {\bibinfo {volume} {14}},\ \bibinfo {pages} {11} (\bibinfo
  {year} {2018})}\BibitemShut {NoStop}%
\bibitem [{\citenamefont {Ozdemir}\ \emph {et~al.}(2019)\citenamefont
  {Ozdemir}, \citenamefont {Rotter}, \citenamefont {Nori},\ and\ \citenamefont
  {Yang}}]{ozdemirParityTimeSymmetry2019a}%
  \BibitemOpen
  \bibfield  {author} {\bibinfo {author} {\bibfnamefont {S.~K.}\ \bibnamefont
  {Ozdemir}}, \bibinfo {author} {\bibfnamefont {S.}~\bibnamefont {Rotter}},
  \bibinfo {author} {\bibfnamefont {F.}~\bibnamefont {Nori}}, \ and\ \bibinfo
  {author} {\bibfnamefont {L.}~\bibnamefont {Yang}},\ }\href {\doibase
  10.1038/s41563-019-0304-9} {\bibfield  {journal} {\bibinfo  {journal} {Nat.
  Mater.}\ }\textbf {\bibinfo {volume} {18}},\ \bibinfo {pages} {783} (\bibinfo
  {year} {2019})}\BibitemShut {NoStop}%
\bibitem [{\citenamefont {Miri}\ and\ \citenamefont
  {Al{\`u}}(2019)}]{miriExceptionalPointsOptics2019a}%
  \BibitemOpen
  \bibfield  {author} {\bibinfo {author} {\bibfnamefont {M.-A.}\ \bibnamefont
  {Miri}}\ and\ \bibinfo {author} {\bibfnamefont {A.}~\bibnamefont {Al{\`u}}},\
  }\href {\doibase 10.1126/science.aar7709} {\bibfield  {journal} {\bibinfo
  {journal} {Science}\ }\textbf {\bibinfo {volume} {363}},\ \bibinfo {pages}
  {eaar7709} (\bibinfo {year} {2019})}\BibitemShut {NoStop}%
\bibitem [{\citenamefont {Gupta}\ \emph {et~al.}()\citenamefont {Gupta},
  \citenamefont {Zou}, \citenamefont {Zhu}, \citenamefont {Lu}, \citenamefont
  {Zhang}, \citenamefont {Liu},\ and\ \citenamefont
  {Chen}}]{doi:10.1002/adma.201903639}%
  \BibitemOpen
  \bibfield  {author} {\bibinfo {author} {\bibfnamefont {S.~K.}\ \bibnamefont
  {Gupta}}, \bibinfo {author} {\bibfnamefont {Y.}~\bibnamefont {Zou}}, \bibinfo
  {author} {\bibfnamefont {X.-Y.}\ \bibnamefont {Zhu}}, \bibinfo {author}
  {\bibfnamefont {M.-H.}\ \bibnamefont {Lu}}, \bibinfo {author} {\bibfnamefont
  {L.-J.}\ \bibnamefont {Zhang}}, \bibinfo {author} {\bibfnamefont {X.-P.}\
  \bibnamefont {Liu}}, \ and\ \bibinfo {author} {\bibfnamefont {Y.-F.}\
  \bibnamefont {Chen}},\ }\href {\doibase 10.1002/adma.201903639} {\bibfield
  {journal} {\bibinfo  {journal} {Advanced Materials}\ }\textbf {\bibinfo
  {volume} {n/a}},\ \bibinfo {pages} {1903639}}\BibitemShut {NoStop}%
\bibitem [{\citenamefont {Martinez~Alvarez}\ \emph {et~al.}(2018)\citenamefont
  {Martinez~Alvarez}, \citenamefont {Barrios~Vargas}, \citenamefont
  {Berdakin},\ and\ \citenamefont
  {Foa~Torres}}]{martinezalvarezTopologicalStatesNonHermitian2018b}%
  \BibitemOpen
  \bibfield  {author} {\bibinfo {author} {\bibfnamefont {V.~M.}\ \bibnamefont
  {Martinez~Alvarez}}, \bibinfo {author} {\bibfnamefont {J.~E.}\ \bibnamefont
  {Barrios~Vargas}}, \bibinfo {author} {\bibfnamefont {M.}~\bibnamefont
  {Berdakin}}, \ and\ \bibinfo {author} {\bibfnamefont {L.~E.~F.}\ \bibnamefont
  {Foa~Torres}},\ }\href {\doibase 10.1140/epjst/e2018-800091-5} {\bibfield
  {journal} {\bibinfo  {journal} {Eur. Phys. J. Spec. Top.}\ }\textbf {\bibinfo
  {volume} {227}},\ \bibinfo {pages} {1295} (\bibinfo {year}
  {2018})}\BibitemShut {NoStop}%
\bibitem [{\citenamefont {Ghatak}\ and\ \citenamefont
  {Das}(2019)}]{Ghatak_2019}%
  \BibitemOpen
  \bibfield  {author} {\bibinfo {author} {\bibfnamefont {A.}~\bibnamefont
  {Ghatak}}\ and\ \bibinfo {author} {\bibfnamefont {T.}~\bibnamefont {Das}},\
  }\href {\doibase 10.1088/1361-648x/ab11b3} {\bibfield  {journal} {\bibinfo
  {journal} {Journal of Physics: Condensed Matter}\ }\textbf {\bibinfo {volume}
  {31}},\ \bibinfo {pages} {263001} (\bibinfo {year} {2019})}\BibitemShut
  {NoStop}%
\bibitem [{\citenamefont {Bender}\ and\ \citenamefont
  {Boettcher}(1998)}]{PhysRevLett.80.5243}%
  \BibitemOpen
  \bibfield  {author} {\bibinfo {author} {\bibfnamefont {C.~M.}\ \bibnamefont
  {Bender}}\ and\ \bibinfo {author} {\bibfnamefont {S.}~\bibnamefont
  {Boettcher}},\ }\href {\doibase 10.1103/PhysRevLett.80.5243} {\bibfield
  {journal} {\bibinfo  {journal} {Phys. Rev. Lett.}\ }\textbf {\bibinfo
  {volume} {80}},\ \bibinfo {pages} {5243} (\bibinfo {year}
  {1998})}\BibitemShut {NoStop}%
\bibitem [{\citenamefont {Bender}\ \emph {et~al.}(2002)\citenamefont {Bender},
  \citenamefont {Brody},\ and\ \citenamefont {Jones}}]{PhysRevLett.89.270401}%
  \BibitemOpen
  \bibfield  {author} {\bibinfo {author} {\bibfnamefont {C.~M.}\ \bibnamefont
  {Bender}}, \bibinfo {author} {\bibfnamefont {D.~C.}\ \bibnamefont {Brody}}, \
  and\ \bibinfo {author} {\bibfnamefont {H.~F.}\ \bibnamefont {Jones}},\ }\href
  {\doibase 10.1103/PhysRevLett.89.270401} {\bibfield  {journal} {\bibinfo
  {journal} {Phys. Rev. Lett.}\ }\textbf {\bibinfo {volume} {89}},\ \bibinfo
  {pages} {270401} (\bibinfo {year} {2002})}\BibitemShut {NoStop}%
\bibitem [{\citenamefont {Stephanov}(1996)}]{PhysRevLett.76.4472}%
  \BibitemOpen
  \bibfield  {author} {\bibinfo {author} {\bibfnamefont {M.~A.}\ \bibnamefont
  {Stephanov}},\ }\href {\doibase 10.1103/PhysRevLett.76.4472} {\bibfield
  {journal} {\bibinfo  {journal} {Phys. Rev. Lett.}\ }\textbf {\bibinfo
  {volume} {76}},\ \bibinfo {pages} {4472} (\bibinfo {year}
  {1996})}\BibitemShut {NoStop}%
\bibitem [{\citenamefont {Hatano}\ and\ \citenamefont
  {Nelson}(1996)}]{PhysRevLett.77.570}%
  \BibitemOpen
  \bibfield  {author} {\bibinfo {author} {\bibfnamefont {N.}~\bibnamefont
  {Hatano}}\ and\ \bibinfo {author} {\bibfnamefont {D.~R.}\ \bibnamefont
  {Nelson}},\ }\href {\doibase 10.1103/PhysRevLett.77.570} {\bibfield
  {journal} {\bibinfo  {journal} {Phys. Rev. Lett.}\ }\textbf {\bibinfo
  {volume} {77}},\ \bibinfo {pages} {570} (\bibinfo {year} {1996})}\BibitemShut
  {NoStop}%
\bibitem [{\citenamefont {Rudner}\ and\ \citenamefont
  {Levitov}(2009)}]{PhysRevLett.102.065703}%
  \BibitemOpen
  \bibfield  {author} {\bibinfo {author} {\bibfnamefont {M.~S.}\ \bibnamefont
  {Rudner}}\ and\ \bibinfo {author} {\bibfnamefont {L.~S.}\ \bibnamefont
  {Levitov}},\ }\href {\doibase 10.1103/PhysRevLett.102.065703} {\bibfield
  {journal} {\bibinfo  {journal} {Phys. Rev. Lett.}\ }\textbf {\bibinfo
  {volume} {102}},\ \bibinfo {pages} {065703} (\bibinfo {year}
  {2009})}\BibitemShut {NoStop}%
\bibitem [{\citenamefont {Longhi}(2010)}]{PhysRevLett.105.013903}%
  \BibitemOpen
  \bibfield  {author} {\bibinfo {author} {\bibfnamefont {S.}~\bibnamefont
  {Longhi}},\ }\href {\doibase 10.1103/PhysRevLett.105.013903} {\bibfield
  {journal} {\bibinfo  {journal} {Phys. Rev. Lett.}\ }\textbf {\bibinfo
  {volume} {105}},\ \bibinfo {pages} {013903} (\bibinfo {year}
  {2010})}\BibitemShut {NoStop}%
\bibitem [{\citenamefont {Lee}(2016)}]{leeAnomalousEdgeState2016a}%
  \BibitemOpen
  \bibfield  {author} {\bibinfo {author} {\bibfnamefont {T.~E.}\ \bibnamefont
  {Lee}},\ }\href {\doibase 10.1103/PhysRevLett.116.133903} {\bibfield
  {journal} {\bibinfo  {journal} {Phys. Rev. Lett.}\ }\textbf {\bibinfo
  {volume} {116}},\ \bibinfo {pages} {133903} (\bibinfo {year}
  {2016})}\BibitemShut {NoStop}%
\bibitem [{\citenamefont {Leykam}\ \emph {et~al.}(2017)\citenamefont {Leykam},
  \citenamefont {Bliokh}, \citenamefont {Huang}, \citenamefont {Chong},\ and\
  \citenamefont {Nori}}]{leykamEdgeModesDegeneracies2017e}%
  \BibitemOpen
  \bibfield  {author} {\bibinfo {author} {\bibfnamefont {D.}~\bibnamefont
  {Leykam}}, \bibinfo {author} {\bibfnamefont {K.~Y.}\ \bibnamefont {Bliokh}},
  \bibinfo {author} {\bibfnamefont {C.}~\bibnamefont {Huang}}, \bibinfo
  {author} {\bibfnamefont {Y.~D.}\ \bibnamefont {Chong}}, \ and\ \bibinfo
  {author} {\bibfnamefont {F.}~\bibnamefont {Nori}},\ }\href {\doibase
  10.1103/PhysRevLett.118.040401} {\bibfield  {journal} {\bibinfo  {journal}
  {Phys. Rev. Lett.}\ }\textbf {\bibinfo {volume} {118}},\ \bibinfo {pages}
  {040401} (\bibinfo {year} {2017})}\BibitemShut {NoStop}%
\bibitem [{\citenamefont {Molina}\ and\ \citenamefont
  {Gonz\'alez}(2018)}]{PhysRevLett.120.146601}%
  \BibitemOpen
  \bibfield  {author} {\bibinfo {author} {\bibfnamefont {R.~A.}\ \bibnamefont
  {Molina}}\ and\ \bibinfo {author} {\bibfnamefont {J.}~\bibnamefont
  {Gonz\'alez}},\ }\href {\doibase 10.1103/PhysRevLett.120.146601} {\bibfield
  {journal} {\bibinfo  {journal} {Phys. Rev. Lett.}\ }\textbf {\bibinfo
  {volume} {120}},\ \bibinfo {pages} {146601} (\bibinfo {year}
  {2018})}\BibitemShut {NoStop}%
\bibitem [{\citenamefont {{Kozii}}\ and\ \citenamefont
  {{Fu}}(2017)}]{2017arXiv170805841K}%
  \BibitemOpen
  \bibfield  {author} {\bibinfo {author} {\bibfnamefont {V.}~\bibnamefont
  {{Kozii}}}\ and\ \bibinfo {author} {\bibfnamefont {L.}~\bibnamefont {{Fu}}},\
  }\href@noop {} {\bibfield  {journal} {\bibinfo  {journal} {arXiv e-prints}\
  ,\ \bibinfo {eid} {arXiv:1708.05841}} (\bibinfo {year} {2017})},\ \Eprint
  {http://arxiv.org/abs/1708.05841} {arXiv:1708.05841} \BibitemShut {NoStop}%
\bibitem [{\citenamefont {Papaj}\ \emph {et~al.}(2019)\citenamefont {Papaj},
  \citenamefont {Isobe},\ and\ \citenamefont {Fu}}]{PhysRevB.99.201107}%
  \BibitemOpen
  \bibfield  {author} {\bibinfo {author} {\bibfnamefont {M.}~\bibnamefont
  {Papaj}}, \bibinfo {author} {\bibfnamefont {H.}~\bibnamefont {Isobe}}, \ and\
  \bibinfo {author} {\bibfnamefont {L.}~\bibnamefont {Fu}},\ }\href {\doibase
  10.1103/PhysRevB.99.201107} {\bibfield  {journal} {\bibinfo  {journal} {Phys.
  Rev. B}\ }\textbf {\bibinfo {volume} {99}},\ \bibinfo {pages} {201107}
  (\bibinfo {year} {2019})}\BibitemShut {NoStop}%
\bibitem [{\citenamefont {Shen}\ \emph {et~al.}(2018)\citenamefont {Shen},
  \citenamefont {Zhen},\ and\ \citenamefont {Fu}}]{PhysRevLett.120.146402}%
  \BibitemOpen
  \bibfield  {author} {\bibinfo {author} {\bibfnamefont {H.}~\bibnamefont
  {Shen}}, \bibinfo {author} {\bibfnamefont {B.}~\bibnamefont {Zhen}}, \ and\
  \bibinfo {author} {\bibfnamefont {L.}~\bibnamefont {Fu}},\ }\href {\doibase
  10.1103/PhysRevLett.120.146402} {\bibfield  {journal} {\bibinfo  {journal}
  {Phys. Rev. Lett.}\ }\textbf {\bibinfo {volume} {120}},\ \bibinfo {pages}
  {146402} (\bibinfo {year} {2018})}\BibitemShut {NoStop}%
\bibitem [{\citenamefont {Shen}\ and\ \citenamefont
  {Fu}(2018)}]{PhysRevLett.121.026403}%
  \BibitemOpen
  \bibfield  {author} {\bibinfo {author} {\bibfnamefont {H.}~\bibnamefont
  {Shen}}\ and\ \bibinfo {author} {\bibfnamefont {L.}~\bibnamefont {Fu}},\
  }\href {\doibase 10.1103/PhysRevLett.121.026403} {\bibfield  {journal}
  {\bibinfo  {journal} {Phys. Rev. Lett.}\ }\textbf {\bibinfo {volume} {121}},\
  \bibinfo {pages} {026403} (\bibinfo {year} {2018})}\BibitemShut {NoStop}%
\bibitem [{\citenamefont {Yao}\ and\ \citenamefont
  {Wang}(2018)}]{yaoEdgeStatesTopological2018b}%
  \BibitemOpen
  \bibfield  {author} {\bibinfo {author} {\bibfnamefont {S.}~\bibnamefont
  {Yao}}\ and\ \bibinfo {author} {\bibfnamefont {Z.}~\bibnamefont {Wang}},\
  }\href {\doibase 10.1103/PhysRevLett.121.086803} {\bibfield  {journal}
  {\bibinfo  {journal} {Phys. Rev. Lett.}\ }\textbf {\bibinfo {volume} {121}},\
  \bibinfo {pages} {086803} (\bibinfo {year} {2018})}\BibitemShut {NoStop}%
\bibitem [{\citenamefont {Yao}\ \emph {et~al.}(2018)\citenamefont {Yao},
  \citenamefont {Song},\ and\ \citenamefont
  {Wang}}]{yaoNonHermitianChernBands2018b}%
  \BibitemOpen
  \bibfield  {author} {\bibinfo {author} {\bibfnamefont {S.}~\bibnamefont
  {Yao}}, \bibinfo {author} {\bibfnamefont {F.}~\bibnamefont {Song}}, \ and\
  \bibinfo {author} {\bibfnamefont {Z.}~\bibnamefont {Wang}},\ }\href {\doibase
  10.1103/PhysRevLett.121.136802} {\bibfield  {journal} {\bibinfo  {journal}
  {Phys. Rev. Lett.}\ }\textbf {\bibinfo {volume} {121}},\ \bibinfo {pages}
  {136802} (\bibinfo {year} {2018})}\BibitemShut {NoStop}%
\bibitem [{\citenamefont {Song}\ \emph
  {et~al.}(2019{\natexlab{a}})\citenamefont {Song}, \citenamefont {Yao},\ and\
  \citenamefont {Wang}}]{PhysRevLett.123.170401}%
  \BibitemOpen
  \bibfield  {author} {\bibinfo {author} {\bibfnamefont {F.}~\bibnamefont
  {Song}}, \bibinfo {author} {\bibfnamefont {S.}~\bibnamefont {Yao}}, \ and\
  \bibinfo {author} {\bibfnamefont {Z.}~\bibnamefont {Wang}},\ }\href {\doibase
  10.1103/PhysRevLett.123.170401} {\bibfield  {journal} {\bibinfo  {journal}
  {Phys. Rev. Lett.}\ }\textbf {\bibinfo {volume} {123}},\ \bibinfo {pages}
  {170401} (\bibinfo {year} {2019}{\natexlab{a}})}\BibitemShut {NoStop}%
\bibitem [{\citenamefont {Song}\ \emph
  {et~al.}(2019{\natexlab{b}})\citenamefont {Song}, \citenamefont {Yao},\ and\
  \citenamefont {Wang}}]{PhysRevLett.123.246801}%
  \BibitemOpen
  \bibfield  {author} {\bibinfo {author} {\bibfnamefont {F.}~\bibnamefont
  {Song}}, \bibinfo {author} {\bibfnamefont {S.}~\bibnamefont {Yao}}, \ and\
  \bibinfo {author} {\bibfnamefont {Z.}~\bibnamefont {Wang}},\ }\href {\doibase
  10.1103/PhysRevLett.123.246801} {\bibfield  {journal} {\bibinfo  {journal}
  {Phys. Rev. Lett.}\ }\textbf {\bibinfo {volume} {123}},\ \bibinfo {pages}
  {246801} (\bibinfo {year} {2019}{\natexlab{b}})}\BibitemShut {NoStop}%
\bibitem [{\citenamefont {Kunst}\ \emph {et~al.}(2018)\citenamefont {Kunst},
  \citenamefont {Edvardsson}, \citenamefont {Budich},\ and\ \citenamefont
  {Bergholtz}}]{kunstBiorthogonalBulkBoundaryCorrespondence2018a}%
  \BibitemOpen
  \bibfield  {author} {\bibinfo {author} {\bibfnamefont {F.~K.}\ \bibnamefont
  {Kunst}}, \bibinfo {author} {\bibfnamefont {E.}~\bibnamefont {Edvardsson}},
  \bibinfo {author} {\bibfnamefont {J.~C.}\ \bibnamefont {Budich}}, \ and\
  \bibinfo {author} {\bibfnamefont {E.~J.}\ \bibnamefont {Bergholtz}},\ }\href
  {\doibase 10.1103/PhysRevLett.121.026808} {\bibfield  {journal} {\bibinfo
  {journal} {Phys. Rev. Lett.}\ }\textbf {\bibinfo {volume} {121}},\ \bibinfo
  {pages} {026808} (\bibinfo {year} {2018})}\BibitemShut {NoStop}%
\bibitem [{\citenamefont {Yokomizo}\ and\ \citenamefont
  {Murakami}(2019)}]{yokomizoNonBlochBandTheory2019a}%
  \BibitemOpen
  \bibfield  {author} {\bibinfo {author} {\bibfnamefont {K.}~\bibnamefont
  {Yokomizo}}\ and\ \bibinfo {author} {\bibfnamefont {S.}~\bibnamefont
  {Murakami}},\ }\href {\doibase 10.1103/PhysRevLett.123.066404} {\bibfield
  {journal} {\bibinfo  {journal} {Phys. Rev. Lett.}\ }\textbf {\bibinfo
  {volume} {123}},\ \bibinfo {pages} {066404} (\bibinfo {year}
  {2019})}\BibitemShut {NoStop}%
\bibitem [{\citenamefont {{Yang}}\ \emph {et~al.}(2019)\citenamefont {{Yang}},
  \citenamefont {{Zhang}}, \citenamefont {{Fang}},\ and\ \citenamefont
  {{Hu}}}]{2019arXiv191205499Y}%
  \BibitemOpen
  \bibfield  {author} {\bibinfo {author} {\bibfnamefont {Z.}~\bibnamefont
  {{Yang}}}, \bibinfo {author} {\bibfnamefont {K.}~\bibnamefont {{Zhang}}},
  \bibinfo {author} {\bibfnamefont {C.}~\bibnamefont {{Fang}}}, \ and\ \bibinfo
  {author} {\bibfnamefont {J.}~\bibnamefont {{Hu}}},\ }\href@noop {} {\bibfield
   {journal} {\bibinfo  {journal} {arXiv e-prints}\ ,\ \bibinfo {eid}
  {arXiv:1912.05499}} (\bibinfo {year} {2019})},\ \Eprint
  {http://arxiv.org/abs/1912.05499} {arXiv:1912.05499} \BibitemShut {NoStop}%
\bibitem [{\citenamefont {Zhang}\ \emph {et~al.}()\citenamefont {Zhang},
  \citenamefont {Yang},\ and\ \citenamefont
  {Fang}}]{zhangCorrespondenceWindingNumbers2019}%
  \BibitemOpen
  \bibfield  {author} {\bibinfo {author} {\bibfnamefont {K.}~\bibnamefont
  {Zhang}}, \bibinfo {author} {\bibfnamefont {Z.}~\bibnamefont {Yang}}, \ and\
  \bibinfo {author} {\bibfnamefont {C.}~\bibnamefont {Fang}},\ }\href@noop {}
  {\ }\Eprint {http://arxiv.org/abs/1910.01131} {arXiv:1910.01131} \BibitemShut
  {NoStop}%
\bibitem [{\citenamefont {Okuma}\ \emph {et~al.}(2020)\citenamefont {Okuma},
  \citenamefont {Kawabata}, \citenamefont {Shiozaki},\ and\ \citenamefont
  {Sato}}]{PhysRevLett.124.086801}%
  \BibitemOpen
  \bibfield  {author} {\bibinfo {author} {\bibfnamefont {N.}~\bibnamefont
  {Okuma}}, \bibinfo {author} {\bibfnamefont {K.}~\bibnamefont {Kawabata}},
  \bibinfo {author} {\bibfnamefont {K.}~\bibnamefont {Shiozaki}}, \ and\
  \bibinfo {author} {\bibfnamefont {M.}~\bibnamefont {Sato}},\ }\href {\doibase
  10.1103/PhysRevLett.124.086801} {\bibfield  {journal} {\bibinfo  {journal}
  {Phys. Rev. Lett.}\ }\textbf {\bibinfo {volume} {124}},\ \bibinfo {pages}
  {086801} (\bibinfo {year} {2020})}\BibitemShut {NoStop}%
\bibitem [{\citenamefont {Xu}\ \emph {et~al.}(2017{\natexlab{a}})\citenamefont
  {Xu}, \citenamefont {Wang},\ and\ \citenamefont
  {Duan}}]{PhysRevLett.118.045701}%
  \BibitemOpen
  \bibfield  {author} {\bibinfo {author} {\bibfnamefont {Y.}~\bibnamefont
  {Xu}}, \bibinfo {author} {\bibfnamefont {S.-T.}\ \bibnamefont {Wang}}, \ and\
  \bibinfo {author} {\bibfnamefont {L.-M.}\ \bibnamefont {Duan}},\ }\href
  {\doibase 10.1103/PhysRevLett.118.045701} {\bibfield  {journal} {\bibinfo
  {journal} {Phys. Rev. Lett.}\ }\textbf {\bibinfo {volume} {118}},\ \bibinfo
  {pages} {045701} (\bibinfo {year} {2017}{\natexlab{a}})}\BibitemShut
  {NoStop}%
\bibitem [{\citenamefont {Cerjan}\ \emph
  {et~al.}(2018{\natexlab{a}})\citenamefont {Cerjan}, \citenamefont {Xiao},
  \citenamefont {Yuan},\ and\ \citenamefont {Fan}}]{PhysRevB.97.075128}%
  \BibitemOpen
  \bibfield  {author} {\bibinfo {author} {\bibfnamefont {A.}~\bibnamefont
  {Cerjan}}, \bibinfo {author} {\bibfnamefont {M.}~\bibnamefont {Xiao}},
  \bibinfo {author} {\bibfnamefont {L.}~\bibnamefont {Yuan}}, \ and\ \bibinfo
  {author} {\bibfnamefont {S.}~\bibnamefont {Fan}},\ }\href {\doibase
  10.1103/PhysRevB.97.075128} {\bibfield  {journal} {\bibinfo  {journal} {Phys.
  Rev. B}\ }\textbf {\bibinfo {volume} {97}},\ \bibinfo {pages} {075128}
  (\bibinfo {year} {2018}{\natexlab{a}})}\BibitemShut {NoStop}%
\bibitem [{\citenamefont {Carlstr\"om}\ and\ \citenamefont
  {Bergholtz}(2018)}]{PhysRevA.98.042114}%
  \BibitemOpen
  \bibfield  {author} {\bibinfo {author} {\bibfnamefont {J.}~\bibnamefont
  {Carlstr\"om}}\ and\ \bibinfo {author} {\bibfnamefont {E.~J.}\ \bibnamefont
  {Bergholtz}},\ }\href {\doibase 10.1103/PhysRevA.98.042114} {\bibfield
  {journal} {\bibinfo  {journal} {Phys. Rev. A}\ }\textbf {\bibinfo {volume}
  {98}},\ \bibinfo {pages} {042114} (\bibinfo {year} {2018})}\BibitemShut
  {NoStop}%
\bibitem [{\citenamefont {Yang}\ and\ \citenamefont
  {Hu}(2019)}]{PhysRevB.99.081102}%
  \BibitemOpen
  \bibfield  {author} {\bibinfo {author} {\bibfnamefont {Z.}~\bibnamefont
  {Yang}}\ and\ \bibinfo {author} {\bibfnamefont {J.}~\bibnamefont {Hu}},\
  }\href {\doibase 10.1103/PhysRevB.99.081102} {\bibfield  {journal} {\bibinfo
  {journal} {Phys. Rev. B}\ }\textbf {\bibinfo {volume} {99}},\ \bibinfo
  {pages} {081102} (\bibinfo {year} {2019})}\BibitemShut {NoStop}%
\bibitem [{\citenamefont {Yang}\ \emph {et~al.}(2020)\citenamefont {Yang},
  \citenamefont {Chiu}, \citenamefont {Fang},\ and\ \citenamefont
  {Hu}}]{PhysRevLett.124.186402}%
  \BibitemOpen
  \bibfield  {author} {\bibinfo {author} {\bibfnamefont {Z.}~\bibnamefont
  {Yang}}, \bibinfo {author} {\bibfnamefont {C.-K.}\ \bibnamefont {Chiu}},
  \bibinfo {author} {\bibfnamefont {C.}~\bibnamefont {Fang}}, \ and\ \bibinfo
  {author} {\bibfnamefont {J.}~\bibnamefont {Hu}},\ }\href {\doibase
  10.1103/PhysRevLett.124.186402} {\bibfield  {journal} {\bibinfo  {journal}
  {Phys. Rev. Lett.}\ }\textbf {\bibinfo {volume} {124}},\ \bibinfo {pages}
  {186402} (\bibinfo {year} {2020})}\BibitemShut {NoStop}%
\bibitem [{\citenamefont {Kawabata}\ \emph
  {et~al.}(2019{\natexlab{a}})\citenamefont {Kawabata}, \citenamefont
  {Bessho},\ and\ \citenamefont {Sato}}]{PhysRevLett.123.066405}%
  \BibitemOpen
  \bibfield  {author} {\bibinfo {author} {\bibfnamefont {K.}~\bibnamefont
  {Kawabata}}, \bibinfo {author} {\bibfnamefont {T.}~\bibnamefont {Bessho}}, \
  and\ \bibinfo {author} {\bibfnamefont {M.}~\bibnamefont {Sato}},\ }\href
  {\doibase 10.1103/PhysRevLett.123.066405} {\bibfield  {journal} {\bibinfo
  {journal} {Phys. Rev. Lett.}\ }\textbf {\bibinfo {volume} {123}},\ \bibinfo
  {pages} {066405} (\bibinfo {year} {2019}{\natexlab{a}})}\BibitemShut
  {NoStop}%
\bibitem [{\citenamefont {Liu}\ \emph {et~al.}(2019{\natexlab{a}})\citenamefont
  {Liu}, \citenamefont {Zhang}, \citenamefont {Ai}, \citenamefont {Gong},
  \citenamefont {Kawabata}, \citenamefont {Ueda},\ and\ \citenamefont
  {Nori}}]{PhysRevLett.122.076801}%
  \BibitemOpen
  \bibfield  {author} {\bibinfo {author} {\bibfnamefont {T.}~\bibnamefont
  {Liu}}, \bibinfo {author} {\bibfnamefont {Y.-R.}\ \bibnamefont {Zhang}},
  \bibinfo {author} {\bibfnamefont {Q.}~\bibnamefont {Ai}}, \bibinfo {author}
  {\bibfnamefont {Z.}~\bibnamefont {Gong}}, \bibinfo {author} {\bibfnamefont
  {K.}~\bibnamefont {Kawabata}}, \bibinfo {author} {\bibfnamefont
  {M.}~\bibnamefont {Ueda}}, \ and\ \bibinfo {author} {\bibfnamefont
  {F.}~\bibnamefont {Nori}},\ }\href {\doibase 10.1103/PhysRevLett.122.076801}
  {\bibfield  {journal} {\bibinfo  {journal} {Phys. Rev. Lett.}\ }\textbf
  {\bibinfo {volume} {122}},\ \bibinfo {pages} {076801} (\bibinfo {year}
  {2019}{\natexlab{a}})}\BibitemShut {NoStop}%
\bibitem [{\citenamefont {Zhang}\ \emph {et~al.}(2019)\citenamefont {Zhang},
  \citenamefont {Rosendo~L\'opez}, \citenamefont {Cheng}, \citenamefont {Liu},\
  and\ \citenamefont {Christensen}}]{PhysRevLett.122.195501}%
  \BibitemOpen
  \bibfield  {author} {\bibinfo {author} {\bibfnamefont {Z.}~\bibnamefont
  {Zhang}}, \bibinfo {author} {\bibfnamefont {M.}~\bibnamefont
  {Rosendo~L\'opez}}, \bibinfo {author} {\bibfnamefont {Y.}~\bibnamefont
  {Cheng}}, \bibinfo {author} {\bibfnamefont {X.}~\bibnamefont {Liu}}, \ and\
  \bibinfo {author} {\bibfnamefont {J.}~\bibnamefont {Christensen}},\ }\href
  {\doibase 10.1103/PhysRevLett.122.195501} {\bibfield  {journal} {\bibinfo
  {journal} {Phys. Rev. Lett.}\ }\textbf {\bibinfo {volume} {122}},\ \bibinfo
  {pages} {195501} (\bibinfo {year} {2019})}\BibitemShut {NoStop}%
\bibitem [{\citenamefont {Luo}\ and\ \citenamefont
  {Zhang}(2019)}]{PhysRevLett.123.073601}%
  \BibitemOpen
  \bibfield  {author} {\bibinfo {author} {\bibfnamefont {X.-W.}\ \bibnamefont
  {Luo}}\ and\ \bibinfo {author} {\bibfnamefont {C.}~\bibnamefont {Zhang}},\
  }\href {\doibase 10.1103/PhysRevLett.123.073601} {\bibfield  {journal}
  {\bibinfo  {journal} {Phys. Rev. Lett.}\ }\textbf {\bibinfo {volume} {123}},\
  \bibinfo {pages} {073601} (\bibinfo {year} {2019})}\BibitemShut {NoStop}%
\bibitem [{\citenamefont {Lee}\ \emph {et~al.}(2019{\natexlab{a}})\citenamefont
  {Lee}, \citenamefont {Li},\ and\ \citenamefont
  {Gong}}]{PhysRevLett.123.016805}%
  \BibitemOpen
  \bibfield  {author} {\bibinfo {author} {\bibfnamefont {C.~H.}\ \bibnamefont
  {Lee}}, \bibinfo {author} {\bibfnamefont {L.}~\bibnamefont {Li}}, \ and\
  \bibinfo {author} {\bibfnamefont {J.}~\bibnamefont {Gong}},\ }\href {\doibase
  10.1103/PhysRevLett.123.016805} {\bibfield  {journal} {\bibinfo  {journal}
  {Phys. Rev. Lett.}\ }\textbf {\bibinfo {volume} {123}},\ \bibinfo {pages}
  {016805} (\bibinfo {year} {2019}{\natexlab{a}})}\BibitemShut {NoStop}%
\bibitem [{\citenamefont {Gong}\ \emph {et~al.}(2018)\citenamefont {Gong},
  \citenamefont {Ashida}, \citenamefont {Kawabata}, \citenamefont {Takasan},
  \citenamefont {Higashikawa},\ and\ \citenamefont {Ueda}}]{PhysRevX.8.031079}%
  \BibitemOpen
  \bibfield  {author} {\bibinfo {author} {\bibfnamefont {Z.}~\bibnamefont
  {Gong}}, \bibinfo {author} {\bibfnamefont {Y.}~\bibnamefont {Ashida}},
  \bibinfo {author} {\bibfnamefont {K.}~\bibnamefont {Kawabata}}, \bibinfo
  {author} {\bibfnamefont {K.}~\bibnamefont {Takasan}}, \bibinfo {author}
  {\bibfnamefont {S.}~\bibnamefont {Higashikawa}}, \ and\ \bibinfo {author}
  {\bibfnamefont {M.}~\bibnamefont {Ueda}},\ }\href {\doibase
  10.1103/PhysRevX.8.031079} {\bibfield  {journal} {\bibinfo  {journal} {Phys.
  Rev. X}\ }\textbf {\bibinfo {volume} {8}},\ \bibinfo {pages} {031079}
  (\bibinfo {year} {2018})}\BibitemShut {NoStop}%
\bibitem [{\citenamefont {Zhou}\ and\ \citenamefont
  {Lee}(2019)}]{PhysRevB.99.235112}%
  \BibitemOpen
  \bibfield  {author} {\bibinfo {author} {\bibfnamefont {H.}~\bibnamefont
  {Zhou}}\ and\ \bibinfo {author} {\bibfnamefont {J.~Y.}\ \bibnamefont {Lee}},\
  }\href {\doibase 10.1103/PhysRevB.99.235112} {\bibfield  {journal} {\bibinfo
  {journal} {Phys. Rev. B}\ }\textbf {\bibinfo {volume} {99}},\ \bibinfo
  {pages} {235112} (\bibinfo {year} {2019})}\BibitemShut {NoStop}%
\bibitem [{\citenamefont {Kawabata}\ \emph
  {et~al.}(2019{\natexlab{b}})\citenamefont {Kawabata}, \citenamefont
  {Shiozaki}, \citenamefont {Ueda},\ and\ \citenamefont
  {Sato}}]{PhysRevX.9.041015}%
  \BibitemOpen
  \bibfield  {author} {\bibinfo {author} {\bibfnamefont {K.}~\bibnamefont
  {Kawabata}}, \bibinfo {author} {\bibfnamefont {K.}~\bibnamefont {Shiozaki}},
  \bibinfo {author} {\bibfnamefont {M.}~\bibnamefont {Ueda}}, \ and\ \bibinfo
  {author} {\bibfnamefont {M.}~\bibnamefont {Sato}},\ }\href {\doibase
  10.1103/PhysRevX.9.041015} {\bibfield  {journal} {\bibinfo  {journal} {Phys.
  Rev. X}\ }\textbf {\bibinfo {volume} {9}},\ \bibinfo {pages} {041015}
  (\bibinfo {year} {2019}{\natexlab{b}})}\BibitemShut {NoStop}%
\bibitem [{\citenamefont {Liu}\ \emph {et~al.}(2019{\natexlab{b}})\citenamefont
  {Liu}, \citenamefont {Jiang},\ and\ \citenamefont
  {Chen}}]{PhysRevB.99.125103}%
  \BibitemOpen
  \bibfield  {author} {\bibinfo {author} {\bibfnamefont {C.-H.}\ \bibnamefont
  {Liu}}, \bibinfo {author} {\bibfnamefont {H.}~\bibnamefont {Jiang}}, \ and\
  \bibinfo {author} {\bibfnamefont {S.}~\bibnamefont {Chen}},\ }\href {\doibase
  10.1103/PhysRevB.99.125103} {\bibfield  {journal} {\bibinfo  {journal} {Phys.
  Rev. B}\ }\textbf {\bibinfo {volume} {99}},\ \bibinfo {pages} {125103}
  (\bibinfo {year} {2019}{\natexlab{b}})}\BibitemShut {NoStop}%
\bibitem [{\citenamefont {Lee}\ \emph {et~al.}(2019{\natexlab{b}})\citenamefont
  {Lee}, \citenamefont {Ahn}, \citenamefont {Zhou},\ and\ \citenamefont
  {Vishwanath}}]{PhysRevLett.123.206404}%
  \BibitemOpen
  \bibfield  {author} {\bibinfo {author} {\bibfnamefont {J.~Y.}\ \bibnamefont
  {Lee}}, \bibinfo {author} {\bibfnamefont {J.}~\bibnamefont {Ahn}}, \bibinfo
  {author} {\bibfnamefont {H.}~\bibnamefont {Zhou}}, \ and\ \bibinfo {author}
  {\bibfnamefont {A.}~\bibnamefont {Vishwanath}},\ }\href {\doibase
  10.1103/PhysRevLett.123.206404} {\bibfield  {journal} {\bibinfo  {journal}
  {Phys. Rev. Lett.}\ }\textbf {\bibinfo {volume} {123}},\ \bibinfo {pages}
  {206404} (\bibinfo {year} {2019}{\natexlab{b}})}\BibitemShut {NoStop}%
\bibitem [{\citenamefont {Hamazaki}\ \emph {et~al.}(2019)\citenamefont
  {Hamazaki}, \citenamefont {Kawabata},\ and\ \citenamefont
  {Ueda}}]{PhysRevLett.123.090603}%
  \BibitemOpen
  \bibfield  {author} {\bibinfo {author} {\bibfnamefont {R.}~\bibnamefont
  {Hamazaki}}, \bibinfo {author} {\bibfnamefont {K.}~\bibnamefont {Kawabata}},
  \ and\ \bibinfo {author} {\bibfnamefont {M.}~\bibnamefont {Ueda}},\ }\href
  {\doibase 10.1103/PhysRevLett.123.090603} {\bibfield  {journal} {\bibinfo
  {journal} {Phys. Rev. Lett.}\ }\textbf {\bibinfo {volume} {123}},\ \bibinfo
  {pages} {090603} (\bibinfo {year} {2019})}\BibitemShut {NoStop}%
\bibitem [{\citenamefont {Longhi}(2019)}]{PhysRevLett.122.237601}%
  \BibitemOpen
  \bibfield  {author} {\bibinfo {author} {\bibfnamefont {S.}~\bibnamefont
  {Longhi}},\ }\href {\doibase 10.1103/PhysRevLett.122.237601} {\bibfield
  {journal} {\bibinfo  {journal} {Phys. Rev. Lett.}\ }\textbf {\bibinfo
  {volume} {122}},\ \bibinfo {pages} {237601} (\bibinfo {year}
  {2019})}\BibitemShut {NoStop}%
\bibitem [{\citenamefont {H\"ockendorf}\ \emph {et~al.}(2019)\citenamefont
  {H\"ockendorf}, \citenamefont {Alvermann},\ and\ \citenamefont
  {Fehske}}]{PhysRevLett.123.190403}%
  \BibitemOpen
  \bibfield  {author} {\bibinfo {author} {\bibfnamefont {B.}~\bibnamefont
  {H\"ockendorf}}, \bibinfo {author} {\bibfnamefont {A.}~\bibnamefont
  {Alvermann}}, \ and\ \bibinfo {author} {\bibfnamefont {H.}~\bibnamefont
  {Fehske}},\ }\href {\doibase 10.1103/PhysRevLett.123.190403} {\bibfield
  {journal} {\bibinfo  {journal} {Phys. Rev. Lett.}\ }\textbf {\bibinfo
  {volume} {123}},\ \bibinfo {pages} {190403} (\bibinfo {year}
  {2019})}\BibitemShut {NoStop}%
\bibitem [{\citenamefont {Kawabata}\ \emph {et~al.}(2018)\citenamefont
  {Kawabata}, \citenamefont {Ashida}, \citenamefont {Katsura},\ and\
  \citenamefont
  {Ueda}}]{kawabataParitytimesymmetricTopologicalSuperconductor2018}%
  \BibitemOpen
  \bibfield  {author} {\bibinfo {author} {\bibfnamefont {K.}~\bibnamefont
  {Kawabata}}, \bibinfo {author} {\bibfnamefont {Y.}~\bibnamefont {Ashida}},
  \bibinfo {author} {\bibfnamefont {H.}~\bibnamefont {Katsura}}, \ and\
  \bibinfo {author} {\bibfnamefont {M.}~\bibnamefont {Ueda}},\ }\href {\doibase
  10.1103/PhysRevB.98.085116} {\bibfield  {journal} {\bibinfo  {journal} {Phys.
  Rev. B}\ }\textbf {\bibinfo {volume} {98}},\ \bibinfo {pages} {085116}
  (\bibinfo {year} {2018})}\BibitemShut {NoStop}%
\bibitem [{\citenamefont {{Yi}}\ and\ \citenamefont
  {{Yang}}(2020)}]{2020arXiv200302219Y}%
  \BibitemOpen
  \bibfield  {author} {\bibinfo {author} {\bibfnamefont {Y.}~\bibnamefont
  {{Yi}}}\ and\ \bibinfo {author} {\bibfnamefont {Z.}~\bibnamefont {{Yang}}},\
  }\href@noop {} {\bibfield  {journal} {\bibinfo  {journal} {arXiv e-prints}\
  ,\ \bibinfo {eid} {arXiv:2003.02219}} (\bibinfo {year} {2020})},\ \Eprint
  {http://arxiv.org/abs/2003.02219} {arXiv:2003.02219} \BibitemShut {NoStop}%
\bibitem [{\citenamefont {Harari}\ \emph {et~al.}(2018)\citenamefont {Harari},
  \citenamefont {Bandres}, \citenamefont {Lumer}, \citenamefont {Rechtsman},
  \citenamefont {Chong}, \citenamefont {Khajavikhan}, \citenamefont
  {Christodoulides},\ and\ \citenamefont {Segev}}]{harari_topological_2018}%
  \BibitemOpen
  \bibfield  {author} {\bibinfo {author} {\bibfnamefont {G.}~\bibnamefont
  {Harari}}, \bibinfo {author} {\bibfnamefont {M.~A.}\ \bibnamefont {Bandres}},
  \bibinfo {author} {\bibfnamefont {Y.}~\bibnamefont {Lumer}}, \bibinfo
  {author} {\bibfnamefont {M.~C.}\ \bibnamefont {Rechtsman}}, \bibinfo {author}
  {\bibfnamefont {Y.~D.}\ \bibnamefont {Chong}}, \bibinfo {author}
  {\bibfnamefont {M.}~\bibnamefont {Khajavikhan}}, \bibinfo {author}
  {\bibfnamefont {D.~N.}\ \bibnamefont {Christodoulides}}, \ and\ \bibinfo
  {author} {\bibfnamefont {M.}~\bibnamefont {Segev}},\ }\href {\doibase
  10.1126/science.aar4003} {\bibfield  {journal} {\bibinfo  {journal}
  {Science}\ ,\ \bibinfo {pages} {eaar4003}} (\bibinfo {year}
  {2018})}\BibitemShut {NoStop}%
\bibitem [{\citenamefont {Bandres}\ \emph {et~al.}(2018)\citenamefont
  {Bandres}, \citenamefont {Wittek}, \citenamefont {Harari}, \citenamefont
  {Parto}, \citenamefont {Ren}, \citenamefont {Segev}, \citenamefont
  {Christodoulides},\ and\ \citenamefont
  {Khajavikhan}}]{bandres_topological_2018}%
  \BibitemOpen
  \bibfield  {author} {\bibinfo {author} {\bibfnamefont {M.~A.}\ \bibnamefont
  {Bandres}}, \bibinfo {author} {\bibfnamefont {S.}~\bibnamefont {Wittek}},
  \bibinfo {author} {\bibfnamefont {G.}~\bibnamefont {Harari}}, \bibinfo
  {author} {\bibfnamefont {M.}~\bibnamefont {Parto}}, \bibinfo {author}
  {\bibfnamefont {J.}~\bibnamefont {Ren}}, \bibinfo {author} {\bibfnamefont
  {M.}~\bibnamefont {Segev}}, \bibinfo {author} {\bibfnamefont {D.~N.}\
  \bibnamefont {Christodoulides}}, \ and\ \bibinfo {author} {\bibfnamefont
  {M.}~\bibnamefont {Khajavikhan}},\ }\href {\doibase 10.1126/science.aar4005}
  {\bibfield  {journal} {\bibinfo  {journal} {Science}\ ,\ \bibinfo {pages}
  {eaar4005}} (\bibinfo {year} {2018})}\BibitemShut {NoStop}%
\bibitem [{\citenamefont {Bahari}\ \emph {et~al.}(2017)\citenamefont {Bahari},
  \citenamefont {Ndao}, \citenamefont {Vallini}, \citenamefont {El~Amili},
  \citenamefont {Fainman},\ and\ \citenamefont {Kant{\'e}}}]{Bahari636}%
  \BibitemOpen
  \bibfield  {author} {\bibinfo {author} {\bibfnamefont {B.}~\bibnamefont
  {Bahari}}, \bibinfo {author} {\bibfnamefont {A.}~\bibnamefont {Ndao}},
  \bibinfo {author} {\bibfnamefont {F.}~\bibnamefont {Vallini}}, \bibinfo
  {author} {\bibfnamefont {A.}~\bibnamefont {El~Amili}}, \bibinfo {author}
  {\bibfnamefont {Y.}~\bibnamefont {Fainman}}, \ and\ \bibinfo {author}
  {\bibfnamefont {B.}~\bibnamefont {Kant{\'e}}},\ }\href {\doibase
  10.1126/science.aao4551} {\bibfield  {journal} {\bibinfo  {journal}
  {Science}\ }\textbf {\bibinfo {volume} {358}},\ \bibinfo {pages} {636}
  (\bibinfo {year} {2017})}\BibitemShut {NoStop}%
\bibitem [{\citenamefont {Yoshida}\ \emph {et~al.}(2019)\citenamefont
  {Yoshida}, \citenamefont {Peters}, \citenamefont {Kawakami},\ and\
  \citenamefont {Hatsugai}}]{PhysRevB.99.121101}%
  \BibitemOpen
  \bibfield  {author} {\bibinfo {author} {\bibfnamefont {T.}~\bibnamefont
  {Yoshida}}, \bibinfo {author} {\bibfnamefont {R.}~\bibnamefont {Peters}},
  \bibinfo {author} {\bibfnamefont {N.}~\bibnamefont {Kawakami}}, \ and\
  \bibinfo {author} {\bibfnamefont {Y.}~\bibnamefont {Hatsugai}},\ }\href
  {\doibase 10.1103/PhysRevB.99.121101} {\bibfield  {journal} {\bibinfo
  {journal} {Phys. Rev. B}\ }\textbf {\bibinfo {volume} {99}},\ \bibinfo
  {pages} {121101} (\bibinfo {year} {2019})}\BibitemShut {NoStop}%
\bibitem [{\citenamefont {Michishita}\ and\ \citenamefont
  {Peters}(2020)}]{PhysRevLett.124.196401}%
  \BibitemOpen
  \bibfield  {author} {\bibinfo {author} {\bibfnamefont {Y.}~\bibnamefont
  {Michishita}}\ and\ \bibinfo {author} {\bibfnamefont {R.}~\bibnamefont
  {Peters}},\ }\href {\doibase 10.1103/PhysRevLett.124.196401} {\bibfield
  {journal} {\bibinfo  {journal} {Phys. Rev. Lett.}\ }\textbf {\bibinfo
  {volume} {124}},\ \bibinfo {pages} {196401} (\bibinfo {year}
  {2020})}\BibitemShut {NoStop}%
\bibitem [{\citenamefont {Lieu}\ \emph {et~al.}(2020)\citenamefont {Lieu},
  \citenamefont {McGinley},\ and\ \citenamefont
  {Cooper}}]{PhysRevLett.124.040401}%
  \BibitemOpen
  \bibfield  {author} {\bibinfo {author} {\bibfnamefont {S.}~\bibnamefont
  {Lieu}}, \bibinfo {author} {\bibfnamefont {M.}~\bibnamefont {McGinley}}, \
  and\ \bibinfo {author} {\bibfnamefont {N.~R.}\ \bibnamefont {Cooper}},\
  }\href {\doibase 10.1103/PhysRevLett.124.040401} {\bibfield  {journal}
  {\bibinfo  {journal} {Phys. Rev. Lett.}\ }\textbf {\bibinfo {volume} {124}},\
  \bibinfo {pages} {040401} (\bibinfo {year} {2020})}\BibitemShut {NoStop}%
\bibitem [{\citenamefont {Kanki}\ \emph {et~al.}(2017)\citenamefont {Kanki},
  \citenamefont {Garmon}, \citenamefont {Tanaka},\ and\ \citenamefont
  {Petrosky}}]{kankiExactDescriptionCoalescing2017}%
  \BibitemOpen
  \bibfield  {author} {\bibinfo {author} {\bibfnamefont {K.}~\bibnamefont
  {Kanki}}, \bibinfo {author} {\bibfnamefont {S.}~\bibnamefont {Garmon}},
  \bibinfo {author} {\bibfnamefont {S.}~\bibnamefont {Tanaka}}, \ and\ \bibinfo
  {author} {\bibfnamefont {T.}~\bibnamefont {Petrosky}},\ }\href {\doibase
  10.1063/1.5002689} {\bibfield  {journal} {\bibinfo  {journal} {Journal of
  Mathematical Physics}\ }\textbf {\bibinfo {volume} {58}},\ \bibinfo {pages}
  {092101} (\bibinfo {year} {2017})}\BibitemShut {NoStop}%
\bibitem [{\citenamefont {Heiss}(2012)}]{heissPhysicsExceptionalPoints2012}%
  \BibitemOpen
  \bibfield  {author} {\bibinfo {author} {\bibfnamefont {W.~D.}\ \bibnamefont
  {Heiss}},\ }\href@noop {} {\bibfield  {journal} {\bibinfo  {journal} {Journal
  of Physics A: Mathematical and Theoretical}\ }\textbf {\bibinfo {volume}
  {45}},\ \bibinfo {pages} {444016} (\bibinfo {year} {2012})}\BibitemShut
  {NoStop}%
\bibitem [{\citenamefont
  {Heiss}(2004)}]{heissExceptionalPointsNonHermitian2004}%
  \BibitemOpen
  \bibfield  {author} {\bibinfo {author} {\bibfnamefont {W.~D.}\ \bibnamefont
  {Heiss}},\ }\href@noop {} {\bibfield  {journal} {\bibinfo  {journal} {Journal
  of Physics A: Mathematical and General}\ }\textbf {\bibinfo {volume} {37}},\
  \bibinfo {pages} {2455} (\bibinfo {year} {2004})}\BibitemShut {NoStop}%
\bibitem [{\citenamefont {Dembowski}\ \emph {et~al.}(2001)\citenamefont
  {Dembowski}, \citenamefont {Gr\"af}, \citenamefont {Harney}, \citenamefont
  {Heine}, \citenamefont {Heiss}, \citenamefont {Rehfeld},\ and\ \citenamefont
  {Richter}}]{PhysRevLett.86.787}%
  \BibitemOpen
  \bibfield  {author} {\bibinfo {author} {\bibfnamefont {C.}~\bibnamefont
  {Dembowski}}, \bibinfo {author} {\bibfnamefont {H.-D.}\ \bibnamefont
  {Gr\"af}}, \bibinfo {author} {\bibfnamefont {H.~L.}\ \bibnamefont {Harney}},
  \bibinfo {author} {\bibfnamefont {A.}~\bibnamefont {Heine}}, \bibinfo
  {author} {\bibfnamefont {W.~D.}\ \bibnamefont {Heiss}}, \bibinfo {author}
  {\bibfnamefont {H.}~\bibnamefont {Rehfeld}}, \ and\ \bibinfo {author}
  {\bibfnamefont {A.}~\bibnamefont {Richter}},\ }\href {\doibase
  10.1103/PhysRevLett.86.787} {\bibfield  {journal} {\bibinfo  {journal} {Phys.
  Rev. Lett.}\ }\textbf {\bibinfo {volume} {86}},\ \bibinfo {pages} {787}
  (\bibinfo {year} {2001})}\BibitemShut {NoStop}%
\bibitem [{\citenamefont {Klaiman}\ \emph {et~al.}(2008)\citenamefont
  {Klaiman}, \citenamefont {G\"unther},\ and\ \citenamefont
  {Moiseyev}}]{PhysRevLett.101.080402}%
  \BibitemOpen
  \bibfield  {author} {\bibinfo {author} {\bibfnamefont {S.}~\bibnamefont
  {Klaiman}}, \bibinfo {author} {\bibfnamefont {U.}~\bibnamefont {G\"unther}},
  \ and\ \bibinfo {author} {\bibfnamefont {N.}~\bibnamefont {Moiseyev}},\
  }\href {\doibase 10.1103/PhysRevLett.101.080402} {\bibfield  {journal}
  {\bibinfo  {journal} {Phys. Rev. Lett.}\ }\textbf {\bibinfo {volume} {101}},\
  \bibinfo {pages} {080402} (\bibinfo {year} {2008})}\BibitemShut {NoStop}%
\bibitem [{\citenamefont {Liertzer}\ \emph {et~al.}(2012)\citenamefont
  {Liertzer}, \citenamefont {Ge}, \citenamefont {Cerjan}, \citenamefont
  {Stone}, \citenamefont {T\"ureci},\ and\ \citenamefont
  {Rotter}}]{PhysRevLett.108.173901}%
  \BibitemOpen
  \bibfield  {author} {\bibinfo {author} {\bibfnamefont {M.}~\bibnamefont
  {Liertzer}}, \bibinfo {author} {\bibfnamefont {L.}~\bibnamefont {Ge}},
  \bibinfo {author} {\bibfnamefont {A.}~\bibnamefont {Cerjan}}, \bibinfo
  {author} {\bibfnamefont {A.~D.}\ \bibnamefont {Stone}}, \bibinfo {author}
  {\bibfnamefont {H.~E.}\ \bibnamefont {T\"ureci}}, \ and\ \bibinfo {author}
  {\bibfnamefont {S.}~\bibnamefont {Rotter}},\ }\href {\doibase
  10.1103/PhysRevLett.108.173901} {\bibfield  {journal} {\bibinfo  {journal}
  {Phys. Rev. Lett.}\ }\textbf {\bibinfo {volume} {108}},\ \bibinfo {pages}
  {173901} (\bibinfo {year} {2012})}\BibitemShut {NoStop}%
\bibitem [{\citenamefont
  {Wiersig}(2014)}]{wiersigEnhancingSensitivityFrequency2014}%
  \BibitemOpen
  \bibfield  {author} {\bibinfo {author} {\bibfnamefont {J.}~\bibnamefont
  {Wiersig}},\ }\href {\doibase 10.1103/PhysRevLett.112.203901} {\bibfield
  {journal} {\bibinfo  {journal} {Phys. Rev. Lett.}\ }\textbf {\bibinfo
  {volume} {112}},\ \bibinfo {pages} {203901} (\bibinfo {year}
  {2014})}\BibitemShut {NoStop}%
\bibitem [{\citenamefont {Tanaka}\ \emph {et~al.}(2014)\citenamefont {Tanaka},
  \citenamefont {Kim},\ and\ \citenamefont
  {Cheon}}]{tanakaExoticQuantumHolonomy2014}%
  \BibitemOpen
  \bibfield  {author} {\bibinfo {author} {\bibfnamefont {A.}~\bibnamefont
  {Tanaka}}, \bibinfo {author} {\bibfnamefont {S.~W.}\ \bibnamefont {Kim}}, \
  and\ \bibinfo {author} {\bibfnamefont {T.}~\bibnamefont {Cheon}},\ }\href
  {\doibase 10.1103/PhysRevE.89.042904} {\bibfield  {journal} {\bibinfo
  {journal} {Phys Rev E Stat Nonlin Soft Matter Phys}\ }\textbf {\bibinfo
  {volume} {89}},\ \bibinfo {pages} {042904} (\bibinfo {year}
  {2014})}\BibitemShut {NoStop}%
\bibitem [{\citenamefont {Doppler}\ \emph {et~al.}(2016)\citenamefont
  {Doppler}, \citenamefont {Mailybaev}, \citenamefont {B{\"o}hm}, \citenamefont
  {Kuhl}, \citenamefont {Girschik}, \citenamefont {Libisch}, \citenamefont
  {Milburn}, \citenamefont {Rabl}, \citenamefont {Moiseyev},\ and\
  \citenamefont {Rotter}}]{dopplerDynamicallyEncirclingExceptional2016}%
  \BibitemOpen
  \bibfield  {author} {\bibinfo {author} {\bibfnamefont {J.}~\bibnamefont
  {Doppler}}, \bibinfo {author} {\bibfnamefont {A.~A.}\ \bibnamefont
  {Mailybaev}}, \bibinfo {author} {\bibfnamefont {J.}~\bibnamefont {B{\"o}hm}},
  \bibinfo {author} {\bibfnamefont {U.}~\bibnamefont {Kuhl}}, \bibinfo {author}
  {\bibfnamefont {A.}~\bibnamefont {Girschik}}, \bibinfo {author}
  {\bibfnamefont {F.}~\bibnamefont {Libisch}}, \bibinfo {author} {\bibfnamefont
  {T.~J.}\ \bibnamefont {Milburn}}, \bibinfo {author} {\bibfnamefont
  {P.}~\bibnamefont {Rabl}}, \bibinfo {author} {\bibfnamefont {N.}~\bibnamefont
  {Moiseyev}}, \ and\ \bibinfo {author} {\bibfnamefont {S.}~\bibnamefont
  {Rotter}},\ }\href {\doibase 10.1038/nature18605} {\bibfield  {journal}
  {\bibinfo  {journal} {Nature}\ }\textbf {\bibinfo {volume} {537}},\ \bibinfo
  {pages} {76} (\bibinfo {year} {2016})}\BibitemShut {NoStop}%
\bibitem [{\citenamefont {Hodaei}(2016)}]{hodaeiDarkstateLasersMode2016}%
  \BibitemOpen
  \bibfield  {author} {\bibinfo {author} {\bibfnamefont {H.}~\bibnamefont
  {Hodaei}},\ }\href@noop {} {\bibfield  {journal} {\bibinfo  {journal} {Opt.
  Lett.}\ }\textbf {\bibinfo {volume} {41}},\ \bibinfo {pages} {3049} (\bibinfo
  {year} {2016})}\BibitemShut {NoStop}%
\bibitem [{\citenamefont {Kang}\ \emph {et~al.}(2016)\citenamefont {Kang},
  \citenamefont {Chen},\ and\ \citenamefont
  {Chong}}]{kangChiralExceptionalPoints2016}%
  \BibitemOpen
  \bibfield  {author} {\bibinfo {author} {\bibfnamefont {M.}~\bibnamefont
  {Kang}}, \bibinfo {author} {\bibfnamefont {J.}~\bibnamefont {Chen}}, \ and\
  \bibinfo {author} {\bibfnamefont {Y.~D.}\ \bibnamefont {Chong}},\ }\href@noop
  {} {\bibfield  {journal} {\bibinfo  {journal} {Phys. Rev. A}\ }\textbf
  {\bibinfo {volume} {94}} (\bibinfo {year} {2016})}\BibitemShut {NoStop}%
\bibitem [{\citenamefont {Kim}\ \emph {et~al.}(2016)\citenamefont {Kim},
  \citenamefont {Hwang}, \citenamefont {Kim}, \citenamefont {Choi},
  \citenamefont {No},\ and\ \citenamefont
  {Park}}]{kimDirectObservationExceptional2016}%
  \BibitemOpen
  \bibfield  {author} {\bibinfo {author} {\bibfnamefont {K.-H.}\ \bibnamefont
  {Kim}}, \bibinfo {author} {\bibfnamefont {M.-S.}\ \bibnamefont {Hwang}},
  \bibinfo {author} {\bibfnamefont {H.-R.}\ \bibnamefont {Kim}}, \bibinfo
  {author} {\bibfnamefont {J.-H.}\ \bibnamefont {Choi}}, \bibinfo {author}
  {\bibfnamefont {Y.-S.}\ \bibnamefont {No}}, \ and\ \bibinfo {author}
  {\bibfnamefont {H.-G.}\ \bibnamefont {Park}},\ }\href {\doibase
  10.1038/ncomms13893} {\bibfield  {journal} {\bibinfo  {journal} {Nat Commun}\
  }\textbf {\bibinfo {volume} {7}},\ \bibinfo {pages} {13893} (\bibinfo {year}
  {2016})}\BibitemShut {NoStop}%
\bibitem [{\citenamefont {Gao}\ \emph {et~al.}(2015)\citenamefont {Gao},
  \citenamefont {Estrecho}, \citenamefont {Bliokh}, \citenamefont {Liew},
  \citenamefont {Fraser}, \citenamefont {Brodbeck}, \citenamefont {Kamp},
  \citenamefont {Schneider}, \citenamefont {H{\"o}fling}, \citenamefont
  {Yamamoto}, \citenamefont {Nori}, \citenamefont {Kivshar}, \citenamefont
  {Truscott}, \citenamefont {Dall},\ and\ \citenamefont
  {Ostrovskaya}}]{gaoObservationNonHermitianDegeneracies2015}%
  \BibitemOpen
  \bibfield  {author} {\bibinfo {author} {\bibfnamefont {T.}~\bibnamefont
  {Gao}}, \bibinfo {author} {\bibfnamefont {E.}~\bibnamefont {Estrecho}},
  \bibinfo {author} {\bibfnamefont {K.~Y.}\ \bibnamefont {Bliokh}}, \bibinfo
  {author} {\bibfnamefont {T.~C.~H.}\ \bibnamefont {Liew}}, \bibinfo {author}
  {\bibfnamefont {M.~D.}\ \bibnamefont {Fraser}}, \bibinfo {author}
  {\bibfnamefont {S.}~\bibnamefont {Brodbeck}}, \bibinfo {author}
  {\bibfnamefont {M.}~\bibnamefont {Kamp}}, \bibinfo {author} {\bibfnamefont
  {C.}~\bibnamefont {Schneider}}, \bibinfo {author} {\bibfnamefont
  {S.}~\bibnamefont {H{\"o}fling}}, \bibinfo {author} {\bibfnamefont
  {Y.}~\bibnamefont {Yamamoto}}, \bibinfo {author} {\bibfnamefont
  {F.}~\bibnamefont {Nori}}, \bibinfo {author} {\bibfnamefont {Y.~S.}\
  \bibnamefont {Kivshar}}, \bibinfo {author} {\bibfnamefont {A.~G.}\
  \bibnamefont {Truscott}}, \bibinfo {author} {\bibfnamefont {R.~G.}\
  \bibnamefont {Dall}}, \ and\ \bibinfo {author} {\bibfnamefont {E.~A.}\
  \bibnamefont {Ostrovskaya}},\ }\href {\doibase 10.1038/nature15522}
  {\bibfield  {journal} {\bibinfo  {journal} {Nature}\ }\textbf {\bibinfo
  {volume} {526}},\ \bibinfo {pages} {554} (\bibinfo {year}
  {2015})}\BibitemShut {NoStop}%
\bibitem [{\citenamefont {Ding}\ \emph {et~al.}(2016)\citenamefont {Ding},
  \citenamefont {Ma}, \citenamefont {Xiao}, \citenamefont {Zhang},\ and\
  \citenamefont {Chan}}]{PhysRevX.6.021007}%
  \BibitemOpen
  \bibfield  {author} {\bibinfo {author} {\bibfnamefont {K.}~\bibnamefont
  {Ding}}, \bibinfo {author} {\bibfnamefont {G.}~\bibnamefont {Ma}}, \bibinfo
  {author} {\bibfnamefont {M.}~\bibnamefont {Xiao}}, \bibinfo {author}
  {\bibfnamefont {Z.~Q.}\ \bibnamefont {Zhang}}, \ and\ \bibinfo {author}
  {\bibfnamefont {C.~T.}\ \bibnamefont {Chan}},\ }\href {\doibase
  10.1103/PhysRevX.6.021007} {\bibfield  {journal} {\bibinfo  {journal} {Phys.
  Rev. X}\ }\textbf {\bibinfo {volume} {6}},\ \bibinfo {pages} {021007}
  (\bibinfo {year} {2016})}\BibitemShut {NoStop}%
\bibitem [{\citenamefont {Shi}\ \emph {et~al.}(2016)\citenamefont {Shi},
  \citenamefont {Dubois}, \citenamefont {Chen}, \citenamefont {Cheng},
  \citenamefont {Ramezani}, \citenamefont {Wang},\ and\ \citenamefont
  {Zhang}}]{shiAccessingExceptionalPoints2016}%
  \BibitemOpen
  \bibfield  {author} {\bibinfo {author} {\bibfnamefont {C.}~\bibnamefont
  {Shi}}, \bibinfo {author} {\bibfnamefont {M.}~\bibnamefont {Dubois}},
  \bibinfo {author} {\bibfnamefont {Y.}~\bibnamefont {Chen}}, \bibinfo {author}
  {\bibfnamefont {L.}~\bibnamefont {Cheng}}, \bibinfo {author} {\bibfnamefont
  {H.}~\bibnamefont {Ramezani}}, \bibinfo {author} {\bibfnamefont
  {Y.}~\bibnamefont {Wang}}, \ and\ \bibinfo {author} {\bibfnamefont
  {X.}~\bibnamefont {Zhang}},\ }\href {\doibase 10.1038/ncomms11110} {\bibfield
   {journal} {\bibinfo  {journal} {Nat Commun}\ }\textbf {\bibinfo {volume}
  {7}},\ \bibinfo {pages} {11110} (\bibinfo {year} {2016})}\BibitemShut
  {NoStop}%
\bibitem [{\citenamefont {Xu}\ \emph {et~al.}(2017{\natexlab{b}})\citenamefont
  {Xu}, \citenamefont {Du}, \citenamefont {Huang},\ and\ \citenamefont
  {Zhang}}]{xuDetectingTopologicalExceptional2017}%
  \BibitemOpen
  \bibfield  {author} {\bibinfo {author} {\bibfnamefont {J.}~\bibnamefont
  {Xu}}, \bibinfo {author} {\bibfnamefont {Y.-X.}\ \bibnamefont {Du}}, \bibinfo
  {author} {\bibfnamefont {W.}~\bibnamefont {Huang}}, \ and\ \bibinfo {author}
  {\bibfnamefont {D.-W.}\ \bibnamefont {Zhang}},\ }\href {\doibase
  10.1364/OE.25.015786} {\bibfield  {journal} {\bibinfo  {journal} {Opt
  Express}\ }\textbf {\bibinfo {volume} {25}},\ \bibinfo {pages} {15786}
  (\bibinfo {year} {2017}{\natexlab{b}})}\BibitemShut {NoStop}%
\bibitem [{\citenamefont {Hodaei}\ \emph {et~al.}(2017)\citenamefont {Hodaei},
  \citenamefont {Hassan}, \citenamefont {Wittek}, \citenamefont
  {{Garcia-Gracia}}, \citenamefont {{El-Ganainy}}, \citenamefont
  {Christodoulides},\ and\ \citenamefont
  {Khajavikhan}}]{hodaeiEnhancedSensitivityHigherorder2017a}%
  \BibitemOpen
  \bibfield  {author} {\bibinfo {author} {\bibfnamefont {H.}~\bibnamefont
  {Hodaei}}, \bibinfo {author} {\bibfnamefont {A.~U.}\ \bibnamefont {Hassan}},
  \bibinfo {author} {\bibfnamefont {S.}~\bibnamefont {Wittek}}, \bibinfo
  {author} {\bibfnamefont {H.}~\bibnamefont {{Garcia-Gracia}}}, \bibinfo
  {author} {\bibfnamefont {R.}~\bibnamefont {{El-Ganainy}}}, \bibinfo {author}
  {\bibfnamefont {D.~N.}\ \bibnamefont {Christodoulides}}, \ and\ \bibinfo
  {author} {\bibfnamefont {M.}~\bibnamefont {Khajavikhan}},\ }\href {\doibase
  10.1038/nature23280} {\bibfield  {journal} {\bibinfo  {journal} {Nature}\
  }\textbf {\bibinfo {volume} {548}},\ \bibinfo {pages} {187} (\bibinfo {year}
  {2017})}\BibitemShut {NoStop}%
\bibitem [{\citenamefont {Pick}\ \emph {et~al.}(2017)\citenamefont {Pick},
  \citenamefont {Zhen}, \citenamefont {Miller}, \citenamefont {Hsu},
  \citenamefont {Hernandez}, \citenamefont {Rodriguez}, \citenamefont {Solja{\v
  c}i{\'c}},\ and\ \citenamefont {Johnson}}]{pickGeneralTheorySpontaneous2017}%
  \BibitemOpen
  \bibfield  {author} {\bibinfo {author} {\bibfnamefont {A.}~\bibnamefont
  {Pick}}, \bibinfo {author} {\bibfnamefont {B.}~\bibnamefont {Zhen}}, \bibinfo
  {author} {\bibfnamefont {O.~D.}\ \bibnamefont {Miller}}, \bibinfo {author}
  {\bibfnamefont {C.~W.}\ \bibnamefont {Hsu}}, \bibinfo {author} {\bibfnamefont
  {F.}~\bibnamefont {Hernandez}}, \bibinfo {author} {\bibfnamefont {A.~W.}\
  \bibnamefont {Rodriguez}}, \bibinfo {author} {\bibfnamefont {M.}~\bibnamefont
  {Solja{\v c}i{\'c}}}, \ and\ \bibinfo {author} {\bibfnamefont {S.~G.}\
  \bibnamefont {Johnson}},\ }\href {\doibase 10.1364/OE.25.012325} {\bibfield
  {journal} {\bibinfo  {journal} {Opt Express}\ }\textbf {\bibinfo {volume}
  {25}},\ \bibinfo {pages} {12325} (\bibinfo {year} {2017})}\BibitemShut
  {NoStop}%
\bibitem [{\citenamefont {Jing}\ \emph {et~al.}(2017)\citenamefont {Jing},
  \citenamefont {Ozdemir}, \citenamefont {L{\"u}},\ and\ \citenamefont
  {Nori}}]{jingHighorderExceptionalPoints2017}%
  \BibitemOpen
  \bibfield  {author} {\bibinfo {author} {\bibfnamefont {H.}~\bibnamefont
  {Jing}}, \bibinfo {author} {\bibfnamefont {S.~K.}\ \bibnamefont {Ozdemir}},
  \bibinfo {author} {\bibfnamefont {H.}~\bibnamefont {L{\"u}}}, \ and\ \bibinfo
  {author} {\bibfnamefont {F.}~\bibnamefont {Nori}},\ }\href@noop {} {\bibfield
   {journal} {\bibinfo  {journal} {Sci. Rep.}\ }\textbf {\bibinfo {volume} {7}}
  (\bibinfo {year} {2017})}\BibitemShut {NoStop}%
\bibitem [{\citenamefont {L\"u}\ \emph {et~al.}(2017)\citenamefont {L\"u},
  \citenamefont {\"Ozdemir}, \citenamefont {Kuang}, \citenamefont {Nori},\ and\
  \citenamefont {Jing}}]{PhysRevApplied.8.044020}%
  \BibitemOpen
  \bibfield  {author} {\bibinfo {author} {\bibfnamefont {H.}~\bibnamefont
  {L\"u}}, \bibinfo {author} {\bibfnamefont {S.~K.}\ \bibnamefont {\"Ozdemir}},
  \bibinfo {author} {\bibfnamefont {L.-M.}\ \bibnamefont {Kuang}}, \bibinfo
  {author} {\bibfnamefont {F.}~\bibnamefont {Nori}}, \ and\ \bibinfo {author}
  {\bibfnamefont {H.}~\bibnamefont {Jing}},\ }\href {\doibase
  10.1103/PhysRevApplied.8.044020} {\bibfield  {journal} {\bibinfo  {journal}
  {Phys. Rev. Applied}\ }\textbf {\bibinfo {volume} {8}},\ \bibinfo {pages}
  {044020} (\bibinfo {year} {2017})}\BibitemShut {NoStop}%
\bibitem [{\citenamefont {Cerjan}\ \emph
  {et~al.}(2018{\natexlab{b}})\citenamefont {Cerjan}, \citenamefont {Xiao},
  \citenamefont {Yuan},\ and\ \citenamefont
  {Fan}}]{cerjanEffectsNonHermitianPerturbations2018}%
  \BibitemOpen
  \bibfield  {author} {\bibinfo {author} {\bibfnamefont {A.}~\bibnamefont
  {Cerjan}}, \bibinfo {author} {\bibfnamefont {M.}~\bibnamefont {Xiao}},
  \bibinfo {author} {\bibfnamefont {L.}~\bibnamefont {Yuan}}, \ and\ \bibinfo
  {author} {\bibfnamefont {S.}~\bibnamefont {Fan}},\ }\href {\doibase
  10.1103/PhysRevB.97.075128} {\bibfield  {journal} {\bibinfo  {journal} {Phys.
  Rev. B}\ }\textbf {\bibinfo {volume} {97}},\ \bibinfo {pages} {075128}
  (\bibinfo {year} {2018}{\natexlab{b}})}\BibitemShut {NoStop}%
\bibitem [{\citenamefont {Lyubarov}\ and\ \citenamefont
  {Poddubny}(2018)}]{lyubarovExceptionalPointsPhoton2018}%
  \BibitemOpen
  \bibfield  {author} {\bibinfo {author} {\bibfnamefont {M.}~\bibnamefont
  {Lyubarov}}\ and\ \bibinfo {author} {\bibfnamefont {A.}~\bibnamefont
  {Poddubny}},\ }\href {\doibase 10.1364/OL.43.005917} {\bibfield  {journal}
  {\bibinfo  {journal} {Opt Lett}\ }\textbf {\bibinfo {volume} {43}},\ \bibinfo
  {pages} {5917} (\bibinfo {year} {2018})}\BibitemShut {NoStop}%
\bibitem [{\citenamefont {Yi}\ \emph {et~al.}(2018)\citenamefont {Yi},
  \citenamefont {Kullig},\ and\ \citenamefont
  {Wiersig}}]{yiPairExceptionalPoints2018}%
  \BibitemOpen
  \bibfield  {author} {\bibinfo {author} {\bibfnamefont {C.-H.}\ \bibnamefont
  {Yi}}, \bibinfo {author} {\bibfnamefont {J.}~\bibnamefont {Kullig}}, \ and\
  \bibinfo {author} {\bibfnamefont {J.}~\bibnamefont {Wiersig}},\ }\href
  {\doibase 10.1103/PhysRevLett.120.093902} {\bibfield  {journal} {\bibinfo
  {journal} {Phys. Rev. Lett.}\ }\textbf {\bibinfo {volume} {120}},\ \bibinfo
  {pages} {093902} (\bibinfo {year} {2018})}\BibitemShut {NoStop}%
\bibitem [{\citenamefont {Zhou}\ \emph {et~al.}(2018)\citenamefont {Zhou},
  \citenamefont {Peng}, \citenamefont {Yoon}, \citenamefont {Hsu},
  \citenamefont {Nelson}, \citenamefont {Fu}, \citenamefont {Joannopoulos},
  \citenamefont {Solja{\v c}i{\'c}},\ and\ \citenamefont
  {Zhen}}]{zhouObservationBulkFermi2018}%
  \BibitemOpen
  \bibfield  {author} {\bibinfo {author} {\bibfnamefont {H.}~\bibnamefont
  {Zhou}}, \bibinfo {author} {\bibfnamefont {C.}~\bibnamefont {Peng}}, \bibinfo
  {author} {\bibfnamefont {Y.}~\bibnamefont {Yoon}}, \bibinfo {author}
  {\bibfnamefont {C.~W.}\ \bibnamefont {Hsu}}, \bibinfo {author} {\bibfnamefont
  {K.~A.}\ \bibnamefont {Nelson}}, \bibinfo {author} {\bibfnamefont
  {L.}~\bibnamefont {Fu}}, \bibinfo {author} {\bibfnamefont {J.~D.}\
  \bibnamefont {Joannopoulos}}, \bibinfo {author} {\bibfnamefont
  {M.}~\bibnamefont {Solja{\v c}i{\'c}}}, \ and\ \bibinfo {author}
  {\bibfnamefont {B.}~\bibnamefont {Zhen}},\ }\href {\doibase
  10.1126/science.aap9859} {\bibfield  {journal} {\bibinfo  {journal}
  {Science}\ }\textbf {\bibinfo {volume} {359}},\ \bibinfo {pages} {1009}
  (\bibinfo {year} {2018})}\BibitemShut {NoStop}%
\bibitem [{\citenamefont {Yoshida}\ \emph {et~al.}(2018)\citenamefont
  {Yoshida}, \citenamefont {Peters},\ and\ \citenamefont
  {Kawakami}}]{yoshidaNonHermitianPerspectiveBand2018a}%
  \BibitemOpen
  \bibfield  {author} {\bibinfo {author} {\bibfnamefont {T.}~\bibnamefont
  {Yoshida}}, \bibinfo {author} {\bibfnamefont {R.}~\bibnamefont {Peters}}, \
  and\ \bibinfo {author} {\bibfnamefont {N.}~\bibnamefont {Kawakami}},\ }\href
  {\doibase 10.1103/PhysRevB.98.035141} {\bibfield  {journal} {\bibinfo
  {journal} {Physical Review B}\ }\textbf {\bibinfo {volume} {98}},\ \bibinfo
  {pages} {035141} (\bibinfo {year} {2018})}\BibitemShut {NoStop}%
\bibitem [{\citenamefont {Wang}\ \emph {et~al.}(2019)\citenamefont {Wang},
  \citenamefont {Hou}, \citenamefont {Lu}, \citenamefont {Chen}, \citenamefont
  {Zhang},\ and\ \citenamefont {Chan}}]{wangArbitraryOrderExceptional2019}%
  \BibitemOpen
  \bibfield  {author} {\bibinfo {author} {\bibfnamefont {S.}~\bibnamefont
  {Wang}}, \bibinfo {author} {\bibfnamefont {B.}~\bibnamefont {Hou}}, \bibinfo
  {author} {\bibfnamefont {W.}~\bibnamefont {Lu}}, \bibinfo {author}
  {\bibfnamefont {Y.}~\bibnamefont {Chen}}, \bibinfo {author} {\bibfnamefont
  {Z.~Q.}\ \bibnamefont {Zhang}}, \ and\ \bibinfo {author} {\bibfnamefont
  {C.~T.}\ \bibnamefont {Chan}},\ }\href {\doibase 10.1038/s41467-019-08826-6}
  {\bibfield  {journal} {\bibinfo  {journal} {Nat Commun}\ }\textbf {\bibinfo
  {volume} {10}},\ \bibinfo {pages} {832} (\bibinfo {year} {2019})}\BibitemShut
  {NoStop}%
\bibitem [{\citenamefont {Sweeney}\ \emph {et~al.}(2019)\citenamefont
  {Sweeney}, \citenamefont {Hsu}, \citenamefont {Rotter},\ and\ \citenamefont
  {Stone}}]{sweeneyPerfectlyAbsorbingExceptional2019}%
  \BibitemOpen
  \bibfield  {author} {\bibinfo {author} {\bibfnamefont {W.~R.}\ \bibnamefont
  {Sweeney}}, \bibinfo {author} {\bibfnamefont {C.~W.}\ \bibnamefont {Hsu}},
  \bibinfo {author} {\bibfnamefont {S.}~\bibnamefont {Rotter}}, \ and\ \bibinfo
  {author} {\bibfnamefont {A.~D.}\ \bibnamefont {Stone}},\ }\href {\doibase
  10.1103/PhysRevLett.122.093901} {\bibfield  {journal} {\bibinfo  {journal}
  {Phys. Rev. Lett.}\ }\textbf {\bibinfo {volume} {122}},\ \bibinfo {pages}
  {093901} (\bibinfo {year} {2019})}\BibitemShut {NoStop}%
\bibitem [{\citenamefont {Hassan}\ \emph {et~al.}(2017)\citenamefont {Hassan},
  \citenamefont {Zhen}, \citenamefont {Solja\ifmmode \check{c}\else
  \v{c}\fi{}i\ifmmode~\acute{c}\else \'{c}\fi{}}, \citenamefont {Khajavikhan},\
  and\ \citenamefont {Christodoulides}}]{PhysRevLett.118.093002}%
  \BibitemOpen
  \bibfield  {author} {\bibinfo {author} {\bibfnamefont {A.~U.}\ \bibnamefont
  {Hassan}}, \bibinfo {author} {\bibfnamefont {B.}~\bibnamefont {Zhen}},
  \bibinfo {author} {\bibfnamefont {M.}~\bibnamefont {Solja\ifmmode
  \check{c}\else \v{c}\fi{}i\ifmmode~\acute{c}\else \'{c}\fi{}}}, \bibinfo
  {author} {\bibfnamefont {M.}~\bibnamefont {Khajavikhan}}, \ and\ \bibinfo
  {author} {\bibfnamefont {D.~N.}\ \bibnamefont {Christodoulides}},\ }\href
  {\doibase 10.1103/PhysRevLett.118.093002} {\bibfield  {journal} {\bibinfo
  {journal} {Phys. Rev. Lett.}\ }\textbf {\bibinfo {volume} {118}},\ \bibinfo
  {pages} {093002} (\bibinfo {year} {2017})}\BibitemShut {NoStop}%
\bibitem [{new()}]{newtwoloops}%
  \BibitemOpen
  \href@noop {} {}\bibinfo {note} {We note that in some particular examples,
  e.g. $f_\mu(\bm{k})=(\cos k_x-1)+i(\cos k_y-1)$, there is only one crossing
  point formed by a horizontal loop and a vertical loop in the BZ. However, as
  will be discussed in the following contents, the topological charge of this
  crossing point is zero, which also preserves the doubling
  theorem.}\BibitemShut {Stop}%
\bibitem [{sup()}]{supp_info}%
  \BibitemOpen
  \href@noop {} {}\bibinfo {note} {See Supplemental Material at
  http://link.aps.org/.}\BibitemShut {Stop}%
\bibitem [{Note1()}]{Note1}%
  \BibitemOpen
  \bibinfo {note} {Wo note that this procedure can also be applied to find DPs
  of Hermitian Hamiltonians.}\BibitemShut {Stop}%
\bibitem [{Note2()}]{Note2}%
  \BibitemOpen
  \bibinfo {note} {This is because in a proper basis, all the matrix elements
  of $\protect \mathcal {H}(\protect \bm {k})$ are periodic functions of
  $\protect \bm {k}$, which is equivalent to the single-valued condition.
  Therefore, the corresponding characteristic polynomial $f_E(\protect \bm
  {k})=\protect \qopname \relax m{det}[E-\protect \mathcal {H}(\protect \bm
  {k})]$, whose coefficients are algebraic functions of these matrix elements,
  must also be single-valued.}\BibitemShut {Stop}%
\bibitem [{Note3()}]{Note3}%
  \BibitemOpen
  \bibinfo {note} {Similar to the FPs, it is possible to have single crossing
  between $\protect \operatorname {Re}\left (\protect \operatorname
  {Disc}_{E}[\protect \mathcal {H}](\protect \bm {k})\right )=0$ and $\protect
  \operatorname {Im}\left (\protect \operatorname {Disc}_{E}[\protect \mathcal
  {H}](\protect \bm {k})\right )=0$. However, the topological charge (which
  will be defined in the following contents) at the crossing point must also be
  zero, which is unstable to external weak perturbations.}\BibitemShut {Stop}%
\bibitem [{exa()}]{example2}%
  \BibitemOpen
  \href@noop {} {}\bibinfo {note} {One can easily check that both $F ( {\bk}
  )=G ( {\bk} )=0$ at $(0,0)$ and $(\pi,0)$.}\BibitemShut {Stop}%
\bibitem [{Note4()}]{Note4}%
  \BibitemOpen
  \bibinfo {note} {Here, we implicitly assume that there is no skin effect, in
  which case the bulk spectrum would drastically depend on the boundary
  conditions. Note that in the presence of certain symmetry (e.g., reflection
  in spinless systems~\cite {2020arXiv200302219Y}), the skin effect is always
  absent.}\BibitemShut {Stop}%
\bibitem [{\citenamefont {{Denner}}\ \emph {et~al.}(2020)\citenamefont
  {{Denner}}, \citenamefont {{Skurativska}}, \citenamefont {{Schindler}},
  \citenamefont {{Fischer}}, \citenamefont {{Thomale}}, \citenamefont
  {{Bzdu{\v{s}}ek}},\ and\ \citenamefont {{Neupert}}}]{2020arXiv200801090D}%
  \BibitemOpen
  \bibfield  {author} {\bibinfo {author} {\bibfnamefont {M.~M.}\ \bibnamefont
  {{Denner}}}, \bibinfo {author} {\bibfnamefont {A.}~\bibnamefont
  {{Skurativska}}}, \bibinfo {author} {\bibfnamefont {F.}~\bibnamefont
  {{Schindler}}}, \bibinfo {author} {\bibfnamefont {M.~H.}\ \bibnamefont
  {{Fischer}}}, \bibinfo {author} {\bibfnamefont {R.}~\bibnamefont
  {{Thomale}}}, \bibinfo {author} {\bibfnamefont {T.}~\bibnamefont
  {{Bzdu{\v{s}}ek}}}, \ and\ \bibinfo {author} {\bibfnamefont {T.}~\bibnamefont
  {{Neupert}}},\ }\href@noop {} {\bibfield  {journal} {\bibinfo  {journal}
  {arXiv e-prints}\ ,\ \bibinfo {eid} {arXiv:2008.01090}} (\bibinfo {year}
  {2020})},\ \Eprint {http://arxiv.org/abs/2008.01090} {arXiv:2008.01090}
  \BibitemShut {NoStop}%
\bibitem [{\citenamefont {Wojcik}\ \emph {et~al.}(2019)\citenamefont {Wojcik},
  \citenamefont {Sun}, \citenamefont {Bzdu{\v s}ek},\ and\ \citenamefont
  {Fan}}]{wojcik2019topological}%
  \BibitemOpen
  \bibfield  {author} {\bibinfo {author} {\bibfnamefont {C.~C.}\ \bibnamefont
  {Wojcik}}, \bibinfo {author} {\bibfnamefont {X.-Q.}\ \bibnamefont {Sun}},
  \bibinfo {author} {\bibfnamefont {T.}~\bibnamefont {Bzdu{\v s}ek}}, \ and\
  \bibinfo {author} {\bibfnamefont {S.}~\bibnamefont {Fan}},\ }\href@noop {}
  {\bibfield  {journal} {\bibinfo  {journal} {ArXiv e-prints}\ } (\bibinfo
  {year} {2019})},\ \Eprint {http://arxiv.org/abs/1911.12748}
  {arXiv:1911.12748} \BibitemShut {NoStop}%
\bibitem [{\citenamefont {I.M.~Gelfand}(1994)}]{Discriminant}%
  \BibitemOpen
  \bibfield  {author} {\bibinfo {author} {\bibfnamefont {A.~Z.}\ \bibnamefont
  {I.M.~Gelfand}, \bibfnamefont {M.M.~Kapranov}},\ }\href@noop {} {\emph
  {\bibinfo {title} {Discriminants, Resultants and Multidimen}}}\ (\bibinfo
  {publisher} {Birkhauser},\ \bibinfo {year} {1994})\BibitemShut {NoStop}%
\bibitem [{\citenamefont {Woody}()}]{Resultant1}%
  \BibitemOpen
  \bibfield  {author} {\bibinfo {author} {\bibfnamefont {H.}~\bibnamefont
  {Woody}},\ }\href@noop {} {\enquote {\bibinfo {title} {Polynomial
  resultants},}\ }\BibitemShut {NoStop}%
\bibitem [{\citenamefont {JANSON}(2010)}]{Resultant2}%
  \BibitemOpen
  \bibfield  {author} {\bibinfo {author} {\bibfnamefont {S.}~\bibnamefont
  {JANSON}},\ }\href@noop {} {\enquote {\bibinfo {title} {Resultant and
  discriminant of polynomials},}\ } (\bibinfo {year} {2010})\BibitemShut
  {NoStop}%
\end{thebibliography}%
\bibliographystyle{apsrev4-1}

\setcounter{secnumdepth}{3}
\setcounter{equation}{0}
\setcounter{figure}{0}
\renewcommand{\theequation}{S-\arabic{equation}}
\renewcommand{\thefigure}{S\arabic{figure}}
\newcommand\Scite[1]{[S\citealp{#1}]}

\makeatletter 

\makeatother


 \clearpage
\newpage

\onecolumngrid
\widetext

\begin{center}
\textbf{
\large{Supplementary Materials for}}
\vspace{0.4cm}

\makeatletter \renewcommand\@biblabel[1]{[S#1]} \makeatother

\textbf{
\hspace{-0.25cm} \large{
``Fermion doubling theorems in 2D non-Hermitian systems for Fermi points and exceptional points" } 
}
\end{center}

\vspace{0.1cm}

\begin{center}
\textbf{Authors:} 
ZhesenYang, A.\ P.\ Schnyder, Jiangping Hu, and Ching-Kai Chiu
\end{center}

\vspace{0.8cm}

These supplementary materials (SM) are organized as follows. In Sec.~\ref{I}  we review some mathematical properties of the discriminant, which are important for calculating the discriminant number $\nu$ and for proving the doubling theorems. We also give some examples on how to use the discriminant to find DPs.  In Sec.~\ref{II} the splitting of three-fold degenerate EPs into two-fold degenerate EPs by generic perturbations is discussed in terms of an example. In Sec.~\ref{III} we derive the relation between the discriminant number~\eqref{invariant disc} and the vorticity invariant~\eqref{Fu}. In Sec.~\ref{cut} we show a two-fold degenerate EP is always a branch cut termination of the energy spectrum. In Sec.~\ref{IV}, we show that any NDP and high-ordered EP are unstable by computing the number of conditions that must be satisfied to form NDPs and EPs with any discriminant number $\nu$. In Sec.~\ref{surface violation} we present two examples to
study 3D non-Hermitian systems with anomalous surfaces that break the EP/FP doubling theorems. We show that in the presence of inversion or reflection symmetry,
3D systems with such anomalous surfaces must have at least one exceptional/Fermi line in the bulk. In the absence of inversion and reflection symmetry, however, there can exist
3D Hamiltonians with a bulk point gap and anomalous FPs on the surface.


\maketitle
\section{Definition and properties of the discriminant} \label{I} 

The discriminant $ \operatorname{Disc}_{E}[\mathcal{H}](\bm{k})$ of the characteristic polynomial $f_E(\bm{k})$ of $\mathcal{H} (\bm{k} )$ is the key quantity that we use in the main text to define the discriminant number $\nu$ and to characterize DPs. The characteristic polynomial of the Bloch Hamiltonian  $\mathcal{H} (\bm{k} )$ is defined as
\begin{equation}
f_E (\bm{k})=\det[E-\mathcal{H}(\bm{k})]=\prod_{i=1}^n [E-E_i(\bm{k})], 
\end{equation} 
where $E_i(\bm{k})$ is the $i$th eigenvalue of the non-Hermitian Hamiltonian $\mathcal{H} (\bm{k} )$. The band structure of $\mathcal{H} (\bm{k} )$ has a DP at $\bm{k}_D$ whenever $E_i(\bm{k}_D)=E_j(\bm{k}_D)$ for some $i\neq j$. In this section we show that such a band degeneracy occurs  if and only if  $\operatorname{Disc}_{E}[\mathcal{H}](\bm{k}_D)=0$. We also show how the zeros of the discriminant, and hence the DPs of $\mathcal{H} (\bm{k})$,  can be computed in an efficient manner from the Sylvester matrix~\cite{Discriminant,Resultant1,Resultant2}. We apply these results to several examples. To simplify the notation we relabel in this SM the discriminant $ \operatorname{Disc}_{E}[\mathcal{H}](\bm{k})$ by $\Delta_f ( \bm{k} ) $, where $f$ is the characteristic polynomial of $\mathcal{H}(\bm{k} )$.

\subsection{Resultant and Sylvester matrix}

Before discussing the discriminant, we start by introducing the resultant and the Sylvester matrix of two polynomials and by reviewing their properties~\cite{Resultant1,Resultant2}.

\hspace{-1.3em} {\bf Definition A.1} (Polynomial). A polynomial $f(x)\in F[x]$ is defined as 
\begin{equation}
f(x)=\prod_{i=1}^n (x-\xi_i)=a_nx^n+a_{n-1}x^{n-1}+...+a_1x+a_0,\quad a_n\neq0 ,
\end{equation}
where each coefficient $a_i$ belongs to the field $F$ and each root $\xi_i$ belongs to the extension of $F$. For example, if $a_n,...,a_0$ are real numbers, $\xi_1,...,\xi_n$ are complex numbers.  \\

For characteristic polynomials of non-Hermitian Hamiltonians $\mathcal{H}$, we choose the field $F$ to be the complex number system $\mathbb{C}$. \\

\hspace{-1.3em} {\bf Definition A.2} (Resultant). Given two polynomials $f(x)=a_nx^n+...+a_0$, $g(x)=b_mx^m+...+b_0\in F[x]$, their resultant relative to the variable $x$ is a polynomial over the field of coefficients of $f(x)$ and $g(x)$, and is defined as 
\begin{equation}
R(f,g)=a_n^mb_m^n\prod_{i,j} (\xi_i-\eta_j),
\end{equation}
where $f(\xi_i)=0$ for $1\leq i\leq n$ and $g(\eta_j)=0$ for $1\leq j\leq m$. \\

\hspace{-1.3em} {\bf Theorem A.3}. {\it Let $f(x)=a_nx^n+...+a_0$, $g(x)=b_mx^m+...+b_0\in F[x]$, 
\begin{enumerate}
	\item Suppose that $f$ has $n$ roots $\xi_1,...,\xi_n$ in some extension of $F$. Then 
	\begin{equation}
	R(f,g)=a_n^m\prod_{i=1}^n g(\xi_i).
	\end{equation}
	\item Suppose that $g$ has $m$ roots $\eta_1,...,\eta_m$ in some extension of $F$. Then 
	\begin{equation}
	R(f,g)=(-1)^{mn}b_m^n\prod_{j=1}^m f(\eta_j). 
	\end{equation}
\end{enumerate} }
\hspace{-1.3em} The proof can be found in Refs.~\onlinecite{Resultant1,Resultant2}. \\

\hspace{-1.3em} {\bf Theorem A.4}. {\it Let $f$ and $g$ be two non-zero polynomials with coefficients in a field $F$. Then $f$ and $g$ have a common root in some extension of $F$ if and only if their resultant $R(f,g)$ is equal to zero. } \\

\hspace{-1.3em} {\it Proof}: Suppose $\gamma$ is their common root, $R(f,g)\propto (\gamma-\gamma)=0$. Conversely, if $R(f,g)=0$, at least one of the factors of $R(f,g)$ must be zero, say $\xi_i-\eta_j=0$, then, $\xi_i=\eta_j$ is their common root. \\

Hence, the resultant can be applied to determine whether two polynomials share a common root. However, by Definition A.2, obtaining the value of the resultant requires to know the roots of each polynomial. The following theorem enables us to calculate the resultant directly according to the coefficients of $f$ and $g$ by using the Sylvester matrix.  \\

\hspace{-1.3em} {\bf Definition A.5} (Sylvester matrix). The Sylvester matrix of two polynomials $f(x)=a_nx^n+...+a_0$, $g(x)=b_mx^m+...+b_0\in F[x]$ is defined by 
\begin{eqnarray}
\mathrm{Syl}(f, g)=\left(\begin{array}{ccccccc}{a_{n}} & {a_{n-1}} & {a_{n-2}} & {\dots} & {0} & {0} & {0} \\ {0} & {a_{n}} & {a_{n-1}} & {\dots} & {0} & {0} & {0} \\ {\vdots} & {\vdots} & {\vdots} & {} & {\vdots} & {\vdots} & {\vdots} \\ {0} & {0} & {0} & {\dots} & {a_{1}} & {a_{0}} & {0} \\ {0} & {0} & {0} & {\cdots} & {a_{2}} & {a_{1}} & {a_{0}} \\ {b_{m}} & {b_{m-1}} & {b_{m-2}} & {\dots} & {0} & {0} & {0} \\ {0} & {b_{m}} & {b_{m-1}} & {\cdots} & {0} & {0} & {0} \\ {\vdots} & {\vdots} & {\vdots} & {} & {\vdots} & {\vdots} & {\vdots} \\ {0} & {0} & {0} & {\cdots} & {b_{1}} & {b_{0}} & {0} \\ {0} & {0} & {0} & {\cdots} & {b_{2}} & {b_{1}} & {b_{0}}\end{array}\right), \label{Syl}
\end{eqnarray}
where $a_n,...,a_0$ are the coefficients of $f$ and $b_m,...,b_0$ are the coefficients of $g$. \\

\hspace{-1.3em} {\bf Theorem A.6}. {\it The resultant of two polynomials $f,g$ equals the determinant of their Sylvester matrix, namely 
\begin{equation}
R(f,g)=\det[\mathrm{Syl}(f, g)]
\end{equation}   }
For example, if $n=3, m=2$, 
\begin{equation}
R(f, g)=\det\left(\begin{array}{lllll}{a_{3}} & {a_{2}} & {a_{1}} & {a_{0}} & {0} \\ {0} & {a_{3}} & {a_{2}} & {a_{1}} & {a_{0}} \\ {b_{2}} & {b_{1}} & {b_{0}} & {0} & {0} \\ {0} & {b_{2}} & {b_{1}} & {b_{0}} & {0} \\ {0} & {0} & {b_{2}} & {b_{1}} & {b_{0}}\end{array}\right). 
\end{equation} 
The proof the this theorem can be found in Refs.~\onlinecite{Resultant1,Resultant2}. 

\subsection{Discriminant}

\hspace{-1.3em} {\bf Definition B.1} (Discriminant). Let $f=a_nx^n+...+a_0$ be a polynomial with coefficients in an arbitrary field $F$. Then the (standard) discriminant of $f$ is defined as 
\begin{equation}
\Delta(f):=a_{n}^{2 n-2} \Delta_{0}(f)=a_{n}^{2 n-2} \prod_{1 \leq i<j \leq n}\left(\xi_{i}-\xi_{j}\right)^{2},
\end{equation}
where $\xi_1,...,\xi_n$ are the roots of $f$ in some extension of $F$. \\

With this definition, we can use the discriminant to directly determine whether the polynomial $f$ has double roots. Namely, we have the following theorem.	\\

\hspace{-1.3em} {\bf Theorem B.2}. {\it Let $f$ be a polynomial of degree $n\geq 1$ with coefficients in a field $F$. Then $f$ has a double root in some extension of $F$ if and only if $\Delta(f)=0$. } \\

	Furthermore, since the resultant can be computed by using its Sylvester matrix, the theorem below provides the connection between the discriminant and the resultant. \\

\hspace{-1.3em} {\bf Theorem B.3}. {\it The discriminant of $f=a_nx^n+...+a_0$ can be expressed by the resultant of $f$ and its derivative $f':=\partial_x f$, namely
\begin{equation}
\Delta(f)=(-1)^{n(n-1)/2}a_n^{-1}R(f,f'). 
\end{equation}}

\hspace{-1.3em} {\it Proof}: From $f(x)=a_{n} \prod_{i=1}^{n}(x-\xi_i)$ and $f^{\prime}\left(\xi_{i}\right)=a_{n} \prod_{j \neq i}\left(\xi_{i}-\xi_{j}\right)$, one can  obtain the following equation by using Theorem A.3,
\begin{equation}
\begin{aligned} R\left(f, f^{\prime}\right) &=a_{n}^{n-1+n} \prod_{i=1}^{n} \prod_{j \neq i}\left(\xi_{i}-\xi_{j}\right)=a_{n}^{2 n-1} \prod_{1 \leq i<j \leq n}\left(\xi_{i}-\xi_{j}\right)\left(\xi_{j}-\xi_{i}\right) \\ &=(-1)^{n(n-1) / 2} a_{n}^{2 n-1} \Delta_{0}(f)=(-1)^{n(n-1) / 2} a_{n} \Delta(f). \end{aligned}
\end{equation}
Without knowing any root values, the explicit form of the discriminant can be obtained directly from the coefficients of the polynomial. 
For non-Hermitian Hamiltonians,   the characteristic polynomial has coefficient $a_n=1$, so that we have 
\begin{equation}
\Delta(f)=(-1)^{n(n-1) / 2}\det[\mathrm{Syl}(f, f' )]. \label{disc compute}
\end{equation}  
This is the main equation, which we use in the main text to compute the discriminant.

\subsection{Some examples}

We  use  three example polynomials to show how the discriminant is computed in practice. 

\subsubsection{$n=2$ case}

If $f(x)=ax^2+bx+c$, then 
\begin{equation}
\Delta(f)=-a^{-1} R\left(f, f^{\prime}\right)=-a^{-1}\det\left(\begin{array}{ccc}{a} & {b} & {c} \\ {2 a} & {b} & {0} \\ {0} & {2 a} & {b}\end{array}\right)=b^{2}-4 a c.
\end{equation}

\subsubsection{$n=3$ case}

If $f(x)=ax^3+bx^2+cx+d$, then 
\begin{equation}
\begin{aligned} \Delta(f) &=-a^{-1} R\left(f, f^{\prime}\right)
=
-a^{-1}\det\left(\begin{array}{ccccc}
{a} & {b} & {c} & {d} & {0} \\ 
{0} & {a} & {b} & {c} & {d} \\ 
 {3 a} & {2 b} & {c} & {0}  & {0} \\ 
{0} & {3 a} & {2 b} & {c} & {0} \\ 
{0} & {0} & {3 a} & {2 b} & {c}
\end{array}\right) \\ 
&=
b^{2} c^{2}-4 a c^{3}-4 b^{3} d+18 a b c d-27 a^{2} d^{2}. \end{aligned}
\end{equation} 

\subsubsection{$n=4$ case}

If $f(x)=ax^4+bx^3+cx^2+dx+e$, then 
\begin{equation}
\begin{aligned}
\Delta(f)&=a^{-1} R\left(f, f^{\prime}\right)=a^{-1}\det\left(\begin{array}{ccccccc}
{a} & {b} & {c} & {d} & {e} & {0} & {0} \\ 
{0} & {a} & {b} & {c} & {d} & {e} & {0} \\ 
{0} & {0} & {a} & {b} & {c} & {d} & {e} \\ 
{4 a} & {3 b} & {2 c} & {d} & {0} & {0}  & {0} \\ 
{0} & {4 a} & {3 b} & {2 c} & {d} & {0} & {0} \\ 
{0} & {0} & {4 a} & {3 b} & {2 c} & {d} & {0} \\ 
{0} & {0} & {0} & {4 a} & {3 b} & {2 c} & {d}
\end{array}\right)\\
=& 
b^{2} c^{2} d^{2}-4 b^{2} c^{3} e-4 b^{3} d^{3}+18 b^{3} c d e-27 b^{4} e^{2}-4 a c^{3} d^{2}+16 a c^{4} e \\ &+18 a b c d^{3}-80 a b c^{2} d e-6 a b^{2} d^{2} e+144 a b^{2} c e^{2}-27 a^{2} d^{4}\\
&+144 a^{2} c d^{2} e-128 a^{2} c^{2} e^{2}-192 a^{2} b d e^{2}+256 a^{3} e^{3}.
\end{aligned}
\end{equation} 

\subsection{Application to non-Hermitian Hamiltonians}

Next, we apply the above considerations to characteristic polynomials $f_E ( \bm{k} )$ of non-Hermitian Hamiltonians $\mathcal{H} (  \bm{k} ) $, which have a complex energy spectrum.
We consider generic two-band and three-band Hamiltonians, compute their discriminants using Eqs.~\eqref{Syl} and \eqref{disc compute}, and derive the criteria for the existence of degeneracy points (DPs).

\subsubsection{Two-band example}

Consider a generic two-band model 
\begin{equation}
\mathcal{H}(\bm{k})=h_0(\bm{k}) \sigma_0 +h_x(\bm{k})\sigma_x+h_y(\bm{k})\sigma_y+h_z(\bm{k})\sigma_z, 
\label{twoband}
\end{equation}
where $h_\mu(\bm{k})=h_\mu^r(\bm{k})+i h_\mu^i(\bm{k})$ are complex functions of $\bm{k}$. The characteristic polynomial of the two-band model can be written as 
\begin{equation}
f_E(\bm{k})=E^2+b(\bm{k}) E+c(\bm{k}),
\label{Ex1a}
\end{equation}
where $b(\bm{k})=-2h_0(\bm{k})$ and $c(\bm{k})=h_0^2(\bm{k})-h_x^2(\bm{k})-h_y^2(\bm{k})-h_z^2(\bm{k})$. Computing the discriminant of polynomial~\eqref{Ex1a} 
with respect to the energy $E$,  we obtain the following condition for the existence of DPs
\begin{equation}
\Delta_f(\bm{k})=b^2(\bm{k})-4c(\bm{k})=4[h_x^2(\bm{k})+h_y^2(\bm{k})+h_z^2(\bm{k})]=0.
\label{Ex1b}
\end{equation} 
This condition can also be obtained from the energy spectrum. I.e., the two bands $E_\pm = h_0(\bm{k}) \pm ([h_x^2(\bm{k})+h_y^2(\bm{k})+h_z^2(\bm{k}))^{1/2}$
are degenerate, whenever the square root is vanishing.

\subsubsection{Three-band model}

A generic three-band model is given by
\begin{equation} \label{supp_three_band_model}
\mathcal{H}(\bm{k})=\sum_{\rho=1}^8g_\rho(\bm{k})\lambda_\rho,
\end{equation}
where we have neglected a term proportional to the identity matrix, since it does not affect the band degeneracies.
The eight Gell-Mann matrices $\lambda_\rho$ are
\begin{equation}
\lambda_{1}=\left(\begin{array}{ccc}{0} & {1} & {0} \\ {1} & {0} & {0} \\ {0} & {0} & {0}\end{array}\right), \quad \lambda_{2}=\left(\begin{array}{ccc}{0} & {-i} & {0} \\ {i} & {0} & {0} \\ {0} & {0} & {0}\end{array}\right), \quad \lambda_{3}=\left(\begin{array}{ccc}{1} & {0} & {0} \\ {0} & {-1} & {0} \\ {0} & {0} & {0}\end{array}\right), \quad 
\lambda_{4}=\left(\begin{array}{lll}{0} & {0} & {1} \\ {0} & {0} & {0} \\ {1} & {0} & {0}\end{array}\right),
\end{equation}
\begin{equation}
\lambda_{5}=\left(\begin{array}{ccc}{0} & {0} & {-i} \\ {0} & {0} & {0} \\ {i} & {0} & {0}\end{array}\right), \quad 
\lambda_{6}=\left(\begin{array}{ccc}{0} & {0} & {0} \\ {0} & {0} & {1} \\ {0} & {1} & {0}\end{array}\right), \quad \lambda_{7}=\left(\begin{array}{ccc}{0} & {0} & {0} \\ {0} & {0} & {-i} \\ {0} & {i} & {0}\end{array}\right), \quad \lambda_{8}=\frac{1}{\sqrt{3}}\left(\begin{array}{ccc}{1} & {0} & {0} \\ {0} & {1} & {0} \\ {0} & {0} & {-2}\end{array}\right).
\end{equation}
The characteristic polynomial of Hamiltonian~\eqref{supp_three_band_model} can be written in the short form  
\begin{equation}
f_E (\bm{k})=E^3+c(\bm{k})E+d(\bm{k}) ,
\label{Ex2a}
\end{equation}
where $c=-\sum_{s=1}^{8}g_s^2$ and $d= g_8 \left(-6 g_1^2-6 g_2^2-6 g_3^2+2 g_8^2+3 \left(g_4^2+g_5^2+g_6^2+g_7^2\right)\right)/3^{3/2}-2 g_1 \left(g_4 g_6+g_5 g_7\right)+2 g_2(g_4 g_7-g_5 g_6)+g_3 \left(-g_4^2-g_5^2+g_6^2+g_7^2\right)
$. 
From the discriminant of $f_E (\bm{k})$ we obtain the following condition for the existence of degeneracy points
\begin{equation}
\Delta_f(\bm{k})=-4 c^3(\bm{k})-27 d^2(\bm{k})=0.
\label{Ex1b}
\end{equation} 
This is a relatively simple equation, which can be analyzed analytically in 2D systems.

\begin {figure}[tb!]
\centerline{\includegraphics[height=9cm]{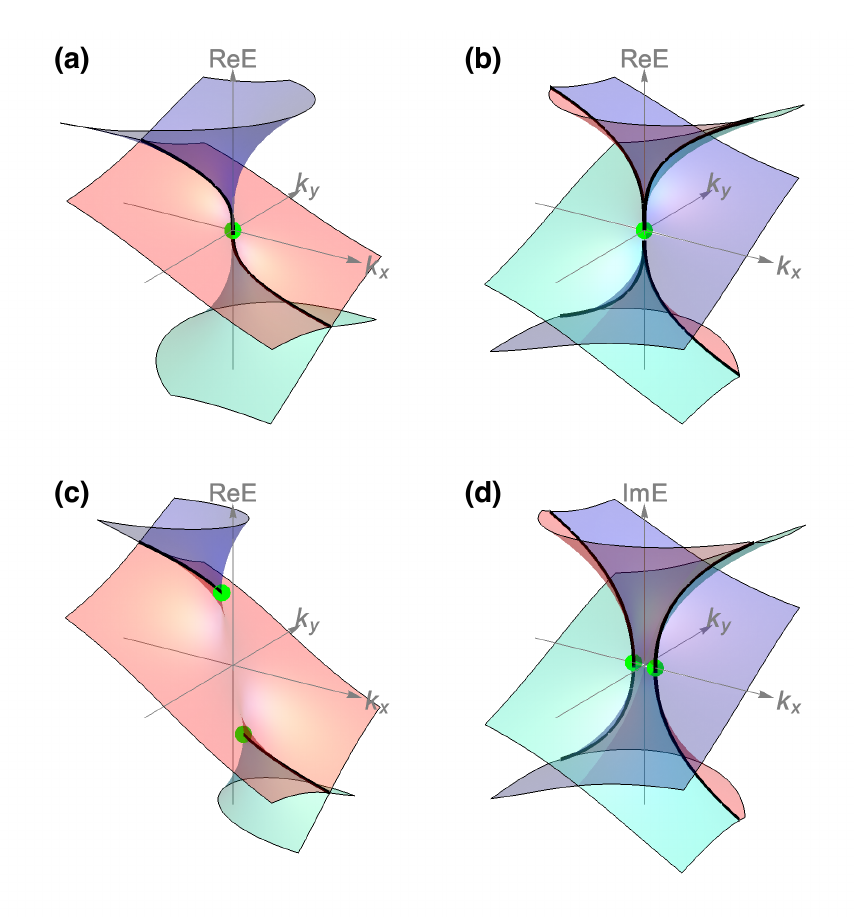}}
\caption{ 
Complex energy dispersions of the  three-band model (\ref{3EP}) 
for $\lambda = 0$ [panels (a) and (b)] and $\lambda = + 1/2$ [panels (c) and (d)].
EPs and branch cuts are indicated by green dots and black arcs, respectively. 
For $\lambda = 0$ there is a three-fold degenerate EP at $\bm{k}=0$. 
For $\lambda = + 1/2$ this EP splits into two two-fold degenerate EPs. 
	\label{P3}}
\end{figure}

\section{Splitting   a three-fold degenerate EP } \label{II}

In this section we use an example to discuss how a three-fold degenerate EP can be split into two-fold degenerate EPs by
 perturbations.  
For this purpose, we consider the following low-energy Hamiltonian with three bands 
\begin{equation}
\mathcal{H}(\delta\bm{k})=\left(\begin{array}{ccc}{0} & {1} & {0} \\ {0} & {0} & {1} \\ {\delta k_{+}} & {0} & {0}\end{array}\right)+\lambda \left(\begin{array}{ccc}{i} & {1} & {0} \\ {1} & {-i} & {0} \\ {0} & {0} & {0}\end{array}\right), \label{3EP}
\end{equation} 
where $\delta k_+= \delta k_x + i\delta k_y$.  The first matrix has a three-fold degenerate EP at $\delta \bm{k} =0$, as shown in Figs.~\ref{P3}(a) and~\ref{P3}(b).
The second matrix, with $|\lambda| \ll 1$, represents a  small perturbation of this EP.
To determine the effects of this perturbation we compute the characteristic polynomial $f_E ( \delta \bm{k} ) =  E^3 - \lambda E -  (1 +  \lambda)  \delta k_+  $
and the discriminant $\Delta_f ( \delta  \bm{k} )= 4\lambda^3-27(1+\lambda)^2\delta k_+^2$. 
From the condition  $\Delta_f ( \delta \bm{k} )=0$, we find that the perturbation
splits the three-fold degenerate EP at $\delta \bm{k} =0$ into two two-fold degenerate EPs, as shown in Figs.~\ref{P3}(c) and~\ref{P3}(d).
The two-fold degenerate EPs are located at  $(\pm k_\lambda,0)$
or $(0,\pm k_\lambda)$, for $\lambda > 0$ and $\lambda < 0$, respectively,  where $k_\lambda= \sqrt{4|\lambda^3|(1+\lambda)^{-2}/27}$. 
Hence, we conclude that three-fold (and higher) degenerate EPs are unstable in the presence of generic perturbations.

\section{Relation between discriminant number and vorticity invariant} \label{III}

Here, we derive the relation between the discriminant number $\nu(\bm{k}_D^l)$, Eq.~\eqref{invariant disc}, and the vorticity invariant~\cite{PhysRevLett.120.146402}
$\nu_{ij} (\bm{k}_D^l)$, Eq.~\eqref{Fu}, from the main text. 
The discriminant number of a DP at $\bm{k}_D^l$ reads
\begin{equation}
\nu(\bm{k}_D^l)=\frac{i}{2\pi}\oint_{\Gamma(\bm{k}_D^l)} d\bm{k} \cdot \nabla_{\bm{k}} \ln \Delta_f(\bm{k}),
\label{E14}
\end{equation}
where $\Delta_f(\bm{k})$ is the discriminant of the characteristic polynomial $f_E ( \bm{k} )$ of the $n\times n$ Hamiltonian $\mathcal{H}( \bm{k} )$, i.e.,  
\begin{equation}
\Delta_f(\bm{k})=\prod_{1 \leq i<j \leq n}\left[E_{i}(\bm{k})-E_{j}(\bm{k})\right]^{2}, 
\end{equation}
where $E_i ( \bm{k} )$ are the energy bands of $\mathcal{H} ( \bm{k} ) $.
Combining the above two equations we obtain
\begin{equation}\begin{aligned}
\nu(\bm{k}_D^l)
&=
\frac{i}{2\pi}\oint_{\Gamma(\bm{k}_D^l)} d\bm{k} \cdot \nabla_{\bm{k}} \ln \prod_{1 \leq i<j \leq n}\left[E_{i}(\boldsymbol{k})-E_{j}(\boldsymbol{k})\right]^2\\
&=
\frac{i}{2\pi}\sum_{i\neq j}\oint_{\Gamma(\bm{k}_D^l)} d\bm{k} \cdot \nabla_{\bm{k}} \ln \left[E_{i}(\boldsymbol{k})-E_{j}(\boldsymbol{k})\right]\\
&=
\frac{-1}{2\pi}\sum_{i \neq j}\oint_{\Gamma(\bm{k}_D^l)} d\bm{k} \cdot \nabla_{\bm{k}} \arg \left[E_{i}(\boldsymbol{k})-E_{j}(\boldsymbol{k})\right]
=
\sum_{i \neq j}\nu_{ij} ( \bm{k}_D^l )    .
\label{E14}
\end{aligned}\end{equation}
In going from the second to the third line we have used $\ln (z)=\ln (|z|)+i \arg (z)$.

\section{Terminations of branch cuts} \label{cut}

	To show that a two-fold degenerate EP with $\nu=\pm 1$ is always the termination of a branch cut in the energy spectrum, we start with a generic non-Hermitian $\mathcal{H}(\bk)$ possessing a two-fold degenerate EP located at $\bk_D$ with energy $E_D$. The Hamiltonian $\mathcal{H}(\bk_D)$ in biorthogonal basis ($\bra{\bar{\psi}_i}$ and $\ket{\psi_i}$) can be written in Jordan canonical form; assuming the EP is the only DP at $\bk_D$, we can rearrange the order of the basis so that the in the Hamiltonian the first $2\times 2$ block for the EP reads 
\bee
\mathcal{H}_{2\times 2}(\bk_D ) =
\left(\begin{array}{cc} 
E_D & 1 \\
0 &  E_D 
\end{array}\right),
\ee
and the remaining part of the Hamiltonian is diagonalized and represents the energy levels away from the EP. Consider momentum $\delta \bk$ slightly deviates away from $\bk_D$ as a perturbation for the Hamiltonian. Within first-order perturbation theory, the first block, which is the projection of $\ket{\phi_1}$ and $\ket{\phi_2}$, still decouples from the diagonal part of the Hamiltonian and describes the effective Hamiltonian near the EP in form of  
\bee 
\mathcal{H}_{\rm{eff}}(\bk_D + \delta \bk) \approx h_0(\delta \bk)\sigma_0 + \sum_{l=1}^3 h_l(\delta \bk)\sigma_l. 
\ee
The energy spectrum is simply given by 
\bee
E_\pm( \delta \bk) = h_0(\delta \bk)\pm\Delta E ( \delta \bk),
\ee
where $\Delta E( \delta \bk) =\sqrt{\sum_{l=1}^3 h_l^2(\delta \bk)}$. Each entry of $H_{\rm{eff}}(\bk_D + \delta \bk)$ is single-valued of $\delta \bk$ by inheriting the property of the original Hamiltonian ($\mathcal{H}(\bk)$), which is single-valued in a proper basis. Hence, $h_\alpha(\delta \bk)$ is single-valued, and only $\Delta E$ determines if the energy spectrum possesses a branch cut. 

	To show the branch cut's existence, the definition and value of the discriminant number are used to impose a constraint for the energy spectrum
\begin{align}
\pm 1 &= \frac{i}{2\pi} \oint _{\Gamma(\bk_D)} d\bk \cdot \nabla_{\bk } \ln{ \rm{Disc}}[\mathcal{H}](\bk) \\
&=  \frac{i}{2\pi} \oint _{\Gamma(\bk_D)} d\delta\bk \cdot \nabla_{\bk } \ln{ \rm{Disc}}[\mathcal{H}_{\rm{eff}}](\bk_D + \delta \bk) \\
& =  \frac{i}{\pi} \oint _{\Gamma(\bk_D)} d\delta\bk \cdot \nabla_{\bk } \ln \Delta E ( \delta \bk),
\end{align}
where the integral path is an arbitrary small closed path encircling $\bk_D$. Since the energy levels away from the EP do not make any contribution to the integral, in the integral ${ \rm{Disc}}[\mathcal{H}](\bk)$ can be simplified to ${ \rm{Disc}}[\mathcal{H}_{\rm{eff}}](\bk_D + \delta \bk)$. Thus, the phase of $\Delta E$ moves along the integral path is limited to $\pm \pi $, which indicates the existence of a branch cut. That is, any integral path enclosing the EP must cross over the branch cut once; hence, the EP must be the termination of the branch cut.

\section{Stability of NDP\MakeLowercase{s} and EP\MakeLowercase{s} }\label{IV} 

From the insights gained by the EP examples in the main text, we learned that any NDPs and high ordered EPs are not robust with small perturbations. 
To show the instability, we first proceed to prove that NDPs are unstable, i.e., they can be deformed into EPs by generic perturbations.
To this end, we consider the non-Hermitian Hamiltonian 
\bee
\mathcal{H}(\bk )=\mathcal{H}_0(\bk )+\delta \mathcal{H}(\delta \bk), 
\ee
where $\mathcal{H}_0(\bk)$  has an NDP at $\bk_D$ with energy $E_D$ and  discriminant number~$\nu \ne 0$, formed by the first two bands $E_1 ({\bk} )$ and $E_2 ( {\bk} )$.
Moreover, $\delta \mathcal{H} ( \delta {\bk} )$ represents a generic small perturbation of the NDP, which is nonzero only at $\bm{k}_D$ and in its vicinity
  ${\bk}_D \pm \delta {\bk}$. 
Within first order perturbation theory, $\delta \mathcal{H}$ deforms the two degenerate energy levels around $\delta {\bk}$ as
$H_{\rm{eff}} (\delta {\bk} )=E_d \sigma_0 +   \boldsymbol{\delta}  ( \delta {\bk} )  \cdot \boldsymbol{\sigma}$, 
with the complex 3D vector obeying $\boldsymbol{\delta}(0)\neq 0$,
yielding the perturbed energy levels $E'_{1/2 } (\delta {\bk} ) = E_d \pm  \sqrt{  \delta^2_1 ( \delta {\bk} ) + \delta^2_2 ( \delta {\bk} ) + \delta^2_3 ( \delta {\bk} )  } $.
All other energy levels are distorted to
\bee
E_i' ( \delta {\bk} ) = E_i ( {\bk}_D + \delta {\bk} ) + \Delta_{ii} ( \delta {\bk} ) ,  \qquad  \textrm{for $i > 2$},
\ee
where $\Delta_{ij} ( \delta {\bk} )  =\bra{\bar{\phi}_i} \delta  \mathcal{H} (\delta {\bk} )  \ket{\phi_j}$ and 
$\bra{\bar{\phi}_i} $ and $ \ket{\phi_i} $ are the left and right  biorthogonal eigenstates of $\mathcal{H}(\bk_D)$, respectively.  
Inserting these expressions into Eq.~\eqref{Dis}, we find that $\delta \mathcal{H}$  deforms the discriminant 
around ${\bk}_D$ to 
\begin{align}
& \operatorname{Disc}_{E}[\mathcal{H}](\bm{k}_D+\delta \bk) \nonumber \\
& = 
[ E_1' (\delta {\bk} )-E_2' (\delta {\bk} ) ]^2
 \prod_{2<i<j}\left[E_i -E_j +\Delta_{ii}-\Delta_{jj}\right]^{2} \nonumber \\
& \times \prod_{2<i} \left[E_1'(\delta \bm{k})-E_i -\Delta_{ii}\right]^{2} \left[E_2'(\delta \bm{k})-E_i -\Delta_{ii}\right]^{2} .
\end{align} 
For sufficiently small $\delta \mathcal{H}$, only the first factor can become zero, while the other factors are nonvanishing. 
Hence, in order for a DP (i.e., an NDP or EP) to exist in the presence of $\delta \mathcal{H}$ we must have
\bee  \label{product_of_delta_i_square}
[ E_1' (\delta {\bk} )  -E_2' (\delta {\bk}) ]^2=4 [ \delta_1^2 (\delta {\bk})  +\delta_2^2  (\delta {\bk})  +\delta_3^2  (\delta {\bk})  ] = 0, 
\ee
at some $\delta {\bk}$. 
The corresponding eigenstate at this DP is of the form
	\begin{equation}
		\psi_{\pm}\propto \left(-\frac{-\delta_3  (\delta {\bk})   \pm\sqrt{\delta_1^2  (\delta {\bk}) +\delta_2^2  (\delta {\bk})+\delta_3^2  (\delta {\bk}) }}{\delta_1  (\delta {\bk}) +i\delta_2  (\delta {\bk}) },1\right)^t .
	\end{equation}
From the above two equations we see that the NDP is in general an EP, since for $\delta_1^2 (\delta {\bk})  +\delta_2^2  (\delta {\bk})  +\delta_3^2  (\delta {\bk})  = 0$ also the two 
eigenstates $\psi_+$ and $\psi_-$ coalesce. The only exception to this is when all complex $\delta_i$ are zero, in which case we have an NDP by definition. 
Hence, this gives two conditions for the existence of an EP (real and imaginary parts of Eq.~\eqref{product_of_delta_i_square} must be zero) 
and six conditions for the existence of an NDP (all three complex $\delta_i$ must be zero).
 In two dimensions, with two momenta $\delta k_x$ and $\delta k_y$, only two conditions can be satisfied simultaneously, such
that  only EPs can form, whereas NDPs are in general absent. Furthermore, the NDP of $\mathcal{H}_0$ with $\nu \neq 0$ cannot vanish the generic perturbation $\delta H$;  
thus, the NDP must be deformed into one or several EPs. 

By refining the above arguments, we can derive the precise number of conditions that must be satisfied to form an NDP or an EP with a given $\nu$ in the following.
That is, we determine whether EPs and NDPs with discriminant number $\nu$ are stable in the presence of generic perturbations.
The results of this analysis are summarized in Table~\ref{T2}. We observe that in two dimensions only EPs with $\nu = \pm 1$ are stable. 
NDPs and EPs with $| \nu | >1$ are split into several EPs with $| \nu | =1$ by small perturbations.

\begin{table}
\begin{tabular}{|c|c|c|c|}
\hline 
DP type& $\nu$ & Stability in 2D & No.~of conditions  \\ 
\hline \hline
EP & 0 & split / gapped & $3$ \\\hline
EP & $\pm1$ & stable & $2$ \\\hline
EP & $\pm N$($N\geq2$) & split & $N(N+1)$ \\\hline
NDP  & $0$ & split / gapped & 6 \\\hline
NDP  & $\pm 1$ & split & 12 \\\hline
NDP  & $\pm 2N (N\geq 1)$ & split & $3N(N+1)$ \\\hline
NDP  & $\pm (2N+1) (N \geq 1)$ & split & $(3N+1)(N+2)$ \\\hline
\end{tabular}
\caption{\label{character2} 
Number of conditions that must be satisfied to form stable EPs and NDPs with discriminant number $\nu$.
In two spatial dimensions, without extra symmetry constraints, only EPs with $\nu=\pm 1$ are stable.
EPs are always more stable than NDPs, since for a given $\nu$ more conditions must be satisfied to
from NDPs than EPs.	
  } \label{T2}
\end{table}

To derive the number of conditions that must be satisfied for stable EPs and NDPs with discriminant number $\nu$, we focus only on two-fold degenerate DPs, since three-fold and higher-fold DPs are unstable against generic perturbations, see Sec.~\ref{II}. Hence, to describe these two-fold degenerate EPs or NDPs in $d$-dimensional systems we can use the generic two-band  Hamiltonian $\mathcal{H}(\bm{k})$ of Eq.~\eqref{twoband}, i.e.,
\begin{equation}
\mathcal{H}(\bm{k})= h_x(\bm{k})\sigma_x+h_y(\bm{k})\sigma_y+h_z(\bm{k})\sigma_z, 
\label{twoband_zwei}
\end{equation}
where $h_{l}(\bm{k})=h_{l}^r(\bm{k})+ih_{l}^i(\bm{k})$ ($l=x,y,z$) are complex functions and $\bm{k}=(k_1,...,k_d)$ is the $d$-dimensional momentum.
Here, we have neglected a term proportional to the identity matrix, as it does not change the band degeneracies. 
As discussed in Eq.~\eqref{Ex1b}, the locations of DPs are found by solving for the zeros of the discriminant $\Delta_f ( {\bf k} )$, i.e.,  
\begin{equation} \label{disc_cond_gen_two_band}
\tfrac{1}{4} \Delta_f(\bm{k})=h_x^2(\bm{k})+h_y^2(\bm{k})+h_z^2(\bm{k})=0,
\end{equation}
Let us suppose that $\bm{k}_D$ is a zero of the discriminant, i.e., a DP is located at  $\bm{k}_D$ with energy $E=0$. 
There exists an invertible matrix $P$ so that 
\begin{equation}
P^{-1}H(\bm{k}_D)P= \left(\begin{array}{lll}{0} & {0} \\  {0} & {0}\end{array}\right),
\quad 
\textrm{or}
\quad 
\left(\begin{array}{lll}{0} & {a} \\  {0} & {0}\end{array}\right),
\end{equation}
which correspond to an NDP or an EP respectively.
For the NDP this means that all components of the $h$-vector at $\bm{k}_D$ must be zero, while for the EP not all three $h_l ( {\bf k}_D )$ must be zero, but $\Delta_f ( \bm{k}_D )=0$.
That is, we have
\begin{equation}\begin{aligned}
{\rm NDP} &: \quad  h_l (\bm{k}_D)= 0  , \quad \textrm{for all $l$}, \\
{\rm EP} &: \quad  h_l (\bm{k}_D)  \neq 0 , \quad \textrm{for some $l$}, 
\quad \textrm{and}  \quad
\Delta_f(\bm{k}_D)=0 .
\end{aligned}
\end{equation}
For the NDP this gives six conditions (since $h_l$ are complex functions), while for the EP this gives only two conditions. 
From this we conclude that EPs are stable in 2D BZs, whose two independent momenta can be adjusted such that two
conditions are simultaneously satisfied. NDPs, on the other hand, are unstable in 2D BZs and only become stable   in 6D BZs, where
six momenta can be adjusted independently to satisfy the six conditions. 
These are, however, only the minimal number of conditions. Depending on the order of the EP or NDP (i.e., its discriminant number $\nu$) the number
of conditions  might be even higher.  
So let us ask how many conditions must be satisfied for an EP or NDP of order $\nu$ to be realized?
To answer this question we consider EPs and NDPs separately and use again the generic two-band model~\eqref{twoband_zwei}.  \\

\subsection{Stability of EPs of order $\nu$}
	
To study the stability of EPs of order $\nu$, we assume that the two-band Hamiltonian~\eqref{twoband_zwei} has  a two-fold degenerate EP of order $\nu$ at $\bm{k}_D$. 
The discriminant~\eqref{disc_cond_gen_two_band}, which is zero at $\bm{k}_D$, can be expanded around $\bm{k}_D$ as
\begin{equation} 
\begin{aligned}
\Delta_f(\bm{k}_D+\delta\bm{k})&=\Delta_f^r(\bm{k}_D+\delta\bm{k})+i \Delta_f^i(\bm{k}_D+\delta\bm{k}) 
=
\sum_{n+m=1}^\infty (\alpha_{nm}^r  +i\alpha_{nm}^i)\delta k_x^n  \delta k_y^m , \label{Expansion1} 
\end{aligned}\end{equation}
where $\alpha_{nm}^{r/i}$   are real numbers.
Let us first assume that the linear terms in the expansion of the discriminant are non-vanishing (i.e., $\alpha^{r/i}_{10}  \ne 0$ and $\alpha^{r/i}_{01}  \ne 0$)
and define the matrix    
\begin{equation}
\alpha:=\left(\begin{array}{cc}{\alpha_{10}^r} & {\alpha_{01}^r} \\ {\alpha_{10}^i} & {\alpha_{01}^i}\end{array}\right) ,
\end{equation}
whose entries are all real and non-zero. 
The discriminant number measures the number of windings in the phase structure of $\Delta_f ( \bm{k} )$ around  $\bm{k}_D$, which, up to first order in $\delta \bm{k}$,
is determined by the matrix $\alpha$. When $\det \alpha =0$  the phase winding is zero giving $\nu=0$, since the complex phases  $\alpha^r_{nm} + i \alpha^i_{nm}$ in the 
two $\bm{k}$ directions (10) and (01) are identical.  This corresponds to an EP of order $\nu=0$, which is realized when the three conditions
$\textrm{Re} \Delta_f ( \bm{k}_D)  = 0$, $\textrm{Im} \Delta_f ( \bm{k}_D ) = 0$, and $\det \alpha =0$ are satisfied (first row of Table~\ref{T2}).
When $\det \alpha \ne 0$, however, the phase winds $2 \pi$ around $\bm{k}_D$, yielding $| \nu| =  1$. 
This corresponds to an EP of order $\nu =| 1 |$, for which only two conditions, $\textrm{Re} \Delta_f ( \bm{k}_D)  = 0$ and $\textrm{Im} \Delta_f ( \bm{k}_D )  = 0$,
must be satisfied (second row of Table~\ref{T2}).

For an EP of order $|\nu| = 2$, all four entries of $\alpha$ must be zero, as otherwise   terms linear in $\delta \bm{k}$ dominate in the expansion~\eqref{Expansion1}, giving winding $| \nu | \leq 1$. Similarly, for an EP or order $| \nu | = N$, all  terms $\alpha^{r/i}_{nm}$   with $m+n<N$ must be zero, as otherwise lower-order terms in the expansion~\eqref{Expansion1} dominate, yielding $| \nu | < N$. Hence, in order to realize an EP with $| \nu | = N$ the following conditions must be satisfied
\begin{eqnarray} \label{condition_EP_N}
\textrm{Re} \Delta_f ( \bm{k}_D)  = 0; \quad  \textrm{Im} \Delta_f ( \bm{k}_D ) = 0; \quad
 \alpha^r_{nm} = 0, \;  \alpha^i_{nm}=0,  \quad \textrm{for all $m+n<N$} .
\end{eqnarray}
These are in total $2+2 ( 2+3+...+N )=N(N+1)$ conditions  (third row of Table~\ref{T2}).
We note that these are necessary, but not always sufficient conditions for the existence of an EP with $| \nu | = N$. That is, the discriminant number $\nu$ depends
on the particular values of $\alpha^{r/i}_{nm}$, with $n+m=N$.  In fact, in general one can derive a phase diagram as a function of $\alpha^{r/i}_{nm}$  (with $n+m=N$),
which  contains regions where $|\nu| = N$, but also regions where $|\nu| = N -2l$ with any positive integer $l$.

\subsection{Stability of NDPs of order $\nu$}

To analyze the stability of NDPs of order $\nu$, we assume that  the generic two-band Hamiltonian~\eqref{twoband_zwei} 
has a two-fold degenerate NDP of order $\nu$ at $\bm{k}_D$.    
The three coefficients of the Pauli matrices in Eq.~\eqref{twoband_zwei}, which are zero at $\bm{k}_D$, can be
expanded around  $\bm{k}_D$ as
\begin{equation}
h_{l}(\bm{k}_D+\delta\bm{k})=       \sum_{ n+m=1}^\infty ( \beta_{lnm}^r + i \beta_{lnm}^i  ) \delta k_x ^n\delta k_y^m ,
\label{Exp}
\end{equation}	
with $l=x,y,z$.
Here, the coefficients satisfy $\beta^{r/i}_{lnm} =  \beta^{r/i}_{lmn} $, but are otherwise mutually independent. 
We note that to analyze the stability of NDPs, expansion~\eqref{Exp} must be used instead of expansion~\eqref{Expansion1}, since 
for NDPs the $\alpha_{nm}^{r/i}$  are not independent of each other.  
Inserting expansion~\eqref{Exp}  into Eq.~\eqref{disc_cond_gen_two_band}, we get
\begin{equation} \label{Expansion2}
\begin{aligned}
\Delta_f(\bm{k}_D+\delta\bm{k})
&= 
 4 \sum_{l=x,y,z} \Big [ \sum_{ n+m=1}^\infty ( \beta_{lnm}^r + i \beta_{lnm}^i  ) \delta k_x ^n\delta k_y^m \Big ]^2. 
\end{aligned}
\end{equation}
If all $\beta^{r/i}_{lnm}$ are nonzero in the above equation, the NDP has in general discriminant number $| \nu | = 2  $, as the lowest order
terms are quadratic in $\delta \bm{k}$. 
For an NDP of order $|\nu | = 2N$ ($N >1$), all terms
$\beta^r_{lnm}$ and $\beta^i_{lnm}$ with $m+n <  N$
must be zero, such that the lowest order terms in Eq.~\eqref{Expansion2} are of order $2N$ in $\delta \bm{k}$. 
In other words, the derivatives $(\partial / \partial \bm{k})^q h_{l}(\bm{k}_d)$ must be zero for  
all $q=1,2, \ldots, N-1$ and $l=x,y,z$, which gives $6(2+3+\ldots+N)= 3 (N+2)(N-1)$ conditions. 
Together with the requirement that $h_l ( \bm{k}_D ) = 0$ for $l=x,y,z$, this gives 
$3N(N+1)$ conditions to realize an NDP with $|\nu| = 2N$ (sixth row of Table~\ref{T2}).

For an NDP with odd $| \nu| = (2N+1)$ ($N>0$), the lowest order term in expansion~\eqref{Expansion2} must be of order   $\delta \bm{k}^{2N+1}$.
Hence, in Eq.~\eqref{Expansion2} all terms $\beta_{lnm}^{r/i}$ with $m+n < N$ must be zero (giving $3N(N+1)$ conditions), such that Eq.~\eqref{Expansion2} 
contains terms of order  $\delta \bm{k}^{2N}$, $\delta \bm{k}^{2N+1}$, and higher. 
The requirement that also all terms of order $\delta \bm{k}^{2N}$ vanish, gives another $2(2N+1)$ conditions.
Hence, in total we have $3N(N+1)+2(2N+1)=(3N+1)(N+2)$ conditions to from NDPs of order $| \nu| = (2N+1)$ (seventh row of Table~\ref{T2}). 

As before, we note that these conditions are necessary, but not always sufficient to from NDPs of order $\nu$. That is, $\nu$ depends in general
on the particular values of $\beta_{lnm}^{r/i}$, giving a phase diagram with regions where $|\nu|$  is  2N (or $2N+1$), and other regions
where $|\nu| = N -2l$ with any positive integer $l$. For example, for $| \nu | = 3$ we find that depending on the values of $\beta_{lnm}^{r/i}$ both
NDPs with $| \nu | = 3$ and $| \nu | = 1$ can form. Indeed, this is the simplest way to form NDPs with $| \nu | = 1$, since 
expansion~\eqref{Expansion2} does not contain linear terms. Hence for NDPs with $| \nu | = 1$ a minimum of twelve conditions
must be satisfied (fifth row of Table~\ref{T2}).

\vspace{0.8cm}

The results of this section are summarized in Table~\ref{T2}. We conclude that in two dimensions only EPs with $| \nu |= 1$ are  stable,  whereas EPs of higher order and  NDPs of arbitrary order can be split into EPs  with $|\nu| =1$ by generic petrubations.
To form NDPs or EPs of higher order, additional conditions must be satisfied, which might be guaranteed by the presence of symmetries.
Table \ref{T2} shows that the number of conditions need to from an EP or NDP with discriminant number $\nu$ scales with $\nu^2$. 
We also observe that  for $\nu > 3$, it is more difficult to from EPs rather than NDPs, since a larger number of constraints must be satisfied.

\section{Example of 3D lattice Hamiltonian with surface EP\MakeLowercase{s} and FP\MakeLowercase{s} that break doubling theorems }  \label{surface violation}

In this section we present two examples to study 3D Hamiltonians, which exhibit anomalous EPs and FPs on their 2D surfaces violating the doubling theorems. At the same time, the Fermion doubling theorems for EPs and FPs still hold in the entire system, including the surfaces and the bulk. 

In the presence of the anomalous surfaces, the bulk Hamiltonians have to be {\it gapless} when reflection symmetry or inversion symmetry is preserved. The first example demonstrates the gapless bulk with reflection symmetry. However, when reflection symmetry and inversion symmetry are broken, the bulk Hamiltonian with the anomalous surfaces can be gapped. We use the second example to show that a 3D Hamiltonian can have both a topological point gap in the bulk and  anomalous FPs on its surface. 

\subsection{Anomalous EPs and FPs emerge in a 3D reflection-symmetric Hamiltonian} \label{reflection example}

We demonstrate that these anomalous surface EPs  and FPs
are accompanied by bulk exceptional lines and bulk Fermi lines, respectively, when reflection symmetry is preserved. 
The 3D reflection-symmetric Hamiltonian is defined on the cubic lattice and given by 
\begin{align} \label{3DHam_examp_supp}
H_{\rm{3D}}(\bm{k})=&(m(\hat{\bk} )+\cos k_z)\Gamma_0+\sqrt{2} \sin( k_x+\pi/4) \Gamma_1 
 + (i+\sin k_y ) \Gamma_2 + \sin k_z \Gamma_3-gi\Gamma_1\Gamma_2,
\end{align}
where $m(\hat{\bk})= \cos k_x + \cos k_y -2.7$ and   $g=0.2$.
The gamma matrices are defined as $\Gamma_0=\rho_3 \sigma_0$ and  $\Gamma_i=\rho_1\sigma_i$ and obey the anticommutation relations $\{\Gamma_\alpha, \Gamma_\beta\} = 2\delta_{\alpha \beta} \bm{I}_{4\times 4}$. 
We note that in the absence of the last term, $g i\Gamma_1\Gamma_2$, the energy bands of $H_{\rm{3D}}(\bm{k})$ are two-fold degenerate in the entire BZ.
The term $g i\Gamma_1\Gamma_2$ splits the degeneracy and gives rise to EPs at some isolated points or lines in the BZ.
Hamiltonian $H_{\rm{3D}}(\bm{k})$ is reflection symmetric under $z \to -z$, i.e., it satisfies 
$R_z H_{\rm{3D}}(k_x,k_y,-k_z) R_z^{-1} =H_{\rm{3D}}(\vec{k})$, with the reflection operator $R_z= \rho_3 \sigma_3$.
This reflection symmetry relates the surface Hamiltonians on the ($001$) and ($00\bar{1}$) faces to each other. 
Moreover, it guarantees that the bulk spectrum of $H_{\rm{3D}}(\bm{k})$ does not depend on the boundary conditions along the $z$ direction (open or periodic)
and that there is no skin effect~\cite{2020arXiv200302219Y}. 

\begin {figure}[t]
\centerline{\includegraphics[height=4.5cm]{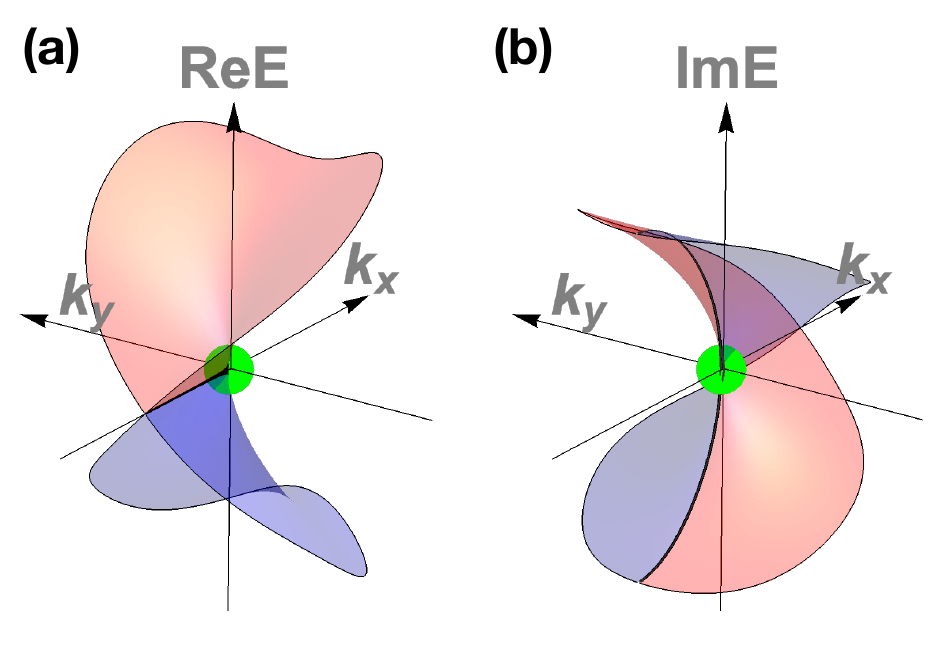}}
\caption{Complex energy dispersion of the surface Hamiltonian~\eqref{surface Hamiltonian} in the region $|m(\hat{\bk})| < 1$. 
The surface spectrum has only one EP (green dot) located at $(-0.02,0) $, which violates the doubling theorem. 
There are two branch cuts (black lines) emanating from the surface EP.}
	\label{Surface} 
\end{figure}

To compute the surface Hamiltonian and surface states on the ($001$) and ($00\bar{1}$) faces, we perform a Fourier transform of Eq.~\eqref{3DHam_examp_supp}
and obtain 
\begin{align}  \label{z open}
\hat{H}_{\rm{3D}} (\hat{\bm{k}} ) 
= &
 \sum_{1 \leq z \leq L } \Big\{ c^\dagger_z \frac{\Gamma_0+i\Gamma_3 }{2}  c_{z+1} +c^\dagger_z  \frac{\Gamma_0-i\Gamma_3 }{2} c_{z-1}  \nonumber \\
&+
c^\dagger_z \big [ m(\hat{\bk} )\Gamma_0 +\sqrt{2} \sin( k_x+\pi/4) \Gamma_1 + (i+\sin k_y ) \Gamma_2-gi\Gamma_1\Gamma_2]  c_z   \Big \}, 
\end{align} 
where  $c^\dagger_z$ and  $c_z$ are creation and annihilation vector operators, respectively, $L$ is the number of layers in the $z$ direction, and $\hat{\bm{k}} = (k_x, k_y)$.
Assuming open boundary conditions in the $z$-direction and using the Harper's equation, we solve for the surface states on the ($00\bar{1}$) face, which gives
\bee  \label{surf_states_z_1}
\ket{1}= \sum_{1\le z \le L} \alpha_{ z} (i\ 0\ 1\ 0)^T c^\dagger_z \ket{0},
\quad
\ket{2}= \sum_{1\le z \le L} \alpha_z (0\ 1\ 0\ i)^T c^\dagger_z \ket{0},
\ee
where $\alpha_z$ 
obeys the recurrence relation $\alpha_{z+1}=-m(\hat{\bk} )\alpha_z $. In the region defined by $|m( \hat{\bk} )|<1$, the above two states are localized at $z=1$, i.e., 
at the ($00\bar{1}$) face, while outside this region there are no surface states. 
By projecting the Hamiltonian onto the subspace spanned by the two surface states~\eqref{surf_states_z_1}, we 
obtain an effective  Hamiltonian describing the ($00\bar{1}$) surface, which is given by
\bee \label{surface Hamiltonian}
\mathcal{H}_{\rm{surf}}(\hat{\bk})=\sqrt{2} \sin( k_x+\pi/4) \sigma_x + (i+\sin k_y ) \sigma _y + g \sigma_z,  
\ee
with surface spectrum $E^\pm_{\rm{surf}} = \pm \sqrt{ \sin (2 k_x )  + \tfrac{1}{2} + g^2 - \tfrac{1}{2}  \cos( 2 k_y) + 2 i \sin k_y } $.
This surface Hamiltonian has a single EP located at $(-0.02,0)$ with energy $E=0$ and discriminant number $\nu=-1$, as illustrated in Fig.~\ref{Surface}.
This clearly violates the doubling theorem, since there is no other EP that could compensate the nonzero $\nu$ of the EP at $(-0.02,0)$.
Assuming chemical potential $\mu =0$, surface Hamiltonian~\eqref{surface Hamiltonian} exhibits also only one FP, namely at $(-0.02,0)$, with winding number $W=-1$,
which violates the doubling theorem for FPs.

\begin {figure}[t]
\centerline{\includegraphics[height=7.5cm]{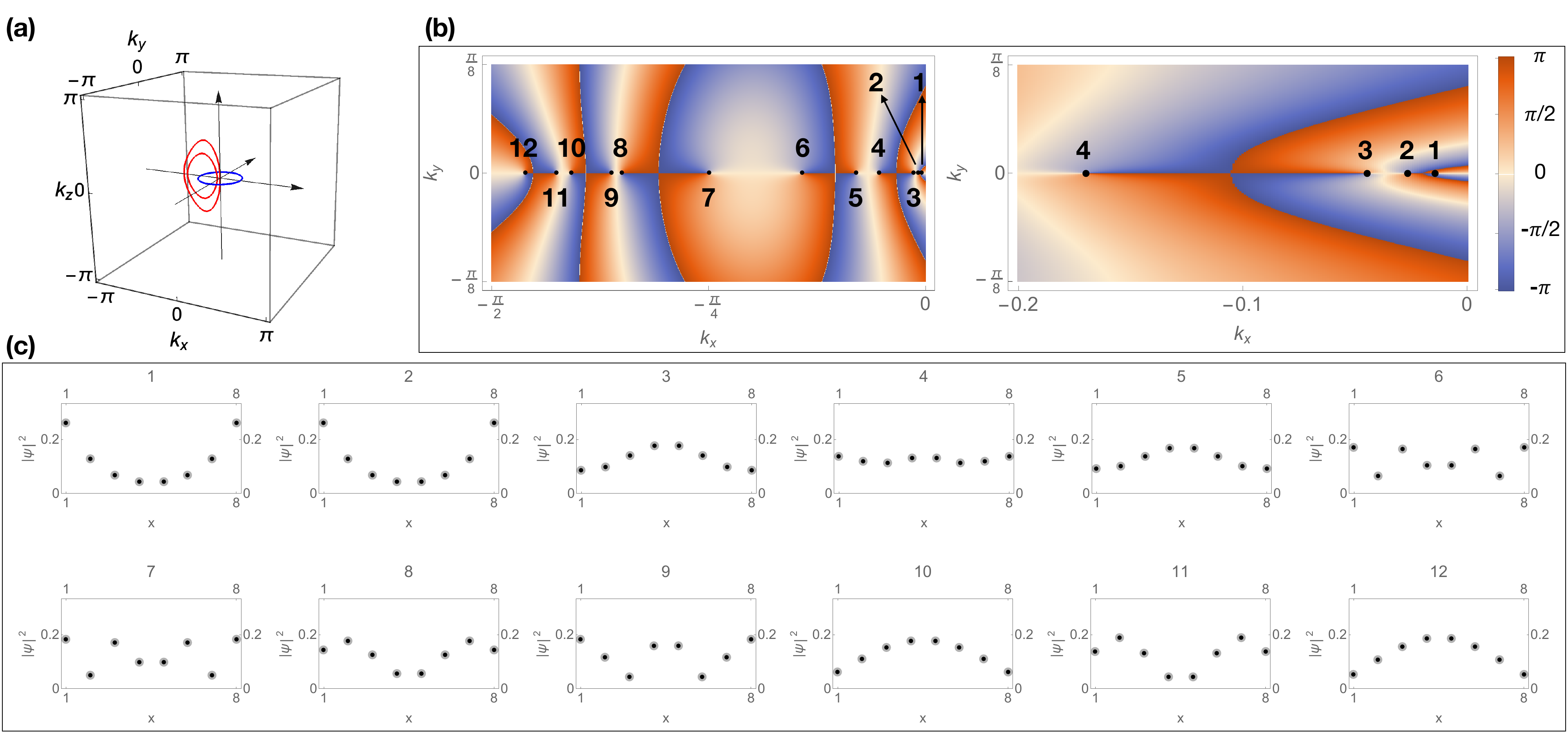}}
\caption{  	\label{boundary_EP}
Exceptional points/lines and Fermi points/lines of the 3D lattice Hamiltonian~\eqref{3DHam_examp_supp}. 
Panel (a) shows the exceptional rings and Fermi rings of Hamiltonian~\eqref{3DHam_examp_supp} in the 3D BZ. The two red rings, located within the $k_y=0$ plane, 
have nonzero $\nu$ and $W$, which cancel the topological charges of the surface EPs and FPs.
The blue ring, within the $k_z=0$ plane, consists of two overlapping exceptional rings, each of which has two-fold degeneracy. The topological charges
of these two overlapping rings cancel each other. 
The panels in (b) present the phase structure of $\det \hat{H}_{\rm{3D}} (\hat{\bm{k}} )$ with $L=8$ layers in the 2D BZ. Vortices in the phase structure correspond
to FPs (and EPs), indicated by the black dots. We note that  FPs and EPs are located at exactly the same position. The right panel is a zoom-in of the left panel around $\hat{\bf k} =0$. 
The panels in (c) show the wave function amplitudes $| \psi |^2$ of the states at the FPs/EPs. The first two FPs/EPs states are located at the surface, while 
the remaining ten are bulk states. }
\end{figure}

Let us now check whether there exists also an EP (FP) at the ($001$) surface, which could potentially cancel the discriminant number of the EP (winding number of the FP) at the ($00\bar{1}$) surface. 
For this purpose, we
repeat similar steps as above 
and obtain the two states
\bee \label{surf_states_top} 
\ket{3}= \sum_{1\le z \le L} \beta_z (i\ 0\ -1\ 0)^T c^\dagger_z \ket{0},
\quad
 \ket{4}= \sum_{1\le z \le L} \beta_z (0\ -1\ 0\ i)^T c^\dagger_z \ket{0},
\ee
where $\beta_z$ obeys the recurrence relation $\beta_{z-1}=-m(\hat{\bk})\beta_z $.
For $| m(\hat{\bk}) | < 1$, the above two states are localized at $z=L$, i.e., at the ($001$) face, whereas outside this region no surface states exist. 
By projecting the Hamiltonian onto the subspace spanned by the two surface states~\eqref{surf_states_top} we obtain
an effective Hamiltonian for the ($001$) surface, which is identical to Eq.~\eqref{surface Hamiltonian}.
The reason for why the two surface Hamiltonians are identical is the reflection symmetry $R_z$, which 
lets $z\rightarrow L -z +1$, thereby  mapping the two surfaces onto each other.
Indeed, we find that $R_z$ maps the states $\ket{1}$ and $\ket{2}$  onto the states $\ket{3}$ and $\ket{4}$,
leading to identical surface Hamiltonians.
As a consequence, both the ($00\bar{1}$) surface and the ($001$) surface have each one EP (FP) with 
discriminant number $\nu = -1$ (winding number $W=-1$).
Therefore, the violation of the doubling theorems by the ($00\bar{1}$) surface is not compensated by the ($001$) surface.
 
This is in apparent contradiction with the doubling theorems~\eqref{FP thm} and~\eqref{nogo degeneracy}, since we can view Hamiltonian~\eqref{z open}  with open boundary conditions
in the $z$ direction  as a periodic 2D systems with $L$ sites per unit cell. 
This contradiction is resolved by the existence of exceptional lines (Fermi lines) in the 3D bulk spectrum, such that the sum over all discriminant numbers (winding numbers)
is zero. 
Indeed, the bulk Hamiltonian $H_{\rm{3D}} ( {\bk} )$ with periodic boundary conditions in all three directions possesses  two exceptional rings within 
the $k_y=0$ plane [red rings in Fig.~\ref{boundary_EP}(a)]. These exceptional rings have nonzero discriminant numbers $\nu$, which cancel the topological
charges of the surface EPs.
For chemical potential $\mu = 0$, the two red rings in Fig.~\ref{boundary_EP}(a)  are simultaneously also Fermi lines, whose winding numbers cancel the winding numbers of the surface FPs.
We note that the bulk Hamiltonian $H_{\rm{3D}} ( \bk )$ exhibits, in addition, two overlapping exceptional rings within the $k_z=0$ plane [blue ring in Fig.~\ref{boundary_EP}(a)]. The topological charges of these two overlapping rings cancel each other out.

To see how precisely the topological charges in the bulk and at the surface neutralize each other, we now consider Hamiltonian $\hat{H}_{\rm{3D}} (\hat{\bm{k}} )$, Eq.~\eqref{z open}, with a fixed number of $L$ layers and open boundary conditions along the $z$ direction. For a small number of layers, the top and bottom surface states couple.
By sandwiching Hamiltonian~\eqref{3DHam_examp_supp} with the states~\eqref{surf_states_z_1} and~\eqref{surf_states_top}, the coupling between the two surfaces
can be described by
\bee
\mathcal{H}_{\rm{2surf}} (\hat{\bk} ) =
\bma 
\mathcal{H}_{\rm{surf}}(\hat{\bk}) & \delta(\hat{\bk}) \sigma_z \\
\delta(\hat{\bk}) \sigma_z & \mathcal{H}_{\rm{surf}}(\hat{\bk})
\ita, \label{two surfaces}
\ee
where $\delta(\hat{\bk}) = (1- m^2(\hat{\bk}))/(m^{-L}(\hat{\bk})-m^L(\hat{\bk}))$. 
From this equation we can see that the hybridization of the top and bottom surfaces leads to
a small shift (i.e., a splitting) of the surface EPs (FPs), but does not change their discriminant number $\nu$ (winding number $W$).
For concreteness we now consider $H_{\rm{3D}} ( {\bk} )$  with $L =8$ layers in the $z$-direction and numerically determine its FPs and EPs.
The phase structure of $\det  \hat{H}_{\rm{3D}} ( \hat{\bk} )$ is shown in Fig.~\ref{boundary_EP}(b). We observe twelve vortices
in the phase structure which correspond to twelve FPs and EPs. (The locations of the FPs and EPs coincide for this particular example.)
The lines across which the color changes abruptly, from red to blue, correspond to branch cuts. 
In Fig.~\ref{boundary_EP}(c) we present the wave function amplitudes of the states at the FPs (EPs), which show that the first
two FPs (EPs), which are closest to $k_x=0$, are surface FPs (EPs), while the other ones are bulk FPs (EPs).
These bulk FPs (EPs) originate from the discretization of the Fermi lines (exceptional lines) shown in Fig.~\ref{boundary_EP}(a).
We find that the two surface FPs (EPs) with Nos.~1,2 have $W=-1$ ($\nu=-1$), the four bulk FPs (EPs) with Nos.~3 to 6 have also $W=-1$ ($\nu=-1$),
whereas the remaining bulk FPs (EPs) (Nos.~7 to 12) have $W=+1$ ($\nu=+1$).
Thus, the winding number (discriminant number) of all FPs (EPs) cancel out, such
that the doubling theorems are satisfied.

\subsection{Bulk energy point gap with anomalous surface FPs}

\begin {figure}
\centerline{\includegraphics[width=1\linewidth]{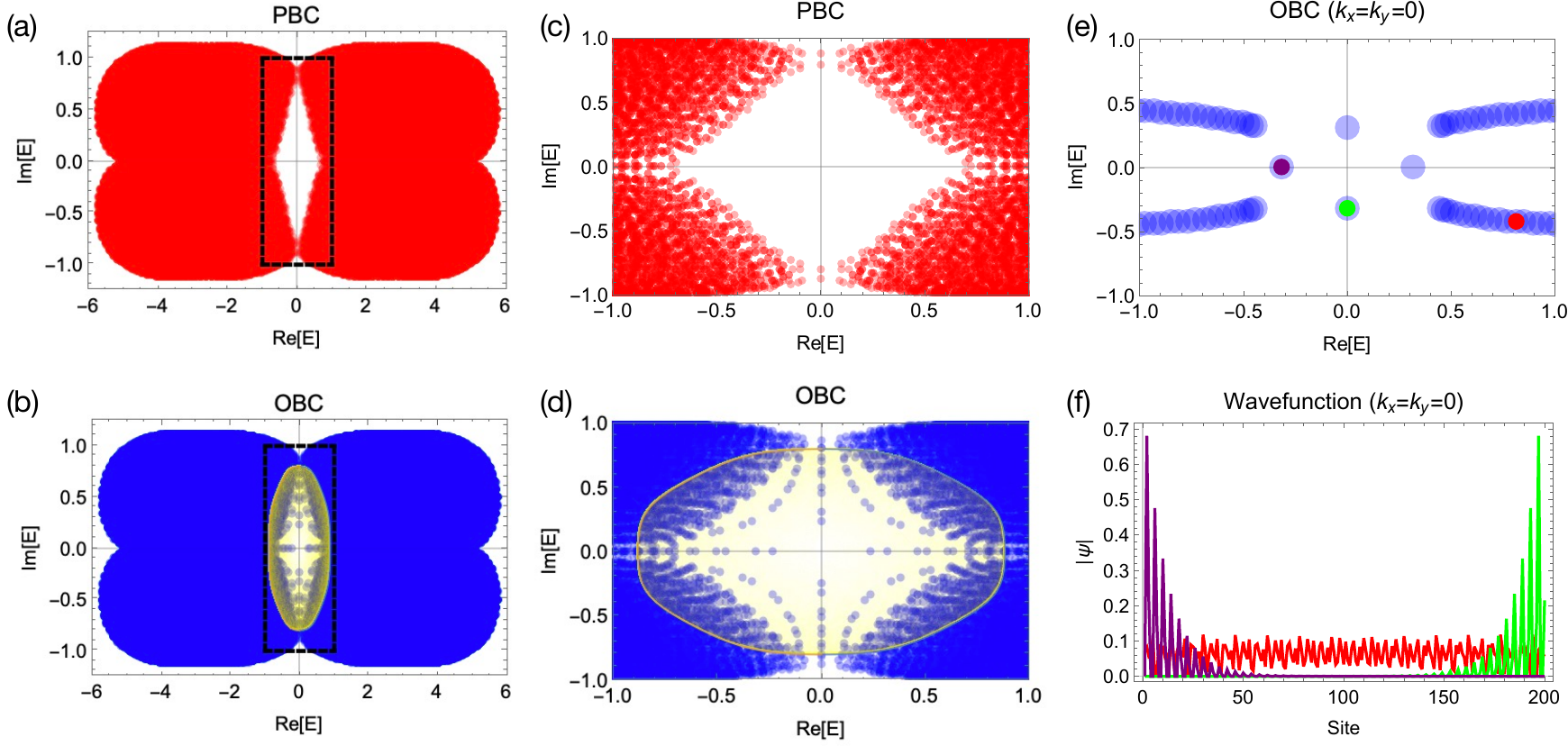}}
\caption{ The energy spectra in the complex plane for periodic and open boundary conditions and the spatial distribution of the selected states. (a,b) show the energy spectrum of $H_{\rm{PBC}}(\bm{k})$ for all $\bk$, which exhibits a bulk point gap around zero energy. (c,d) represent the energy spectrum of $\hat{H}_{\rm{OBC}} (\hat{\bm{k}} )$ with $L=50$ in the complex plane. 
The points within the bulk point gap (yellow region) correspond to surface states. Note that panels (c) and (d) are zoom-ins of panels (a) and (b)  around zero energy,  as indicated
by the dashed region in (a) and (b), respectively. 
Panel (e) shows the energy spectra for  $k_x=k_y=0$  held fixed and open boundary conditions. The four points around zero energy (green and violet) are surface states, whose spatially resolved
wavefunction amplitudes are shown in panel (f). The points further away from zero energy in (e) (red dot) correspond to bulk states, with the wavefunction amplitude shown in (f). 
 \label{FS1}}
\end{figure}

By removing reflection symmetry and inversion symmetry, in principle a 3D non-Hermitian Hamiltonian can possess an anomalous FP on its surface and an energy point gap in its bulk. 
An example of such a Hamiltonian has been proposed in the literature recently~\cite{2020arXiv200801090D}. Here we propose a similar model to demonstrate our doubling theorem and the application of the FP invariants. The Bloch Hamiltonian breaking reflection symmetry and inversion symmetry reads  
\begin{equation}
H_{\rm{PBC}}(\bm{k})=\left(
\begin{array}{cccc}
0 & 0 & s_z -i (c_z+ m(\hat{\bk})) & \Delta  \\
0 & 0 & \Delta  & -s_z -i (c_z+m(\hat{\bk})) \\
s_z +i (c_z+m(\hat{\bk}))  & \sin (k_x+\gamma)-i s_y & 0 & 0 \\
\sin (k_x-\gamma)+i s_y & -s_z +i (c_z+m(\hat{\bk}))  & 0 & 0 \\
\end{array}
\right),
\label{FPmodel}
\end{equation}
where $m(\hat{\bk})=\cos k_x + \cos k_y - 2.7$, $\gamma=1/10$, $s_{y/z}=\sin k_{y/z}$, and $c_{z}=\cos k_{z}$. This Hamiltonian preserves chiral symmetry, since it obeys $SH_{\rm{PBC}}(\bm{k})S^{-1}= - H_{\rm{PBC}}(\bm{k})$ with the operator $S=\rho_3 \sigma_0$. The bulk spectrum in the complex plane  shows an energy point gap near zero energy, 
see Fig.~\ref{FS1}(a,b).
Now consider the open boundary condition in the $z$ direction. We can then treat the system as an effective 1D model by fixing $k_x,k_y$. Since the Hamiltonian obeys 
\begin{equation}
\mathcal{\bar{T}}\mathcal{H}_{\rm{PBC}}(k_x,k_y,k_z)\mathcal{\bar{T}}^{-1}=\mathcal{H}_{\rm{PBC}}(k_x,k_y,-k_z),
\end{equation}
where $\bar{\mathcal{T}}=\tau_x\sigma_x\mathcal{K}^t$ and $\mathcal{K}^t$ is the transpose operator, spinless anomalous time-reversal symmetry is preserved~\cite{PhysRevX.9.041015}. This symmetry rules out the skin effect in the $z$ direction~\cite{PhysRevX.9.041015,2020arXiv200302219Y}. Therefore, the  spectra in open and periodic boundary conditions are almost identical,
with the only difference between the two originating from the surface states.
 To examine the difference, we rewrite the Hamiltonian in the $z$ real space in the form of the second quantization
\begin{align}  \label{HFPforZ}
\hat{H}_{\rm{OBC}} (\hat{\bm{k}} ) 
= &
 \sum_{1 \leq z \leq L } \Big\{ c^\dagger_z \frac{\Lambda_0+i\Lambda_3 }{2}  c_{z+1} +c^\dagger_z  \frac{\Lambda_0-i\Lambda_3 }{2} c_{z-1}  \nonumber \\
&+c^\dagger_z 
\left(
\begin{array}{cccc}
0 & 0 & -i  m(\hat{\bk}) & \Delta  \\
0 & 0 & \Delta  & -i m(\hat{\bk}) \\
i m(\hat{\bk})   & \sin (k_x+\gamma)-i s_y & 0 & 0 \\
\sin (k_x-\gamma)+i s_y & i m(\hat{\bk})  & 0 & 0 \\
\end{array}
\right)
c_z  \Big \},
\end{align} 
where $\Lambda_0=\rho_2\sigma_0$, $\Lambda_3=\rho_1 \sigma_3$, $c^\dagger_z$ and  $c_z$ are creation and annihilation vector operators, respectively, $L$ is the number of layers in the $z$ direction, and $\hat{\bm{k}} = (k_x, k_y)$. Furthermore, the $(001)$ and $(00\bar{1})$ faces are located {\clb at $z=1$ and $z=L$, respectively}. For the Hamiltonian in the open boundary condition, as shown in Fig.~\ref{FS1}(d,e), some energy states appear inside the bulk point gap in the complex energy plane. In particular, Fig.~\ref{FS1}(f) shows that the states inside the bulk gap are located at the ($001$) and ($00\bar{1}$) faces. 

\begin {figure}[tb!]
\centerline{\includegraphics[width=1\linewidth]{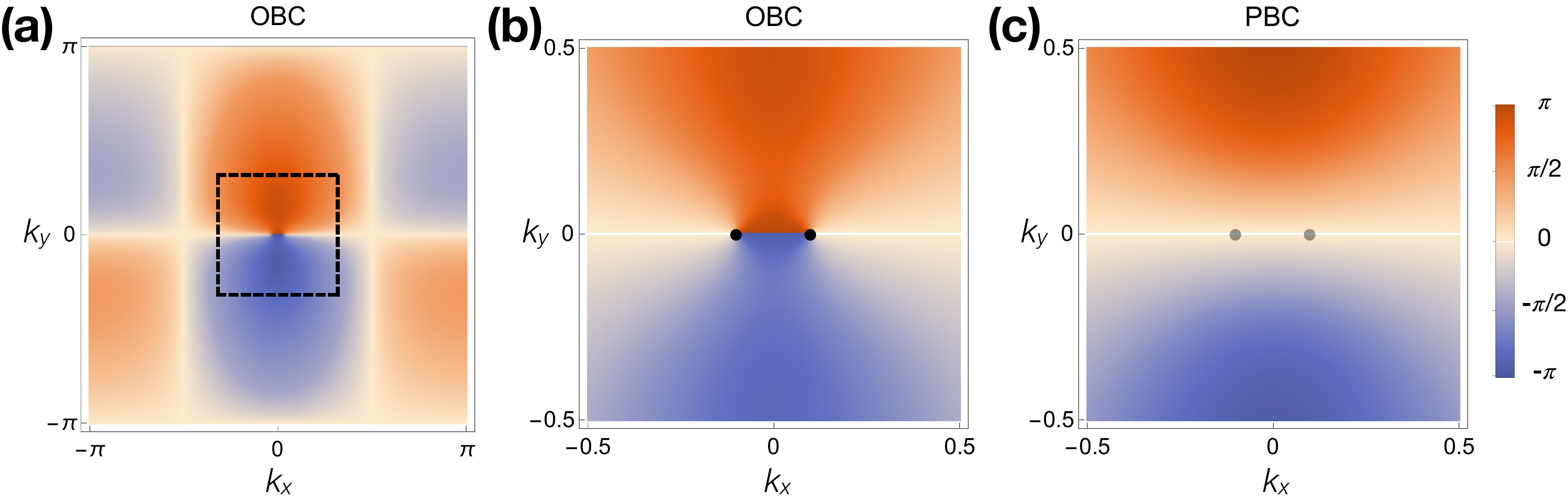}}
\caption{
Phase plots of (a), (b) the determinant of the Hamiltonian with open boundary conditions  $\det[\hat{H}_{\rm{OBC}} (\hat{\bm{k}} )]$ (with $L=25$), 
and (c) the product of the determinant of the Hamiltonian with periodic boundary conditions  $\prod_{k_z} \det[H_{\rm{PBC}}(\bm{k}) ]$,
as a function of $k_x$ and $k_y$.
The black points in (b) represent surface Fermi points   on the (001) surface, while the grey points in (c) indicate the absence of Fermi points in the bulk spectrum (i.e., for periodic boundary conditions). 
	\label{FS2}}
\end{figure}

We are interested in FPs at $E=0$, since chiral symmetry guarantees that zero energy is located in the center of the bulk point gap. Since the zero energy is not an eigenvalue of the PBC Hamiltonian $H_{\rm{PBC}}(\bk)$, its determinant, which the product of the eigenvalues, does not vanish for any $\bk$
\begin{equation}
\det[H_{\rm{PBC}}(\bk)]\neq 0.
\end{equation}	
On the other hand, the Hamiltonian (\ref{HFPforZ}) possesses FPs at $E=0$ at the open boundaries. In order to show the presence of the FPs with non-zero charge, we calculate the real and imaginary parts of the determinant for the OBC Hamiltonian 
\begin{equation}
\left( \Re[\det[\hat{H}_{\rm{OBC}} (\hat{\bm{k}} ) ]],~\Im[\det[\hat{H}_{\rm{OBC}} (\hat{\bm{k}} ) ]]\right). 
\label{SEq4}
\end{equation}
As shown in Fig.~\ref{FS2}(a,b), the phase-plot of $\det[\hat{H}_{\rm{OBC}} (\hat{\bm{k}} )]$ shows that the determinant vanishes at two points located at $(\pm 0.1 ,0)$ as phase singularities [black points in Fig.~\ref{FS2}(b)], which represent the FPs of the OBC Hamiltonian. The phase winding in the plots indicates the winding number $W=\mp 1$ at $(\pm 0.1,0)$. 

For the comparison, we also plotted the phase of $\prod_{k_z} \det[H_{\rm{PBC}}(\bm{k}) ]$ in PBC. The phase-plot in Fig.~\ref{FS2}(c) shows that there is no phase singularity in the entire 2D BZ, which indicates the absence of   FPs.  By comparing the two phase-plots with different boundary conditions, we can conclude that the two FPs stems from the surface states. \\

To understand the physics of the anomalous surface, we analyze the surface states and then obtain the effective surface Hamiltonian from the projection of the bulk Hamiltonian. By solving the Harper's equation in the lattice Hamiltonian (\ref{HFPforZ}), the surface states on the $(00\bar{1})$ surface near $z=1$ are given by  
\bee  
\ket{1}= \sum_{1\le z \le L} \alpha_{ z} (0\ 1\ 0\ 0)^T c^\dagger_z \ket{0},
\quad
\ket{2}= \sum_{1\le z \le L} \alpha_z (0\ 0\ 1\ 0)^T c^\dagger_z \ket{0},
\ee
where $\alpha_z$ normalizes the states and obeys the recurrence relation $\alpha_{z+1}=-m(\hat{\bk} )\alpha_z $. As $|m(\hat{\bk} )|<1$, the states are localized near the surface. By projecting the Hamiltonian (\ref{HFPforZ}) into these two surface states, the effective surface Hamiltonian is written as 
\bee
H_{\rm surf}^{00\bar{1}}(\hat{\bk})=
\bma 0 & \Delta \\
\sin (k_x + \gamma) - i s_y & 0 
\ita.
\ee
Therefore, on the surface the only FP with $E=0$ appears at $(-\gamma,0)$ and is also the only EP. Let us put the focus on the FP. By use of Eq.~\eqref{invariant FP} we find that the single FP on the surface has non-zero winding number ($W=1$), thereby breaking the FP doubling theorem as an anomaly.  Being different from the previous example (\ref{reflection example}), the energy of the surface FP is present in the bulk point gap. Since the surface states appear in the range of $|m(\hat{\bk} )|<1$ and the surface energy spectrum is given by $E_\pm = \pm \sqrt{\Delta (\sin (k_x + \gamma) - i s_y)}$, the surface spectrum covers the entire area of the bulk point gap as shown in Fig.~\ref{FS1}(d). 

The discussion can be easily extended to the $(001)$ face. Likewise, the surface states on the $(001)$ face near $z=L$ are given by  
\bee  
\ket{3}= \sum_{1\le z \le L} \alpha_{ z} (1\ 0\ 0\ 0)^T c^\dagger_z \ket{0},
\quad
\ket{4}= \sum_{1\le z \le L} \alpha_z (0\ 0\ 0\ 1)^T c^\dagger_z \ket{0},
\ee
The $(001)$ surface Hamiltonian is written as
\bee
H_{\rm surf}^{001}(\hat{\bk})=
\bma 0 & \Delta \\
\sin (k_x - \gamma) + i s_y & 0 
\ita.
\ee
The only FP with $E=0$ on the surface is located at $(\gamma,0)$ and carries $W=-1$ winding number. Thus, the two FPs with the opposite winding numbers on the two different surfaces are neutralized and the doubling theorem is still preserved in the entire system in the absence of the bulk FP. 

The violation of the non-Hermitian doubling theorem on the surface is an important indicator of the non-trivial bulk topology. 
Indeed, we find that the gapped bulk has non-trivial bulk topology, which is described by the 3D winding number given by~\cite{2020arXiv200801090D,PhysRevX.8.031079,PhysRevX.9.041015,Ghatak_2019} 
\bee 
\omega_{\rm{3D}} = \int_{\rm{BZ}} \frac{d^3 \bk}{24 \pi^2} \epsilon_{ijk}  {\rm{Tr}} [Q_i(\bk)Q_j(\bk)Q_k(\bk)], 
\ee
where $Q_i(\bk) = \big (  H_{\rm{PBC}}(\bm{k}) - \mu \big )^{-1} \partial_{k_i } \big (  H_{\rm{PBC}}(\bm{k}) - \mu \big )$ and $\mu$ is the chemical point inside the point gap. Since the surface FP is at zero energy, we choose $\mu=0$ inside the point gap so that the winding number is well-defined. For the Hamiltonian (\ref{FPmodel}) we find that $\omega_{\rm{3D}}=-1$, confirming the non-trivial bulk topology. 

\end{document}